\documentclass[%
reprint,
amsmath,amssymb,
aps,
]{revtex4-1}

\usepackage{graphicx}
\usepackage{dcolumn}
\usepackage{bm}
\usepackage{mathrsfs}
\usepackage{mathtools}

\begin{document}

    \preprint{APS/123-QED}

\title{\large\bf Majorana vortex modes in spin-singlet chiral superconductors with noncollinear spin ordering: Local density of states study}

\author{A.\,O.\, Zlotnikov}
\email{zlotn@iph.krasn.ru}

\affiliation{%
    Kirensky Institute of Physics, Federal Research Center KSC SB RAS, 660036 Krasnoyarsk, Russia}


\begin{abstract}

In the present study topologically nontrivial edge and vortex bound states are described in the coexistence phase of chiral spin-singlet superconductivity and noncollinear spin ordering on a triangular lattice in the presence of few (up to four) vortices. We consider the topological phase transition induced by the magnetic order between the phase hosting Majorana modes and the initial phase of the chiral d-wave superconductivity supporting non-Majorana modes which is also topologically nontrivial. The change of the excitation spectrum at the critical point is obtained in both cases of open and periodic boundary conditions in the presence of vortices. It is proved that zero energy Majorana modes localized at vortex cores are caused by noncollinear long-range magnetic ordering. Even though nearby excitation energies of subgap states including the edge-localized and vortex-localized states are very close to each other, the energy difference between different vortex bound states is an order of magnitude higher. This difference determines the energy gap for Majorana vortex modes separating them from other vortex bound states. It is found that even in the presence of noncollinear spin ordering its value can be estimated from the excitation energy of vortex bound states in the pure chiral d-wave state for the nonmagnetic case. By studying local density of states near the vortex cores the possibility to experimentally detect the described Majorana vortex modes by scanning tunneling microscopy is discussed. It is demonstrated that Majorana vortex modes and Majorana antivortex modes induced by noncollinear magnetism have different features in energy and spatially resolved density of states due to the chiral symmetry on the superconducting order parameter.

\end{abstract}

\maketitle


\section{\label{sec1}Introduction}

Majorana modes that are localized at the ends of quasi-1D quantum wires have been actively studied in recent years both theoretically and experimentally~\cite{lutchyn-10, oreg-10, shen-21, valkov-22}. It is believed that such spatially separated modes can be used for the realization of topologically protected quantum computations~\cite{nayak-08}. The similar modes with zero excitation energies propagate
along the edges in 2D and quasi-2D topological superconductors, thus preventing their use in braiding. On the other hand, well separated Majorana modes in 2D topological superconductors
can be localized on inhomogeneities or topological defects, such as Abrikosov vortices~\cite{read-00, ivanov-01, stern-04, gurarie-07, fu-08, elliott-15}, magnetic skyrmions~\cite{yang-16, zlotnikov-21}, corners in higher-order topological superconductors~\cite{volovik-10, zhu-18, zlotnikov-21}, and so on.

Initially the Majorana vortex modes were predicted in $p_x+ip_y$-wave superconductors \cite{read-00, ivanov-01, stern-04, gurarie-07, tewari-07, silaev-08, kraus-09, silaev-13, liu-15, akzyanov-16} in which the presence of a topologically nontrivial phase is caused by the triplet superconductivity. For superconductors with spin-singlet Cooper pairings an additional interaction, mixing electrons with different spins, should exist to induce topological superconductivity with Majorana modes. Therefore Majorana modes were predicted in the vortex state in spin-singlet superconductors with pronounced spin-orbit interaction~\cite{iskin-12, bjornson-13}, topological insulator~/~superconductor heterostructures~\cite{fu-08, chiu-11, akzyanov-14}, iron chalcogenide such as FeTe$_{0.55}$Se$_{0.45}$~\cite{chen-18, wang-18, zhu-20, pathak-21}, superconducting Dirac and Weyl semimetals~\cite{yan-20}.

It should be noted, that the subgap states including Majorana modes in 2D topological superconductors requires the absence of bulk gapless excitations in order to be realized.
For example, in 2D nodal superconductors, such as $d$-wave superconductors and extended $s$-wave superconductors with zeros of the order parameter on nodal lines in the Brillouin zone, the trivial  modes appear near zero energy. Such modes are caused by the nodal bulk excitations and prevent the formation of Majorana modes. Therefore a mechanism for opening an energy gap in the bulk spectrum of nodal superconductors should be suggested. Recently, it was shown that the gap is open for coupled structures of the nodal superconductor and a magnetic texture crystal~\cite{steffensen-21}. In such a system Majorana modes are pinned by a vortex in the magnetization, rather than the superconducting term. Another way to open the gap is the formation of a chiral superconducting state with the complex order parameter, such as, for example, the $d_{x^2-y^2}+id_{xy}$ state (in the following this symmetry is marked as $d_1+id_2$). It should be mentioned that these statements are correct for 2D systems with open boundary conditions in two directions of the lattice. If the strip geometry is considered, when open boundaries exist in one direction of the lattice and the periodic boundary conditions are applied in the other direction, it becomes possible to study Majorana end modes in nodal superconductors~\cite{sato-10}.

The chiral d-wave superconductivity is characterized by spin-singlet pairings and its order parameter vanishes at some points of the Brillouin zone (nodal points). Therefore the chiral superconductor has the gapped excitation spectrum under wide conditions in the homogeneous case with the periodic boundary conditions. The energy gap is closed only for certain parameters when the Fermi contour intersects the set of nodal points~\cite{zhou-08, valkov-17}. It is believed that the $d_1+id_2$-wave superconductivity can be formed in hexagonal lattice systems (as an example, Na$_x$CoO$_2$~\cite{baskaran-03, zhou-08} with triangular lattice and graphene with honeycomb lattice \cite{black-schaffer-14, kagan-15}).

It was proved over 20 years ago that the $d_1+id_2$ superconductor is the topological one \cite{volovik-97} with zero energy edge states which are not Majoranas due to the presence of the only spin-singlet component of the order parameter. The spectrum of vortex bound states in chiral superconductors with the order parameter $\Delta_p \propto (p_x + ip_y)^{N}$ is more complex: it has zero energy for odd $N$ and is gapped for even $N$ \cite{volovik-99}. Since $N=2$ for the $d_1+id_2$-wave superconductor, the vortex modes exist only at finite energy (in contrast to, for example, the $d_{xz}+id_{yz}$-wave pairing symmetry)~\cite{lee-16, volovik-16}.

It is known that a helical long-range magnetic ordering in superconducting structures with the spin-singlet component of pairing can also lead to topological superconductivity and, consequently, to the formation of Majorana modes. Among such structures are spin chains on the surface of a superconductor~\cite{choy-11, nadj-perge-13, klinovaja-13} and 2D magnetic superconductors~\cite{martin-12, lu-13, valkov-18, bedow-20, rex-20, zlotnikov-21, steffensen-22}. Helical magnet-superconductor heterostructures were realized experimentally, specifically, a chain of Fe atoms with 120-degree ordering on a superconducting Ir or Re substrate~\cite{menzel-12, kim-18}, as well as a Fe or Mn monolayer on the Re surface~\cite{palacio-morales-19, spethmann-20}. It is believed that helical ordering of Fe is caused by Dzyaloshinskii-Moriya interaction induced by the spin-orbit interaction in the substrate~\cite{menzel-12}. In the present study it is proposed that structures containing layers of antiferromagnets and superconductors both having a hexagonal lattice can also demonstrate topologically nontrivial phenomena. It is connected with the fact that noncollinear 120$^{\circ}$ order induced by the geometrical frustrations in the triangular lattice is well studied and confirmed experimentally (for example, in quasi-two-dimensional RbFe(MoO$_2$)$_4$~\cite{smirnov-17, soldatov-20}).

It should be noted that in the vast majority of the theoretical studies of topological superconductivity in magnetic superconductors the emphasis has been on the description of Majorana end modes rather than
Majorana vortex modes. The formation of Majorana modes on the vortex-antivortex pair previously was described in \cite{lu-13} for the coexistence state of chiral $d_1+id_2$ superconductivity and stripe magnetic ordering on a triangular lattice with periodic boundary conditions.

In the studied model the chiral superconductivity provides the gapped energy spectrum, while 120-degree spin ordering supports Majorana modes.
Unlike the approach of Ref. \cite{lu-13}, in the present study the different vortex structures, such as single vortex or antivortex, pair of vortices and vortex-antivortex pair, four vortices, are considered on the 2D lattice with the open boundary conditions to study the interplay between vortex and edge bound states. Consideration of open boundary conditions is connected with the fact that in real topological superconductors edges always exist (it can be also a boundary between topologically trivial and non-trivial phases) which can modify the excitation spectrum and complicate the idealized picture for vortex bound states under periodic boundary conditions. It is well known that in the presence of the single vortex in a topological superconductor one Majorana fermion appears in the vortex core, while the other Majorana fermion have to exist near the edges or at some distance from the vortex core which is determined by the magnetic length \cite{read-00, gurarie-07, bjornson-13, akzyanov-16}. As far as we know a study of the vortex bound states in the presence of few vortices in a limited lattice with edges is not carried out for any topological superconductor.

We also study the effects in the limited lattice when the topological phase transition occurs between two topologically nontrivial phases. One phase arises due to noncollinear magnetism and supports Majorana vortex modes, while in the other phase with even topological invariant zero vortex modes are not Majorana. The last phase exist also in the chiral d-wave state in the absence of magnetic ordering~\cite{volovik-97}.

It is widely accepted that Majorana bound states in vortices can be detected by scanning tunneling spectroscopy (for example, \cite{read-00, akzyanov-14, wang-18, pathak-21}). As it is known in the simplest approximation the tunneling conductance is proportional to the local density of states. In this article, we focus on the local density of states of the chiral $d_1+id_2$ superconductor with 120-degree spin ordering on a triangular lattice in the presence of Majorana modes localized at vortex cores. For the open boundary conditions it is shown that the excitation energy difference between Majorana vortex modes with zero energy and other subgap vortex bound states significantly exceeds the one between edge states.
The part of the results is inherent to the coexistent state of chiral superconductivity and noncollinear spin ordering.
It is found that even in the presence of noncollinear spin ordering the value of such energy difference is well estimated by the energy of vortex bound states in chiral d-wave superconductor~\cite{lee-16} that, in your turn, differs from the energy of Caroli-de Gennes-Matricon states in the s-wave case~\cite{caroli-64}.
The experimental detection of the obtained Majorana vortex modes by scanning tunneling microscopy is discussed. We also obtain the difference in the local density of states for Majorana modes localized at the vortex and the antivortex cores due to chiral $d_1+id_2$ symmetry of superconductivity.

The article has been organized in five sections. In Sec. \ref{sec2} the Hamiltonian of the spin-singlet chiral superconductor with noncollinear spin ordering in the presence of vortices is introduced and the methods are described. The relation between the results of previous studies \cite{lu-13} and \cite{lee-16} is discussed in Sec. \ref{sec22}. Namely, appearance of zero energy vortex bound states in $d_1+id_2$-wave superconducting state in the presence of noncollinear spin ordering is noted. In Sec. \ref{sec3} the numerical results for the excitation energies, local density of states around the vortex (or antivortex) cores and near the edges for different vortex structures are presented. The obtained results are discussed in Sec. \ref{sec4}. Conclusions are given in Sec. \ref{sec5}.

\section{\label{sec2}Model and methods}

In the framework of tight-binding approach and mean-field approximation the coexistence
of superconductivity and in-plane noncollinear spin
ordering can be described by the Hamiltonian
\begin{eqnarray}
\label{Ham1}
\mathscr{H} & = & \sum_{f \sigma} (-\mu-\eta_{\sigma}h_z) c_{f \sigma}^{\dag}c_{f \sigma} + \sum_{ff_1 \sigma} t_{ff_1} c_{f \sigma}^{\dag}c_{f_1 \sigma} +
\nonumber \\
& + & h \sum_{f} \left( \exp({i\bf QR }_f) c_{f \uparrow}^{\dag}c_{f \downarrow} + \exp(-i{\bf Q R}_f) c_{f \downarrow}^{\dag}c_{f \uparrow}  \right) +
\nonumber \\
& + & \sum_{ff_1} \left( \Delta_{ff_1} c_{f \uparrow}^{\dag}c_{f_1 \downarrow}^{\dag}  + \Delta_{ff_1}^*  c_{f_1 \downarrow} c_{f \uparrow}   \right),
\end{eqnarray}
where $\mu$ is the chemical potential, $h_z$ is the Zeeman splitting term due to external magnetic field along the $z$ axis, $t_{ff_1}$ is the hopping parameter, $h=JM/2$ is the exchange field. The exchange field is caused by $s-d(f)$ exchange interaction with the parameter $J$ between itinerant electrons and frozen localized spins having a noncollinear magnetic structure for which the vector spin operator is defined as
\begin{equation}
\label{Sf}
\left\langle {\bf S}_f \right\rangle = M \left( \cos({\bf QR}_f), -\sin({\bf QR}_f), 0 \right).
\end{equation}
Here, the spin structure vector is ${\bf Q}$, and $M$ is magnetic order parameter. We suppose that in-plane magnetic anisotropy exist, and ordered magnetic moments are not canted in low magnetic field along $z$-axis. The parameter $\Delta_{ff_1}$ defines the amplitude of superconducting pairings.

In the Bogoliubov---de Gennes representation the Hamiltonian is expressed as
\begin{align}
\mathscr{H} = \frac{1}{2} \Psi^{\dag} H \Psi, \, \, \, \Psi^{\dag} = [c_{\uparrow}^{\dag} \, c_{\downarrow}^{\dag} \, c_{\downarrow} \, c_{\uparrow}],
\end{align}
and
\begin{eqnarray}
H & = & \left( {\begin{array}{*{20}{c}}
{{{ \xi }_{\uparrow}}}&{h}&{{D}}&0\\
{{{ h}^ * }}&{{{ \xi }_{\downarrow}}}&0&{ - D^T}\\
{ D^{\dag}}&0&{ -  \xi _{ \downarrow}^T}&{ - {h}}\\
0&-D^*&{ -  h^*}&{ -  \xi _{ \uparrow}^T}
\end{array}} \right). \nonumber
\end{eqnarray}

The triangular lattice with the edges along the basic translation vectors ${\bf a}_1 = a(\sqrt{3}/2,-1/2)$ and ${\bf a}_2 = a(0,1)$ is considered. The number of sites on the edges of the lattice is the same and denoted as $N_s$. For this case the each element in the matrix $H$ is the matrix of the size $N_s^2$. In the following, different bases defining the radius-vector in the real space are used: ${\bf R}_f = x_f{\bf e}_x + y_f{\bf e}_y = (n-1){\bf a}_1 + (m-1){\bf a}_2$.

The Hamiltonian (\ref{Ham1}) belongs to the D-symmetry class which is described by the integer topological invariant $\mathbb{Z}$. In the present study the topological invariant $\tilde{N}_3$ \cite{ishikawa-87, volovik-09} expressed in terms of Green functions is used
\begin{eqnarray}
&& \label{Z_inv} \tilde{N}_{3} =
\frac{\varepsilon_{\mu\nu\lambda}}{48\pi^{2}} \sum_{\sigma}
\nonumber \\
&&
\int\limits_{-\infty}^{\infty}d\omega\iint\limits_{-\pi}\limits^{\pi}
dk_{1}dk_{2}
\text{Tr}\left(\widehat{G}_{\sigma}\partial_{\mu}\widehat{G}^{-1}_{\sigma}
\widehat{G}_{\sigma}\partial_{\nu}\widehat{G}^{-1}_{\sigma}
\widehat{G}_{\sigma}\partial_{\lambda}\widehat{G}^{-1}_{\sigma} \right).
\nonumber \\
\end{eqnarray}
Here, the repeated indices $\mu$, $\nu$, $\lambda = 1, \, 2, \, 3$
imply summation,  $\varepsilon_{\mu\nu\lambda}$ is the Levi-Civita symbol, $\partial_{1(2)} \equiv
\partial/\partial k_{1(2)}$, $\partial_{3} \equiv
\partial/\partial \omega$, and $\widehat{G}_{\sigma}(i\omega,{\bf k})$ is the matrix
Green function under periodic boundary conditions whose poles determine the bulk fermion spectrum. Details of the derivation of the matrix Green function and the bulk energy spectrum are presented in Appendix A.

In general the parameter $\Delta_{f f_1}$ can define the
amplitude of superconducting pairings between the fermions on the same site ($f=f_1$) leading to s-wave superconductivity. The on-site exchange field (the third sum in \eqref{Ham1}) in this case has a strong pair-breaking effect. Nevertheless, noncollinear spin ordering and superconductivity can coexist if the pairing interaction is non-local \cite{valkov-19}. Therefore in the following the pairings between electrons on nearest sites are considered.
The presence of vortices is parameterized through the dependence on coordinates of the superconducting order parameter:
\begin{eqnarray}
\label{SCOP}
&& \Delta_{<f f_1>} = \Delta \exp\left[ i l \arg(z({\bf R}_f) - z({\bf R}_{f_1}))\right] e^{i\pi/3}
\times
\\ \nonumber
&&
\prod_i \tanh \left( \frac{|{\bf R}_{ff_1} - {\bf R}_{v_i}|}{\xi} \right) \exp\left[ i l_{v_i} \arg(z({\bf R}_{ff_1}) - z({\bf R}_{v_i})) \right],
\nonumber
\end{eqnarray}
where the first exponent determines the symmetry of the superconducting order parameter (for the chiral $d_1+id_2$ symmetry on the triangular lattice $l=2$), $z({\bf R}_f) = x_f+iy_f$ (in the basis of ${\bf e}_x$, ${\bf e}_y$), the additional phase $\exp(i\pi/3)$ is used for convenience. The last product describes different structures of
vortices ($l_{v_i} = +1$ for a vortex, and $l_{v_i} = -1$ for an antivortex), where ${\bf R}_{v_i}$ is a coordinate of the vortex (or antivortex) center and $\xi$ is the characteristic length. Here, ${\bf R}_{ff_1} = ({\bf R}_{f}+{\bf R}_{f_1})/2$.

The energy spectrum and probability densities of subgap states are found using the Bogoliubov transformation with Bogoliubov quasi-particle operators $\alpha_{j}$ ($j=1, \dots, 2N_s^2$):
\begin{eqnarray}
\label{alpha}
\alpha_{j} = \sum_{nm\sigma} \left(u_{jnm\sigma}c_{nm\sigma}+v_{jnm\sigma}c_{nm\sigma}^{\dag}\right).
\end{eqnarray}
Here we explicitly use the coordinates $n$ and $m$ in the basis of ${\bf a}_1$, ${\bf a}_2$ instead of the site index $f$.

From the formula for site-dependent electron concentration $\left\langle c_{nm\sigma}^{\dag} c_{nm\sigma} \right\rangle$ the spatially and frequency-resolved density of states (DOS) is written as
\begin{eqnarray}
\label{rho}
\rho \left( {\bf R}_f, \omega \right) &=& \sum_{\substack{\sigma, j\\(\varepsilon_j > 0)}} \left[ \left| u_{jnm\sigma} \right|^2 \delta\left(\omega - \varepsilon_j\right) + \right.
\nonumber \\
&+&
\left. \left|v_{jnm\sigma} \right|^2 \delta\left(\omega + \varepsilon_j\right) \right],
\end{eqnarray}
where the positive excitation energies $\varepsilon_j$ are chosen.
According to \cite{pathak-21} it is more informative to use the local DOS integrated over a region around a vortex core, edges, etc.:
\begin{eqnarray}
\text{LDOS}_{v}\left(\omega\right) = \frac{1}{N_W} \sum_{|n-n_v|<W} \sum_{|m-m_v|<W} \rho \left( {\bf R}_f, \omega \right).
\nonumber \\
\end{eqnarray}
For $\omega = \varepsilon_j = 0$ and an unbroadened zero-energy level DOS $\rho \left( {\bf R}_f, \omega \right)$ corresponds to the widely used probability density of the $j$-excitation
\begin{eqnarray}
\label{PrD}
\text{PrD}_{j} \left( {\bf R}_f \right) = \sum_{\sigma} \left( \left| u_{jnm\sigma} \right|^2 + \left| v_{jnm\sigma} \right|^2 \right).
\end{eqnarray}

To describe spatially separated Majorana modes in the case of a doubly degenerate ground state two
Majorana quasiparticle operators $b^{\prime}$ and $b^{\prime \prime}$ (following Kitaev \cite{kitaev-01}) are introduced. They are expressed through the initial Majorana operators set $\gamma_{Anm\sigma} = c_{nm\sigma} + c_{nm\sigma}^{\dag}$ and $\gamma_{Bnm\sigma} = i \left(c_{nm\sigma}^{\dag} - c_{nm\sigma} \right)$:
\begin{eqnarray}
\label{b_gamma1}
&& b^{\prime} = \alpha_{1}+\alpha_{1}^{\dag} =
\\
&=& \sum_{nm\sigma} \left( \text{Re}(w_{1nm\sigma})\gamma_{Anm\sigma} - \text{Im}(z_{1nm\sigma})\gamma_{Bnm\sigma}\right),
\nonumber \\
\label{b_gamma2}
&& b^{\prime \prime} = i\left(\alpha_{1}^{\dag}-\alpha_{1}\right) =
\\
&=& \sum_{nm\sigma} \left( \text{Im}(w_{1nm\sigma})\gamma_{Anm\sigma}+\text{Re}(z_{1nm\sigma})\gamma_{Bnm\sigma}\right),
\nonumber
\end{eqnarray}
where $w_{jnm\sigma} = u_{jnm\sigma}+v_{jnm\sigma}$ and $z_{jnm\sigma} = u_{jnm\sigma}-v_{jnm\sigma}$.

\section{\label{sec22} Appearance of zero energy vortex bound states due to 120-degree magnetic order}

\begin{figure}[h]
        \includegraphics[width=0.35\textwidth]{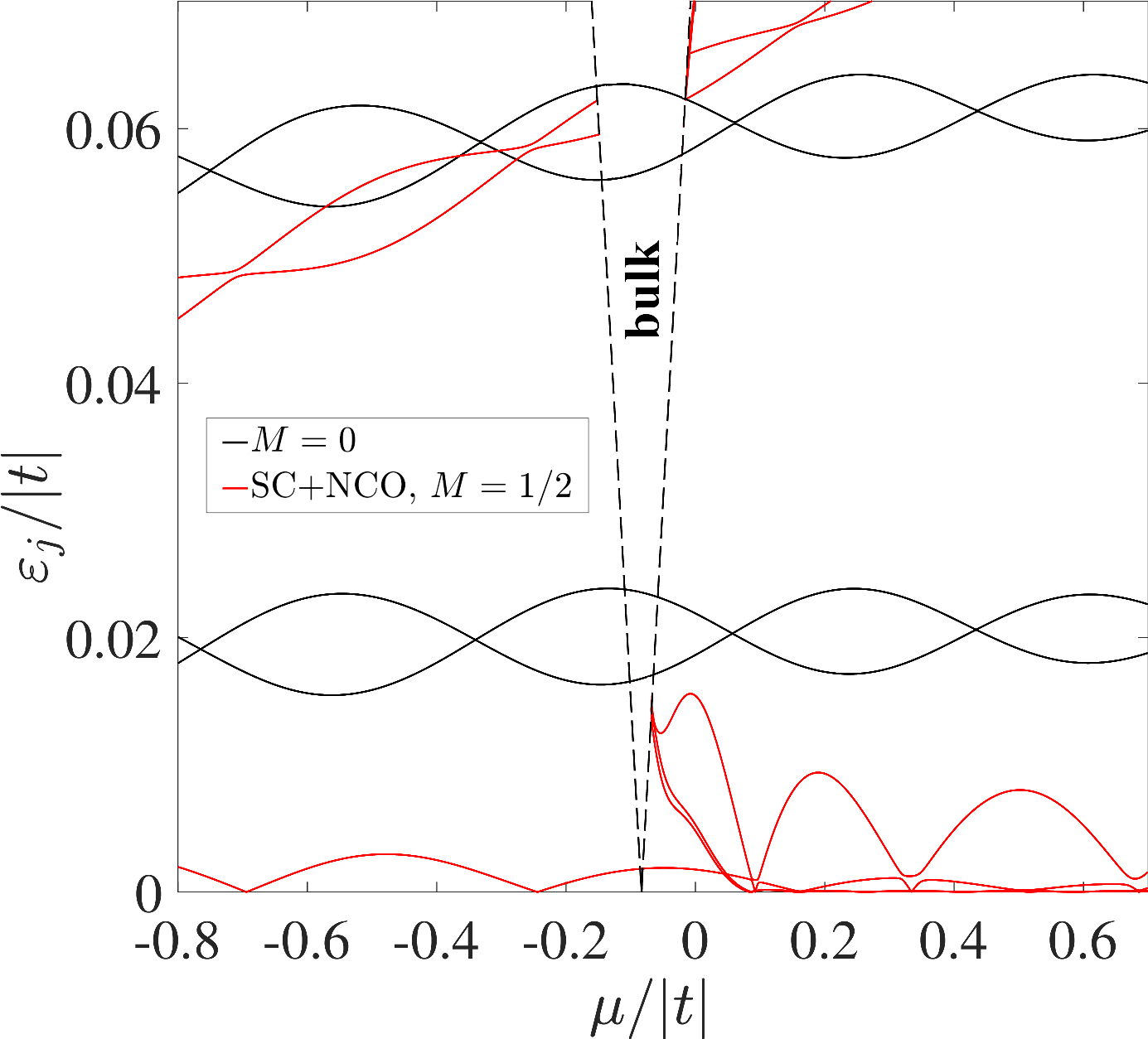}
        \caption{\label{PBC} The chemical-potential dependence of
the lower excitation energies in the presence of the vortex-antivortex pair on the triangular lattice with periodic boundary conditions in two cases: without long-range magnetic order ($M=0$, in agreement with the results of \cite{lee-16}), and for coexistence of superconductivity and noncollinear spin ordering (SC+NCO, $M=0.5$, in agreement with \cite{lu-13}). The dashed line shows the bulk superconducting gap in the presence of NCO. Parameters: $J=4$, $\Delta = 0.2$, $h_z = 0$ (in units of the hopping parameter $|t|$).}
\end{figure}

To study vortex bound states, it is generally accepted to separate them from edge states.
Therefore, in order to suppress edge-localized states periodic boundary conditions can be applied. For the periodic vortex lattice a vortex unit cell with few vortices inside should be properly chosen. Then considering a plenty of such unit cells periodic boundary conditions can be imposed~\cite{takigawa-00}. Another way to deal with periodic boundary conditions in periodic vortex lattices is applying the correct gauge transformation as described in~\cite{franz-00, liu-15, pathak-21}.

On the other hand, for the case of few single vortices in the lattice the periodicity can be realized only for special vortex structures providing continuity of the phase of the order parameter on the opposite edges.
For example, on a square lattice this can be done for the vortex-antivortex pair~\cite{lee-16} which is described by the expression similar to \eqref{SCOP} with the additional factor providing periodic boundary conditions. As mentioned above, in this case zero energy excitations are absent, and vortex bound state has a finite excitation energy.

The authors of~\cite{lu-13} showed that in the presence of the vortex-antivortex pair on the triangular lattice with periodic boundary conditions Majorana vortex modes with zero energy arise. To provide continuity of the phase of the order parameter on the opposite lattice edges the vortex-antivortex pair is parameterized through the Jacobi theta-functions (see Supplemental materials for~\cite{lu-13}). Although it was not explicitly emphasized in Refs.~\cite{lu-13, lee-16}, but it can be concluded that namely noncollinear spin ordering leads to appearance of Majorana vortex modes among the subgap vortex bound states with finite energy. It is connected with the fact that symmetry class C of the $d_1+id_2$-wave state changes to class D in the presence of magnetic order.

In the present section this result is repeated in the tight-binding approach for the triangular lattice of the identical size along ${\bf a}_1$ and ${\bf a}_2$ and with periodic boundary conditions. We consider the vortex-antivortex pair with ${\bf R}_{v_1} = \left( \frac{N_s}{3}, \frac{N_s+1}{2} \right)$ (antivortex) and ${\bf R}_{v_2} = \left( \frac{2N_s}{3}+1, \frac{N_s+1}{2} \right)$ (vortex) on the triangular lattice ($N_s = 63$) in two cases: (i) in the presence of 120$^{\circ}$ (${\bf Q} = \left( 2\pi/3, 2\pi/3 \right)$) long-range magnetic order; (ii) without magnetic ordering. The slightly modified parametrization of the vortex-antivortex pair is used in comparison with the one in~\cite{lu-13}. The chemical-potential dependencies of the lowest excitation energies in both cases are displayed in Fig. \ref{PBC} for the parameters $J=4$, $\Delta = 0.2$, $h_z = 0$ (here and henceforth all energy variables are in units of the hopping parameter $|t|$).

\begin{figure}[b]
        \includegraphics[width=0.22\textwidth]{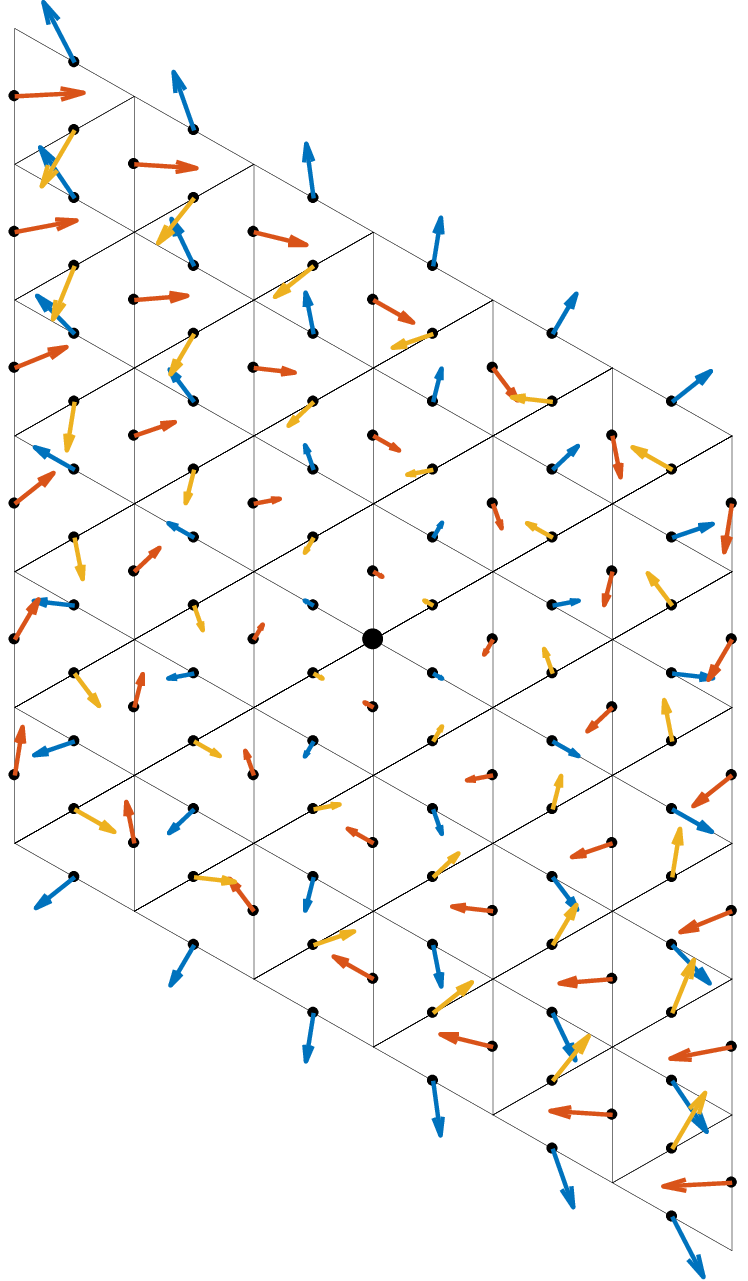}
        \caption{\label{Single_vortex} The chiral d-wave superconducting order parameter \eqref{SCOP} in the presence of the vortex which center is denoted by the large black dot. The arrows on the nearest-neighbor bonds indicate the magnitude and the phase of the order parameter.}
\end{figure}

It is clearly seen that in the absence of magnetic ordering only finite-energy vortex bound states (like Caroli-de Gennes-Matricon states~\cite{caroli-64}) are present inside the bulk superconducting gap ($\sim 2\Delta$). The oscillations of their energies are caused by the presence of the vortex pair. The modes with zero excitation energy appear in the coexistence phase of superconductivity and noncollinear spin ordering (SC+NCO). A finding here is that the bulk superconducting gap is closed at certain conditions and the topological phase transition occurs which is driven by noncollinear magnetic ordering. The closing gap is shown in Fig. \ref{PBC} by the dashed line. On the left of the transition point the topological invariant $\tilde{N}_3 = 1$ (see \eqref{Z_inv}), while on the right of the transition $\tilde{N}_3 = 4$. It is seen that the energy of vortex bound states drops down in the $\tilde{N}_3 = 4$ topological phase. For clarity, we do not show the excitation energies above the bulk gap. The zero modes in the $\tilde{N}_3 = 1$ phase correspond to Majorana vortex modes $b^{\prime}$ and $b^{\prime \prime}$ localized at different cores of the vortex-antivortex pair, while these modes are not separated in the $\tilde{N}_3 = 4$ phase. The detail difference between above mentioned topological phases will be discussed in the next section.

It should be noted that taking into account Zeeman splitting ($h_z \ne 0$) can lead to formation of zero modes even at $M = 0$. Although, such modes are not spatially se\-parated and, consequently, they are not Majorana modes.

In real systems periodic boundary conditions may not be applicable. Therefore in the following, the simultaneous presence of edge and vortex bound states in the SC+NCO phase with open boundary conditions, as well as features of local density of states will be described. Note that topological edge states in the chiral d-wave state can have zero excitation energy even in the absence of magnetic order, in contrast to the vortex bound states.

\begin{figure} [t]
        \includegraphics[width=0.35\textwidth]{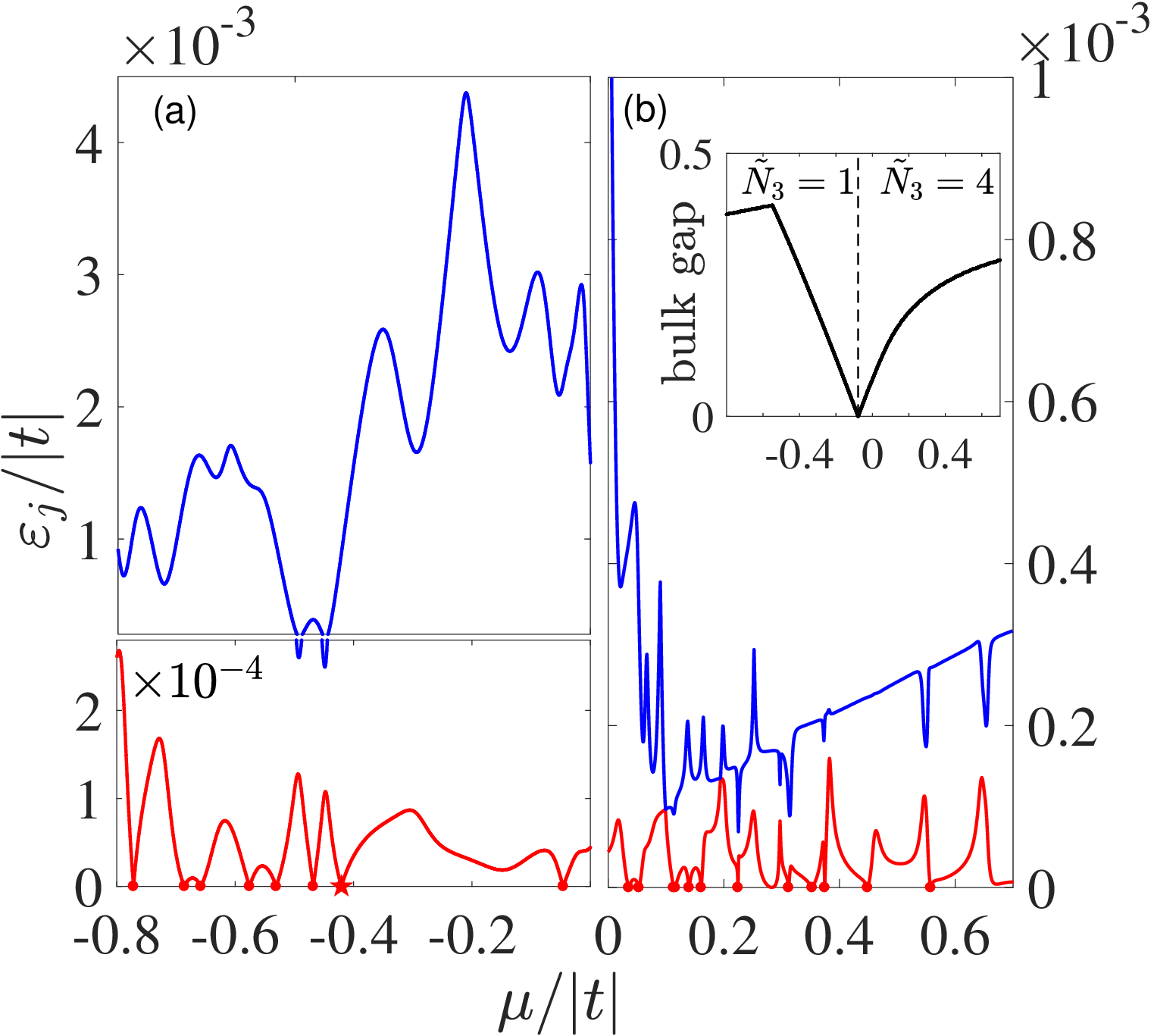}
        \caption{\label{cond_zero} The chemical potential dependence of
the first two excitation energies in the presence of the vortex with the center at $((N_s+1)/2,(N_s+1)/2)$ on the triangular lattice ($N_s = 63$) with open boundary conditions in SC+NCO phase. Parameters: $h=1$, $\Delta = 0.2$, $h_z = 0.1$. The x-axis is divided at value $\mu = 0$ to use different energy scales. Therefore, (a) is for $\mu<0$, and (b) is for $\mu>0$. Inset in (b): The chemical-potential dependence of the bulk gap in the excitation spectrum of the homogeneous system under periodic boundary conditions. The vertical dashed line divide topological phases with different values of the topological invariant.}
\end{figure}

\section{\label{sec3} Local density of states study}

\subsection{\label{secA} Single vortex on the lattice}

The structure of the order parameter~\eqref{SCOP} in the presence of the single vortex with $\xi = 4a$ and ${\bf R}_{v}$ at the center of the triangular lattice is presented in Fig.~\ref{Single_vortex} (the region of the lattice in the vicinity of the vortex core is only shown). The vortex center position is marked  by the large black dot. It is seen that the pairing amplitudes of the bonds nearest to the vortex center are suppressed. As usual, the phase of the order parameter on equivalent bonds denoted by the arrows with the same color changes by $2\pi$ going around the vortex center.

In this section such vortex with the center located at the point $(n_v, m_v) = ((N_s+1)/2,(N_s+1)/2)$ (in the basis of ${\bf a}_1$, ${\bf a}_2$) on the triangular lattice with the parallelogram shape, $N_s = 63$, and open boundaries is considered. The chosen lattice size when $N_s$ is multiple of 3 provides zero in-plane total magnetic moment. In Fig. \ref{cond_zero} the chemical-potential dependence of two lowest excitation energies is presented for the parameters $h=1$, $\Delta = 0.2$, $h_z = 0.1$.

For clarity the chemical potential ranges on Figs. \ref{cond_zero}(a) and \ref{cond_zero}(b) are divided by the value $\mu = 0$ since the energy scales are mismatched on the presented intervals. This value is close to the critical point $\mu \approx -0.0785$ at which topological phase transition between different phases occurs in the homogeneous system (without vortices) with periodic boundary conditions. The whole chemical potential interval corresponds to the case when on-site electron concentration, far from inhomogeneities such as a vortex core and edges, changes from approximately 0.7 to near half filling. Again, as in Fig. \ref{PBC}, the topological transition is manifested on the dependence of the size of the bulk gap on chemical potential (see inset of Fig. \ref{cond_zero}(b)). In the inset it is seen that gapless excitations occur at the critical point separating topological phases with different topological invariant $\tilde{N}_3$. Therefore in Fig. \ref{cond_zero}(a) the results are presented when the system is in the topologically non-trivial phase with topological invariant $\tilde{N}_3 = 1$, excepting the transition region near $\mu \approx -0.0785$, while in Fig. \ref{cond_zero}(b) the results correspond to the $\tilde{N}_3 = 4$ phase.

\begin{figure} [t]
        \includegraphics[width=0.2\textwidth]{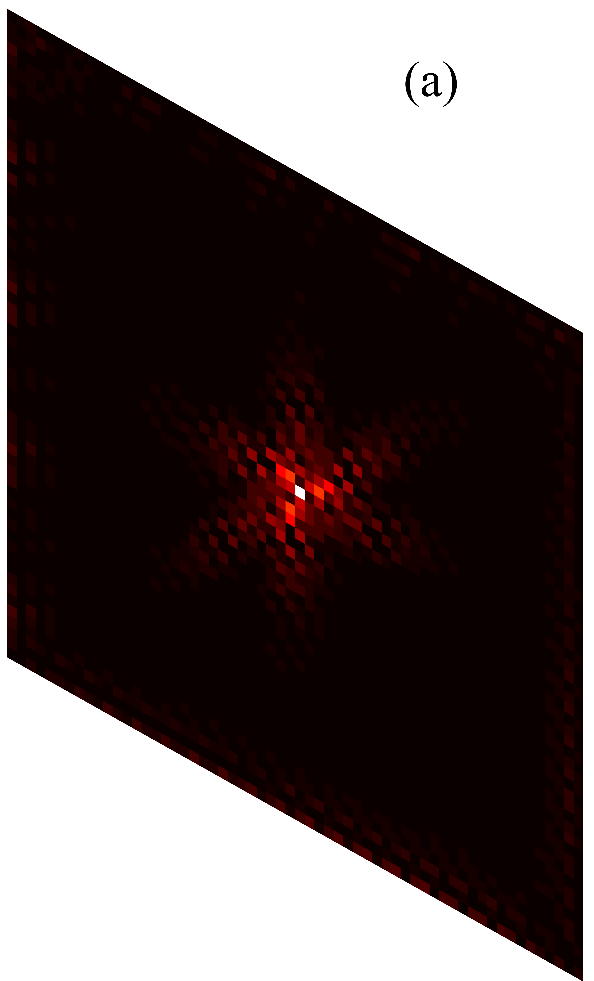}
        \includegraphics[width=0.2\textwidth]{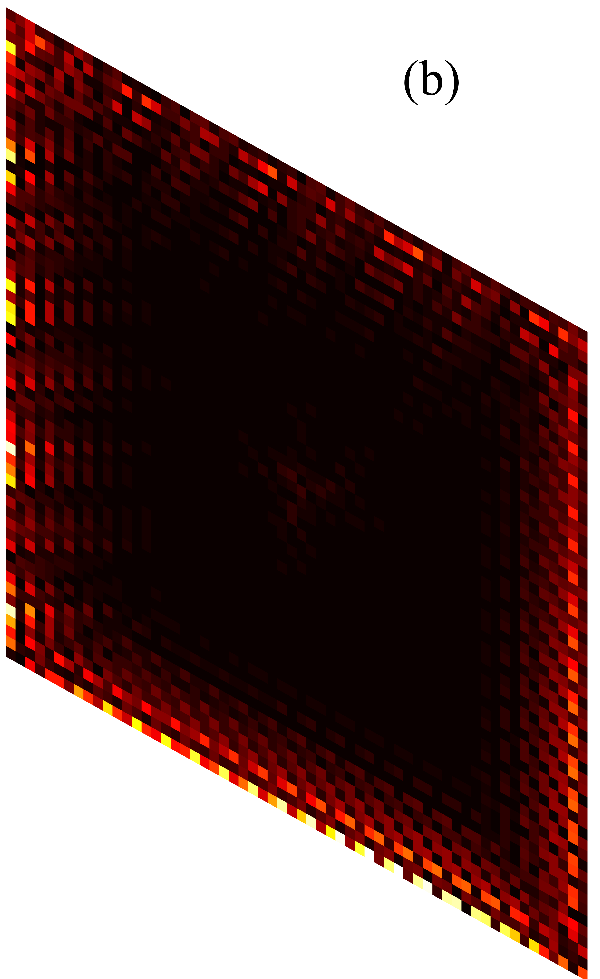}
        \caption{\label{m_modes} Majorana modes in the presence of the single vortex on the triangular lattice (the vortex center coincides with the lattice center) at $\mu \approx - 0.4205$ (this value is marked by the star in Fig. \ref{cond_zero}). The distributions of $b^{\prime}$  (a) and $b^{\prime \prime}$ (b) modes on the lattice are shown. Black color corresponds to near zero values, the brightest color is for maximum values reaching $0.24$. The parameters are the same as in Fig. \ref{cond_zero}.}
\end{figure}

First of all, the general properties of two distinct phases are discussed.  The phase with $\tilde{N}_3 = 4$ is the topologically non-trivial phase which is caused by chiral d-wave superconductivity. Such a phase exist also without long-range magnetic order ($h=0$) and its topological invariant is $Q = 2$, where $Q$ is the winding number of the pseudo-spin vector introduced by Anderson~\cite{volovik-97, zhou-08}. We argue that the relation $\tilde{N}_3(h=0) = 2Q$ between the expression \eqref{Z_inv} and the winding number $Q$ in the absence of spin ordering is obtained analytically.
It is known that the $\tilde{N}_3 = 4$ (or $Q = 2$) phase does not support Majorana modes due to spin-singlet character of $d_1+id_2$ superconductivity. On the other hand, the topologically non-trivial phase with $\tilde{N}_3 = 1$ appears only in the presence of noncollinear spin ordering ($h \ne 0$). It is believed that such phase hosts Majorana modes \cite{lu-13, zlotnikov-21}.

We note that the obtained in Fig. \ref{cond_zero} oscillations of energy near zero are typical for topologically nontrivial phases~\cite{hegde-16, valkov-22}. As it is known, the lowest excitation energy in topological superconductors is determined also by the intervortex distance and the lattice size $N_s$ in consideration of open boundary conditions. The same oscillations of zero modes are obtained in Fig. \ref{PBC} for the vortex-antivortex pair in the lattice with periodic boundary conditions. This excitation energy is infinitesimal for dilute enough vortices and large enough $N_s$ and grows under decreasing $N_s$ or intervortex distance. However, an excitation with zero energy can be also realized by the destructive interference of Majorana modes~\cite{hegde-16, valkov-22}. In this study the parameters, as well as $N_s$ and vortex structures are chosen in such a way that the presented zero-energy solutions are obtained with good accuracy determined only by a numerical error.

\begin{figure} [t]
        \includegraphics[width=0.35\textwidth]{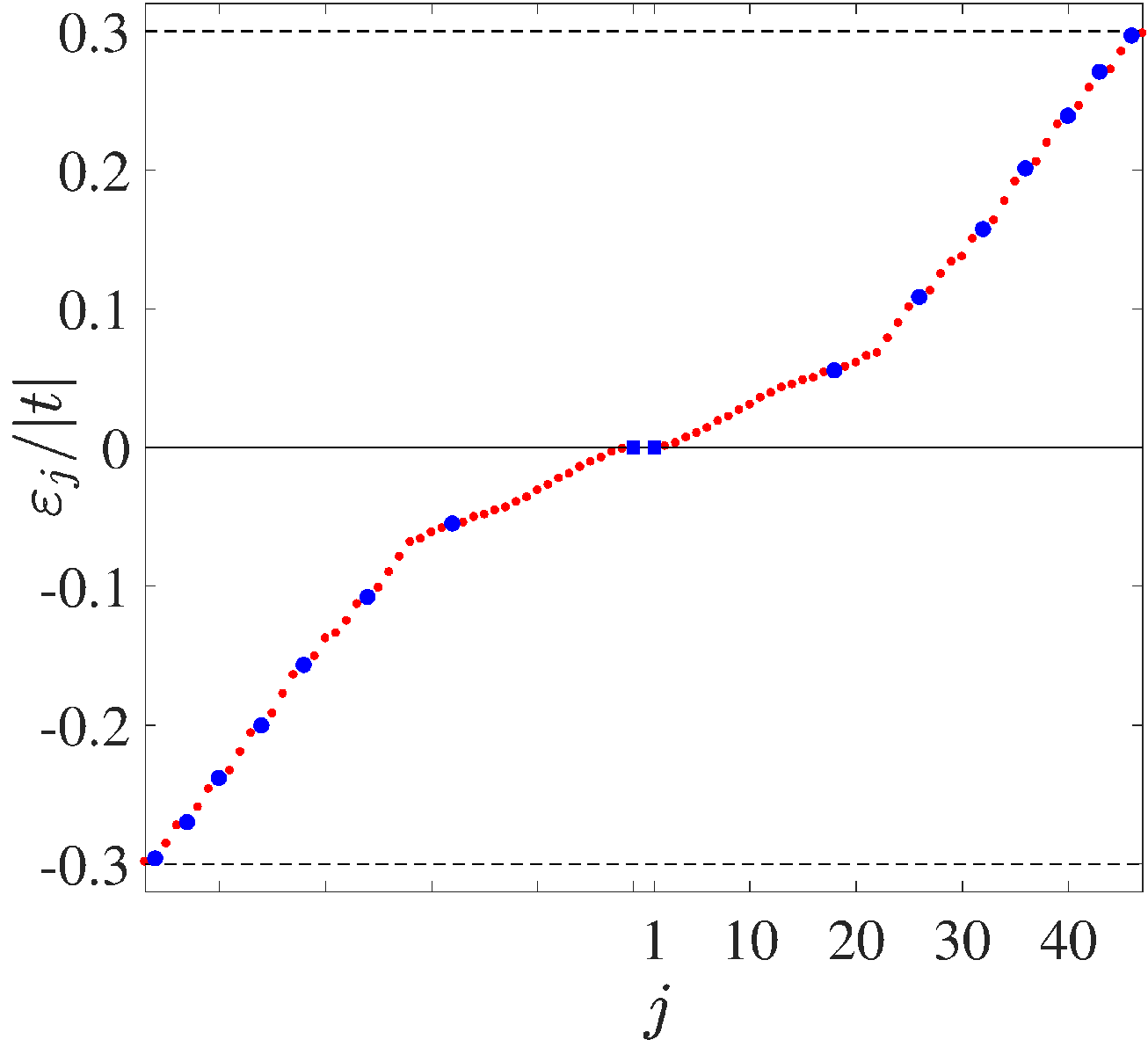}
        \caption{\label{exc_energy} The excitation energies at $\mu \approx - 0.4205$. Square markers denote implementation of Majorana modes, large points --- ingap vortex bound states, small points are for ingap edge bound states. The bulk gap is denoted by the dashed lines.}
\end{figure}

The chemical potential values corresponding to zero modes formation are denoted by dots on the $x$ axis in Figs. \ref{cond_zero}(a) and \ref{cond_zero}(b). Such values are calculated from the system of equations on the coefficients of the Bogoliubov transformation (the algorithm is described in, for example, \cite{nijholt-16}). It should be noted that for the limited lattice a real closure of the bulk gap can not be obtained since in this case the discrete energy levels are realized instead of continuous energies for an infinite lattice with periodic boundary conditions. It means that none of the presented points for zero modes does correspond to vanishing bulk gap.
In the transition region near $\mu \approx -0.0785$ the minimal energy of quasi-bulk states in the limited lattice is $ 0.01$ but not zero and there is still subgap energies corresponding to vortex and edges bound states. Far from the transition point the bulk states are well defined and their energy slightly exceeds the bulk gap shown in the inset of Fig. \ref{cond_zero}(b). This explains why for the limited system a qualitative change of the excitation spectrum occurs not exactly at the critical point, but in the vicinity of it.

\begin{figure*}[t]
        \includegraphics[width=0.98\textwidth]{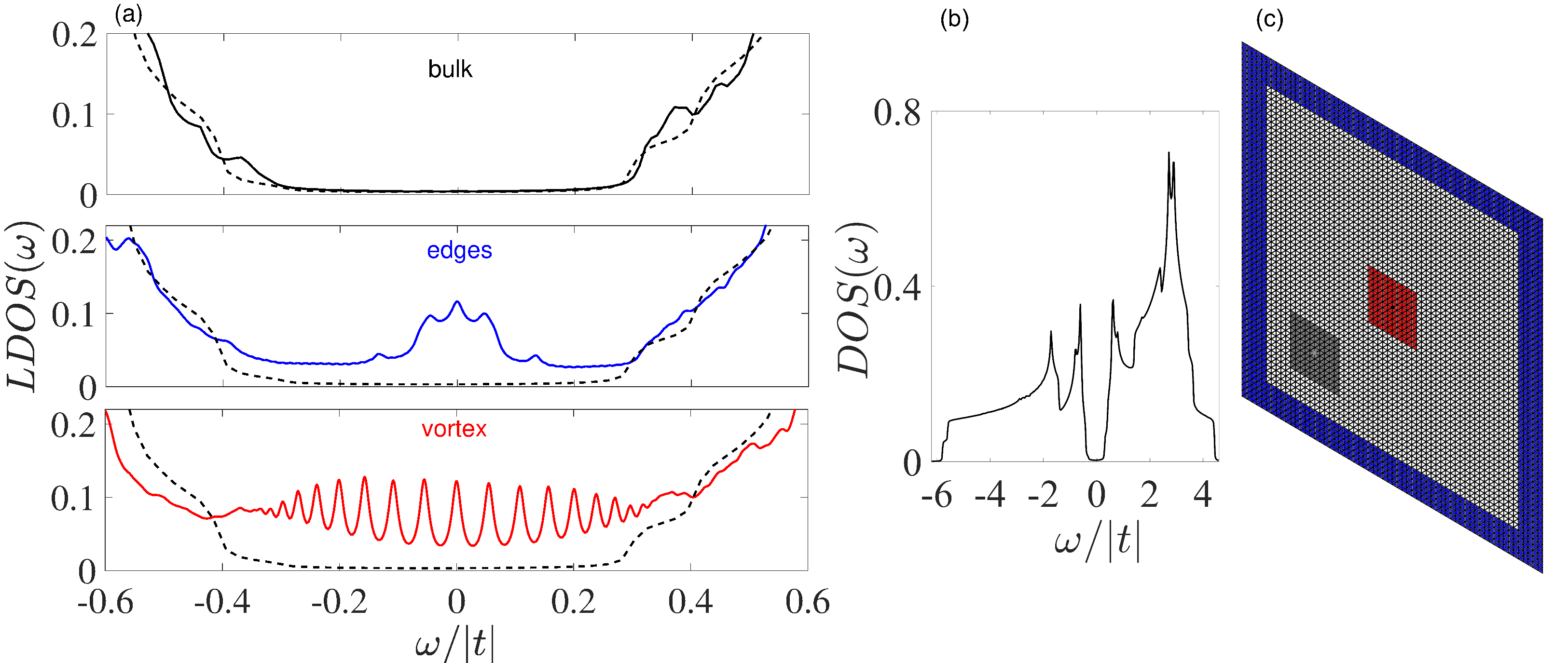}
        \caption{\label{LDOS_vort} (a) Local density of states (LDOS) in the vicinity of the single vortex (solid line on the lower graph), near the edges (solid line on the middle graph), and far from inhomogeneities (solid line on the upper graph) at $\mu \approx - 0.4205$ with existing Majorana modes. The broadening parameter for LDOS is $\delta = \Delta/20$. The dashed line is DOS calculated for comparison in the homogeneous case with periodic boundary conditions. (b) The same DOS as shown by the dashed line in (a) but for all energies. (c) The integration areas on the lattice for calculation of $\text{LDOS}(\omega)$ are highlighted near the vortex core, near the edges, and far from inhomogeneities, respectively.}
\end{figure*}

However, the difference between the phases with various $\tilde{N}_3$ is clearly seen in Fig. \ref{cond_zero}. In the $\tilde{N}_3 = 4$ phase the second excitation energy is much lower than the one in the $\tilde{N}_3 = 1$ phase. It is connected with the fact that in the $\tilde{N}_3 = 4$ phase there appear four zero edge modes with different quantum numbers $k_{i}$ along the edge in the strip geometry. At the same time in the $\tilde{N}_3 = 1$ phase the zero edge mode is single and appears always at $k_{i} = -2\pi/3$. Such edge spectra in different phases are discussed in \cite{valkov-22}.
The solutions in the limited 2D lattice are similar to the solutions in the strip geometry running through a finite set of $k_{i}$. Therefore in the 2D case a number of edge bound states near zero energy grows in the phase with $\tilde{N}_3 = 4$ in comparison with the topologically nontrivial phase with $\tilde{N}_3 = 1$ supporting Majorana modes. The same is true for vortex bound states, as it is shown in Fig. \ref{PBC}.

We also should note that the lowest energy branch in Fig. \ref{cond_zero}(a) in the $\tilde{N}_3 = 1$ phase describe vortex bound states with reduced edge contributions, while the second branch is for only edge states. Due to the closeness of various energies in the $\tilde{N}_3 = 4$ phase the excitations with equal contributions of the vortex core and edges are realized there. This excitation can be the lowest on energy or correspond to the second excitation on magnitude depending on the parameters, while the remaining excitation is for predominantly vortex bound state.

The another difference between two topological phases in Fig. \ref{cond_zero} concerns zero energy modes.
In the $\tilde{N}_3 = 1$ phase the characteristic spatial distributions of $b^{\prime}$ and $b^{\prime \prime}$ modes (see \eqref{b_gamma1} and \eqref{b_gamma2}) with zero energy are shown in Fig. \ref{m_modes}.  This zero-energy solution is marked by the star in Fig. \ref{cond_zero} at $\mu \approx - 0.4205$. It is seen that the first Majorana mode is localized in the vortex core, while the second mode is located near the edges. This localization is general for all zero modes in the $\tilde{N}_3 = 1$ phase. Such distributions confirm the formation of Majorana modes in the $\tilde{N}_3 = 1$ phase and are known for other topological superconducting systems \cite{gurarie-07, silaev-08, iskin-12, bjornson-13, akzyanov-16}. To obtain this result, the Bogoliubov-like coefficients in the definitions of Majorana operators $b^{\prime}$ \eqref{b_gamma1} and $b^{\prime \prime}$ \eqref{b_gamma2}  are calculated. Obviously, these coefficients are real. In Fig. \ref{m_modes}(a) the distribution on the lattice sites of the coefficient $|\text{Re}(w_{1nm\uparrow})|$ from $b^{\prime}$ is shown, while Fig. \ref{m_modes}(b) is for $|\text{Im}(w_{1nm\uparrow})|$ from $b^{\prime \prime}$. All other coefficients have the same localized character, i.e., vortex core localization for $b^{\prime}$ and edge localization for $b^{\prime \prime}$, and, for this reason, they are not shown. This behavior is unique for true Majorana modes. Indeed, the representations \eqref{b_gamma1} and \eqref{b_gamma2} can be introduced for any excitation of any system with particle-hole symmetry. However, even if $b^{\prime}$ and $b^{\prime \prime}$ correspond to zero energy, but they are not spatially separated, they can not be Majorana modes.

Such non-Majorana zero modes exist in the $\tilde{N}_3 = 4$ phase in the presence of 120$^{\circ}$ spin ordering. In this phase edge and vortex bound states with zero excitation energy can also appear. Nevertheless, the modes $b^{\prime}$ and $b^{\prime \prime}$ corresponding to such zero-energy states have simultaneously non-zero values near the vortex core and near the edges.

\begin{figure*}[t]
        \includegraphics[width=0.4\textwidth]{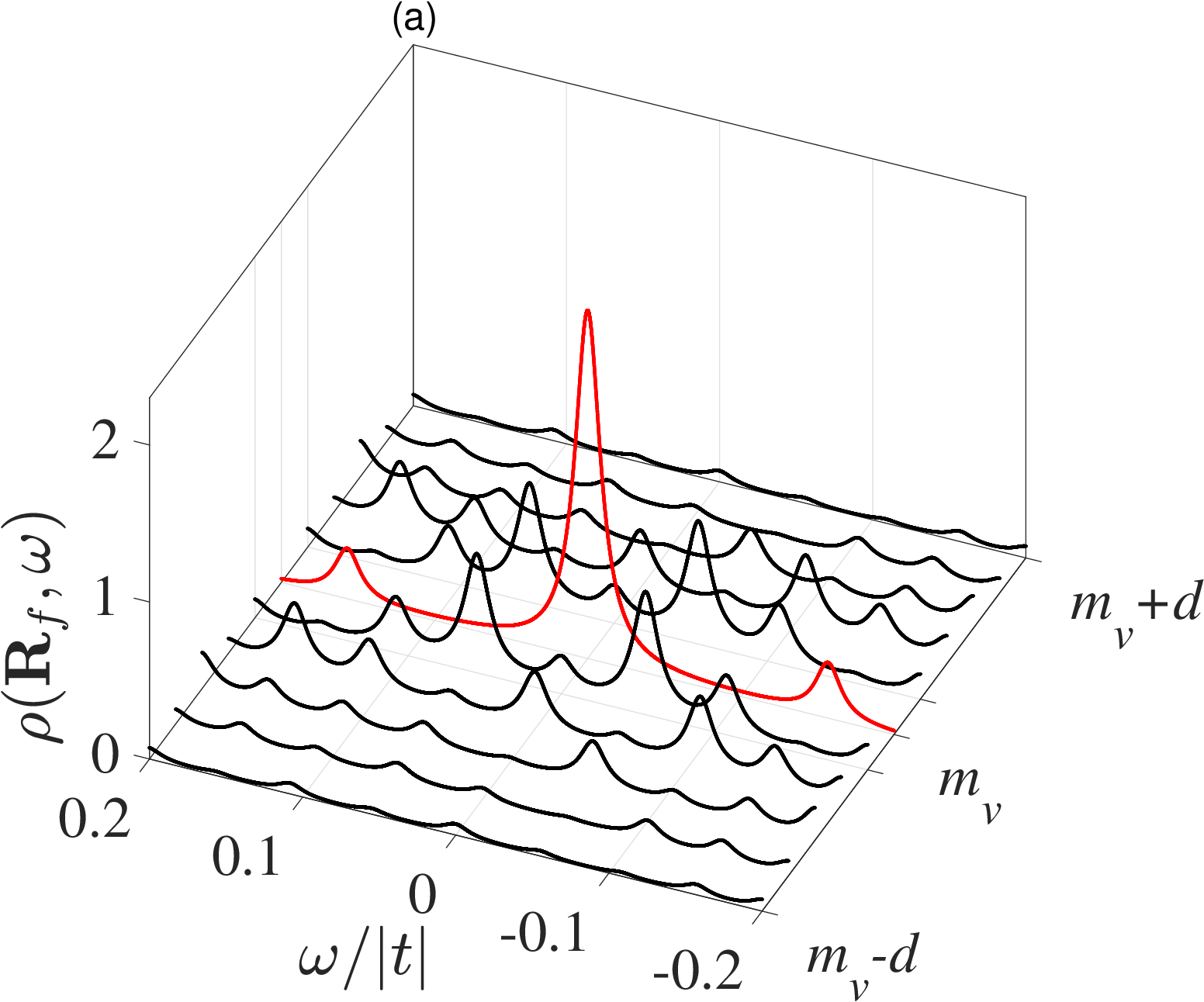}
        \includegraphics[width=0.4\textwidth]{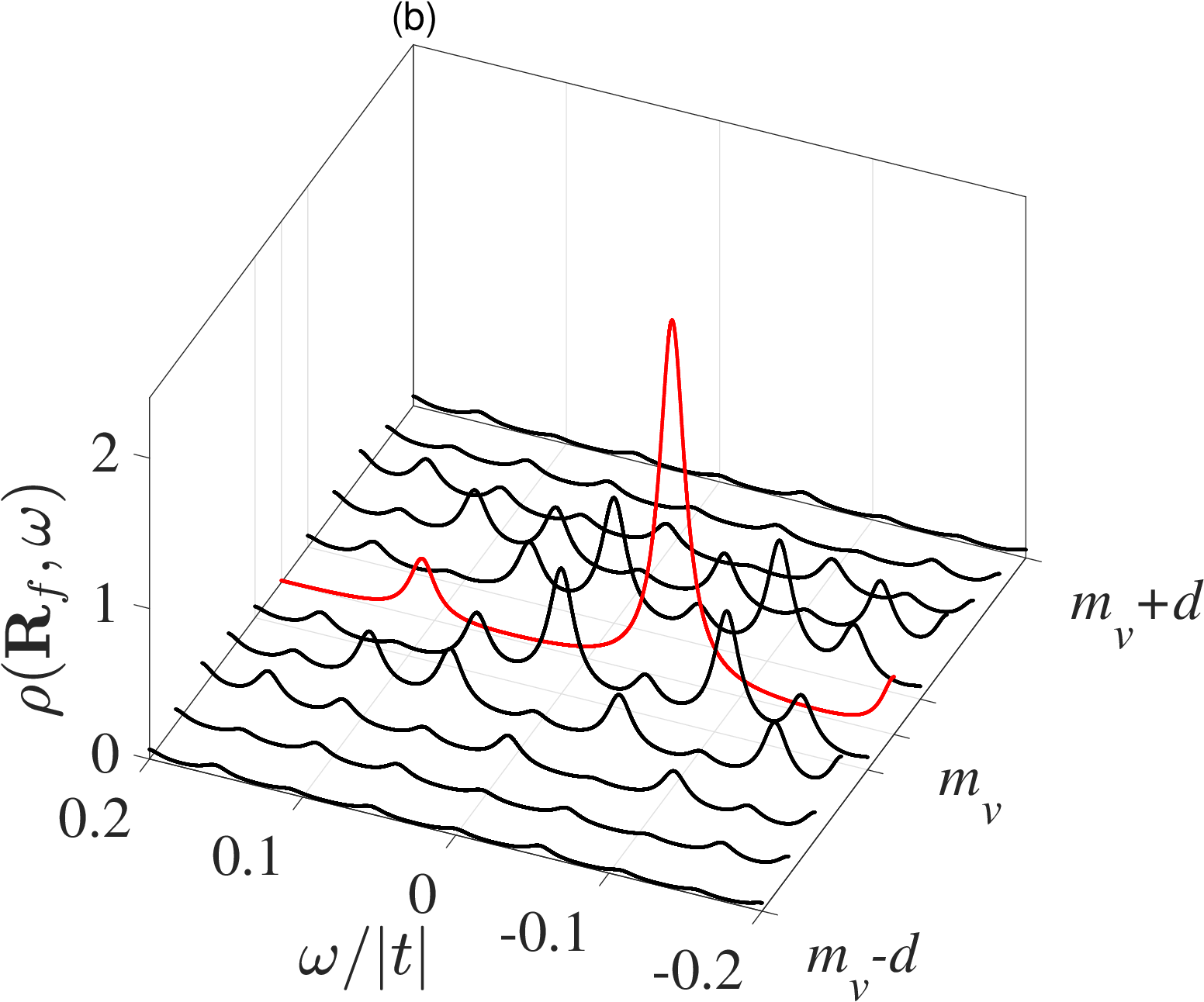}
        \caption{\label{pho_vort_avort} (a) The spatially and energetically resolved density of states $\rho \left( {\bf R}_f = (n, m), \omega \right)$ near the single vortex center $\left(n_v, m_v\right)$ and zero energy at $\mu \approx - 0.4205$ supporting Majorana vortex mode. The coordinate $n = n_v$ is fixed, while $m$ is changed from $m_v-d$ to $m_v+d$ with $d=5$. (b) The same map of $\rho \left( {\bf R}_f, \omega \right)$ for the single antivortex (the antivortex center is in the center of the triangular lattice) at slightly modified $\mu \approx - 0.3962$ (the other parameters do not changed). Tuning $\mu$ is done to obtain Majorana modes for the antivortex case.}
\end{figure*}

The excitation energies at $\mu \approx - 0.4205$ are demonstrated in Fig. \ref{exc_energy}. The square markers are zero energy Majorana modes. The large points denote excitation energies of ingap bound states which are localized near the vortex core.  At finite energy such states are analogues of Caroli-de Gennes-Matricon bound states for the chiral d-wave superconductor with triangular lattice. The small points are for energies of edge states. The bulk gap size is marked by the horizontal dashed lines. It is seen that the energy difference between vortex bound states, including the Majorana vortex mode, is an order of magnitude higher than the difference between the adjacent energies corresponding to the edge states. Therefore the Majorana mode located in the vortex core can be observed by the methods probing LDOS in contrast to edge bound states.

Dependencies of LDOS on energy near the single vortex core, edges, and far from inhomogeneities are shown in Fig. \ref{LDOS_vort}(a). For comparison the energy dependence of DOS without vortices and under periodic boundary conditions is presented in Fig. \ref{LDOS_vort}(b). A well-defined bulk superconducting gap is seen. For the limited lattice in the Shubnikov phase the edge and vortex bound states appear. Notably, the Majorana vortex mode is well separated by energy from finite-energy vortex states. On the other hand, a band of overlapping edge states appears in the superconducting gap for LDOS near the edges with the chosen broadening parameter $\delta$. The integration areas for $\text{LDOS}(\omega)$ are shown in Fig \ref{LDOS_vort}(c). In calculations of $\text{LDOS}(\omega)$ the broadening parameter $\delta = \Delta/20$ for the energy levels is used (to compare with the experimental STM resolution in FeTe$_{0.55}$Se$_{0.45}$ see \cite{chen-18, zhu-20, pathak-21}).
It should be also noted that in the presence of Zeeman splitting $h_z = 0.1$ the average z-component of spin operator $\left\langle S_{f}^{z} \right\rangle$ is non-zero (approximately $0.025$) near the vortex core and edges, while it is close to zero in the bulk of the lattice.

\subsection{\label{secB} Single antivortex on the lattice}

It is known that there is difference between the wave functions of vortex bound states and bound states in the case of an antivortex in chiral superconductors ($p_x+ip_y$-wave and $d_1+id_2$-wave) \cite{kraus-09, lee-16}. The reason is that the phase winding of the order parameter is not equivalent in the case of the vortex and antivortex in chiral superconductors, while for s-wave superconductivity this effect is absent. We show that such difference is also manifested for Majorana bound states localized at the vortex or the antivortex cores in the case of the coexistence of chiral $d_1+id_2$ superconductivity and 120$^{\circ}$ spin ordering on the triangular lattice.

For the antivortex (with the same ${\bf R}_{v}$ in the center of the triangular lattice and $\xi = 4a$, $N_s = 63$) a number of zero modes appears at slightly different values of chemical potential comparing to the case of the vortex in Sec.~\ref{secA}. However, the Majorana modes are also localized near the antivortex core and near the lattice edges in the topological phase with $\tilde{N}_3 = 1$, and the energy spectrum with subgap states is qualitatively the same. As it is mentioned above, $\rho \left( {\bf R}_f, \omega \right)$ in the vicinity of the antivortex center should be distinguished from the one in the case of the vortex. To show it the dependencies of $\rho \left( {\bf R}_f, \omega \right)$ on coordinates and energy are presented in Fig. \ref{pho_vort_avort}(a) for the vortex at $\mu \approx - 0.4205$ and in Fig. \ref{pho_vort_avort}(b) for the antivortex at $\mu \approx - 0.3962$ (the other parameters remain unchanged). The chosen conditions support Majorana modes formation for both cases. The highlighted line on the graphs corresponds to the energy dependence of $\rho \left( {\bf R}_f, \omega \right)$ exactly at the center of the vortex/antivortex ${\bf R}_f = {\bf R}_v = \left(n_v, m_v\right)$ (in the basis of ${\bf a}_1$ and ${\bf a}_2$). To obtain the rest energy dependencies the coordinate $m$ of ${\bf R}_f$ is changed from $m_v-d$ to $m_v+d$ ($d=5$), while $n$ remains fixed $n=n_v$. It is seen that for the Majorana vortex mode  $\rho \left( {\bf R}_f, \omega \right)$ has a maximum exactly at the center of the vortex and at zero energy. On the other hand, the maximum of $\rho \left( {\bf R}_f, \omega \right)$  for the antivortex is shifted from zero energy at $m=m_v$ and from the antivortex center at $\omega = 0$. It should be noted that LDOS for the antivortex case qualitatively does not changed from LDOS shown in Fig. \ref{LDOS_vort}. Therefore, to distinguish the obtained features of $\rho \left( {\bf R}_f, \omega \right)$ for the vortex and the antivortex fairly precise positioning of a tip in STM experiments for larger vortices should be carried out.

\subsection{\label{secC} Pair of vortices or vortex-antivortex pair on the lattice}

\begin{figure} [b]
        \includegraphics[width=0.4\textwidth]{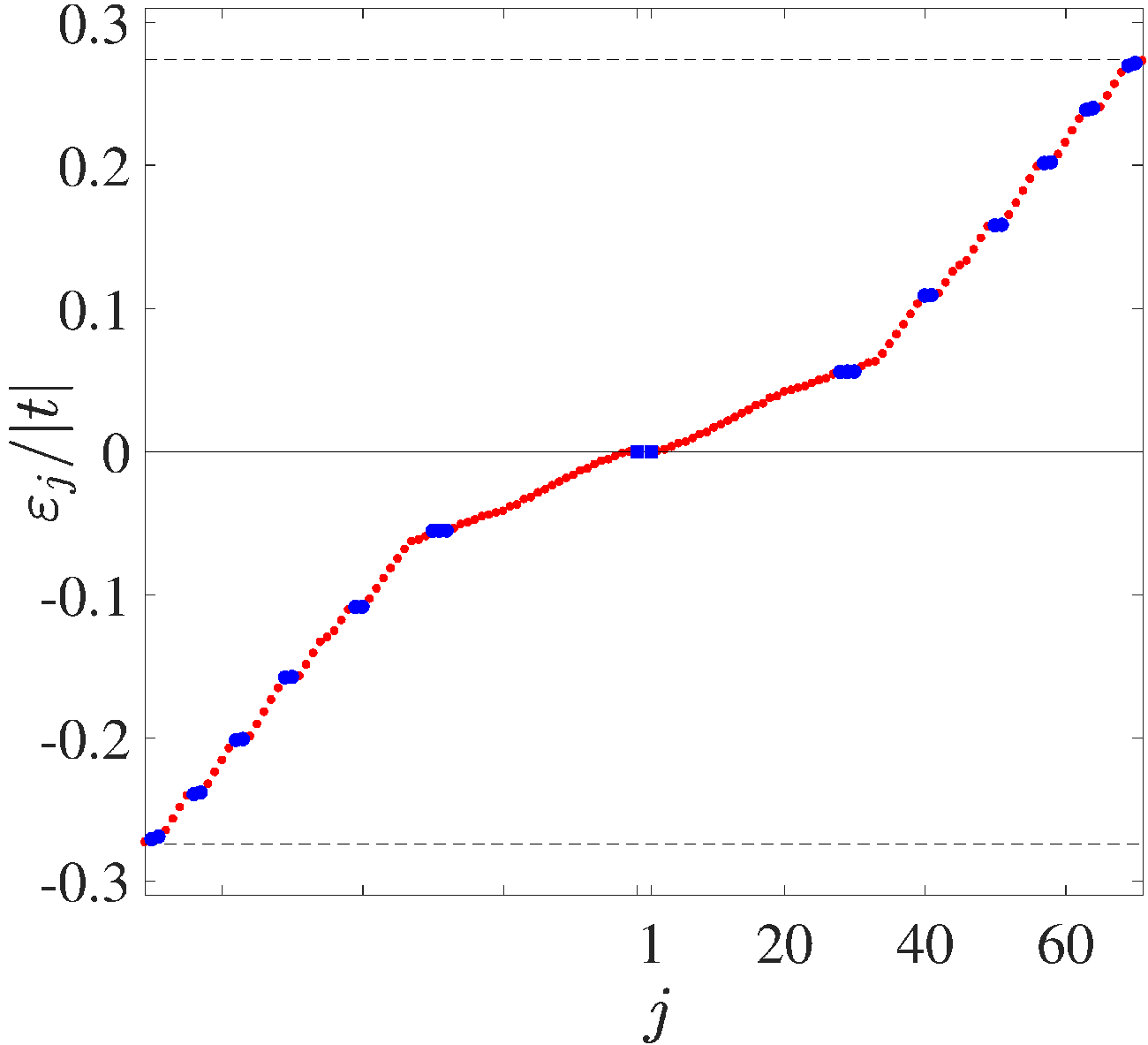}
        \caption{\label{exc_energy_2vort} The excitations energies in the presence of the pair of vortices with the vortex centers at $\left( \frac{N_s}{3}, \frac{N_s+1}{2} \right)$ and $\left( \frac{2N_s}{3}+1, \frac{N_s+1}{2} \right)$ (the lattice size is $N_s = 99$) at $\mu \approx - 0.389$. Highlighting markers at zero energy denote Majorana modes, large points --- ingap vortex bound states, small points are for ingap edge bound states. The bulk gap is denoted by the dashed lines.}
\end{figure}

As it was discussed in Sec. \ref{sec22}, Majorana modes have to be localized at the cores of the vortices considering the vortex-vortex or vortex-antivortex pairs. Usually the corresponding eigenstate problem is solved with periodic boundary conditions. Existence of edges in real systems can influence on zero modes: if the vortex is located close enough to the edge then each zero energy mode is represented as a superposition of vortex bound states and edge bound states. In the  studied model this is the case for the pair of vortices with coordinates of the centers $\left( \frac{N_s}{3}, \frac{N_s+1}{2} \right)$ and $\left( \frac{2N_s}{3}+1, \frac{N_s+1}{2} \right)$ on the triangular lattice with $N_s = 63$. The pair position is the same as in Sec. \ref{sec22}, but here open boundary conditions are applied and the vortices are parameterized by using \eqref{SCOP}. To exclude a mixing of vortex and edge bound states in the lattice with open boundary conditions, the lattice size should be increased. Therefore in this section the triangular lattice with $N_s = 99$ is considered.

Considering the pair of vortices, Majorana vortex modes and zero-energy edge modes can be separated from each other. Therefore in the topological phase with $\tilde{N}_3 = 1$ a part of zero modes at certain values of chemical potential contain two Majorana modes localized at the vortex cores, while for the other values of chemical potential the zero modes are near the edges. This is in contrast with the case of the single vortex, when all the zero energy solutions in the $\tilde{N}_3 = 1$ phase contain simultaneously one Majorana mode inhabiting at the vortex core and the other Majorana mode localized in the vicinity of the edges.

Next we are going to focus on the Majorana vortex modes instead of the edge modes and demonstrate their energy resolution from the other bound states. The result for the excitation spectrum in the presence of the pair of vortices at $\mu \approx - 0.389$ is presented in Fig. \ref{exc_energy_2vort}. As before, the markers at zero energy are for Majorana modes, the large points are vortex bound states, the small points correspond to energies of edge states.

\begin{figure} [t]
        \includegraphics[width=0.2\textwidth]{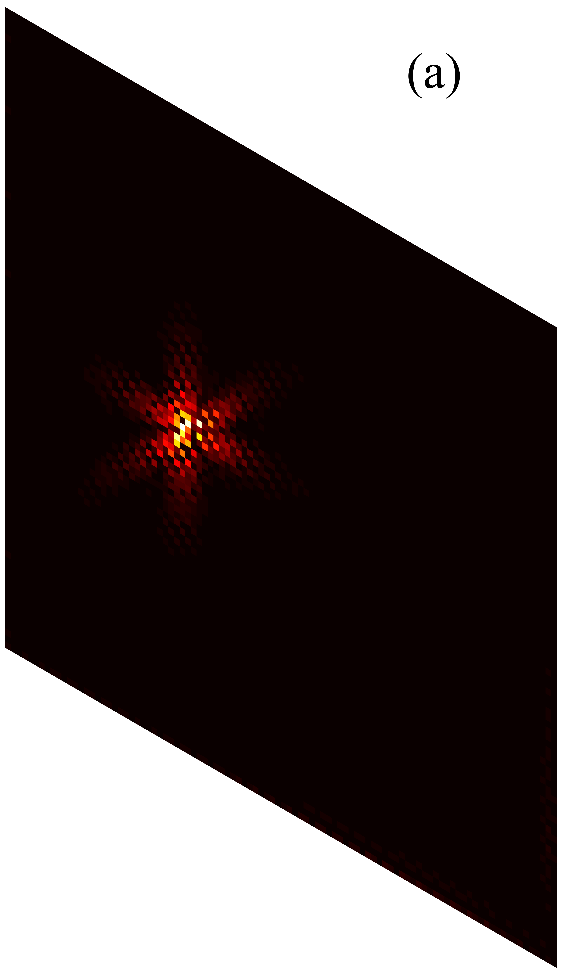}
        \includegraphics[width=0.2\textwidth]{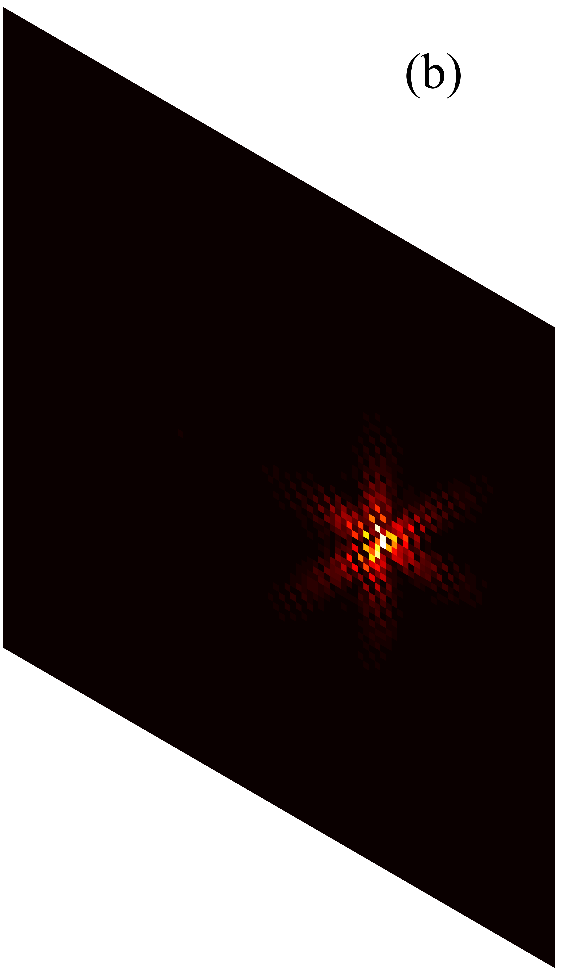}
        \caption{\label{m_modes_2vort} Majorana modes $b^{\prime}$  (a) and $b^{\prime \prime}$ (b) localized on the pair of vortices. Black color corresponds to near zero values, the brightest color is for maximum values $\approx 0.125$. The parameters are the same as in Fig. \ref{exc_energy_2vort}.}
\end{figure}

As it is known, an increase of the lattice size increases a number of subgap states. We can see this result comparing Fig. \ref{exc_energy_2vort} and Fig. \ref{exc_energy}.  It is also seen that the difference between second and first excitation energies is decreased comparing to the case of Fig. \ref{exc_energy}. However, the second excitation energy corresponds to the edge bound states, while the first excitation with zero energy matches to the formation of spatially separated Majorana modes localized on the various vortex cores. The distributions of modes $b^{\prime}$ and $b^{\prime \prime}$ are demonstrated in Fig. \ref{m_modes_2vort}. The probability density near the edges is negligible. For the pair of vortices the energy difference between Majorana vortex modes and the other vortex bound states as well as LDOS behavior remain qualitatively the same as in the case of the single vortex shown in Fig. \ref{LDOS_vort}(a).

When the zero modes are localized only near the edges at certain parameters, the second excitation energy corresponds to the formation of vortex bound states. Further increase in the lattice size and intervortex distance causes the second excitation energy to shift to near zero energy.

For the vortex-antivortex pair the Majorana modes localized at the cores also exist and the excitation energy spectrum with LDOS are qualitatively the same as for the pair of vortices. Although the tiny features of $\rho \left( {\bf R}_f, \omega \right)$ near the antivortex center described in Sec.~\ref{secB} are preserved for the vortex-antivortex pair.

\begin{figure} [t]
        \includegraphics[width=0.36\textwidth]{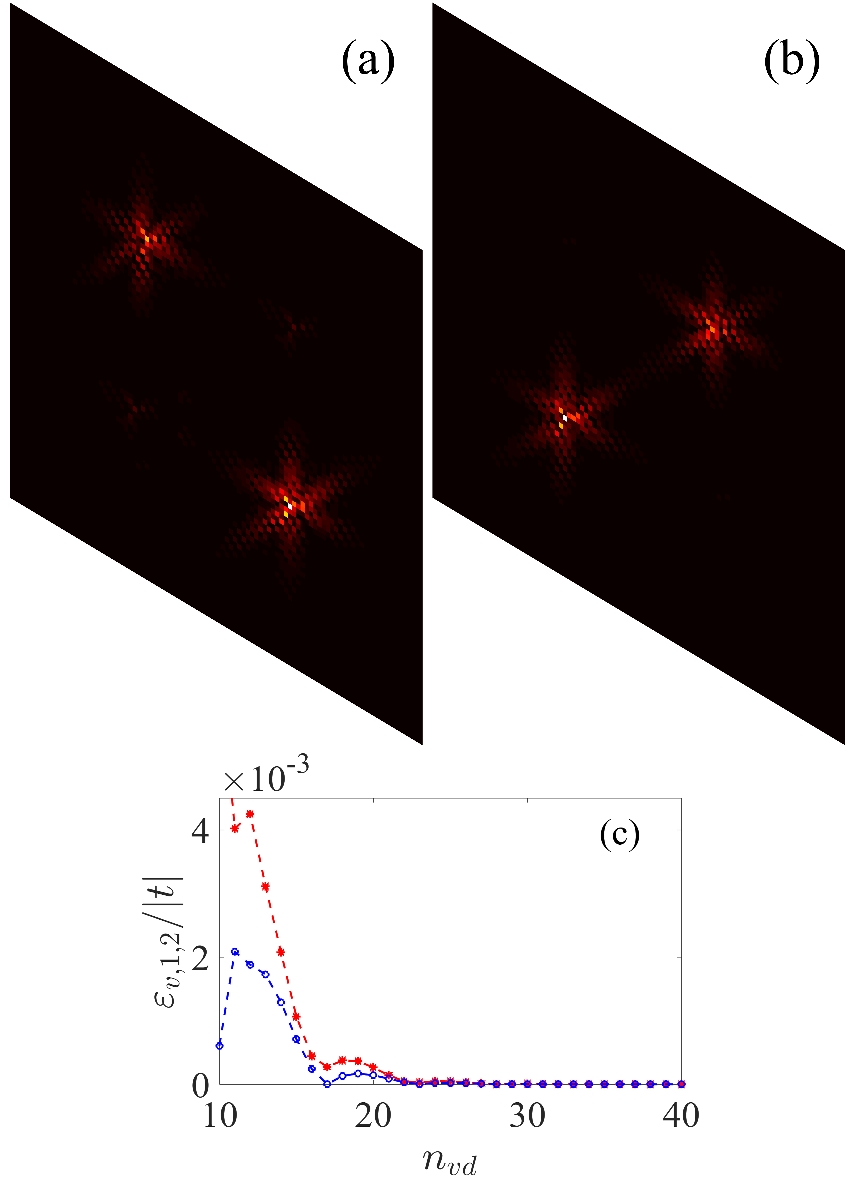}
        \caption{\label{m_modes_4vort} Probability densities $\sqrt{\text{PrD}}_1$ for the first excitation with zero energy (a) and $\sqrt{\text{PrD}}_2$ for the second excitation energy ($\varepsilon_2 \approx 2.7 \, 10^{-4}$) (b) in the presence of four vortices. Majorana modes $b^{\prime}$ and $b^{\prime \prime}$ in (a) are localized on different vortex cores. (c) The dependence of the lowest excitation energies corresponding to vortex bound states on intervortex distance $2n_{vd}$ in the lattice with $N_s = 153$. The energies of edge states are not shown. The point $n_{vd} = 17$ corresponds to the vortex structure depicted at (a) and (b).}
\end{figure}

\subsection{\label{secD} Four vortices on the lattice}

In this section four vortices are described with ${\bf R}_{v_{i=1\dots4}} = \left( \frac{N_s+1}{2} - n_{vd}, \frac{N_s+1}{2}-n_{vd} \right)$, $\left( \frac{N_s+1}{2} + n_{vd}, \frac{N_s+1}{2} + n_{vd} \right)$, $\left( \frac{N_s+1}{2} - n_{vd}, \frac{N_s+1}{2} + n_{vd} \right)$, $\left( \frac{N_s+1}{2} + n_{vd}, \frac{N_s+1}{2} - n_{vd} \right)$ on the triangular lattices with $N_s = 99$ and $N_s = 153$. The zero energy solutions exist on the same interval of chemical potential as in the previous sections. The main difference from the previous results is that vortex bound states appear also for the second excitation nearby with zero-energy Majorana vortex modes.
This result is demonstrated in Fig.~\ref{m_modes_4vort}a where $\sqrt{\text{PrD}}_1$ \eqref{PrD} of the first excitation is shown for $N_s = 99$, $n_{vd} = (N_s+3)/6$, and $\mu = -0.5902$. We take the square root here to resolve sixfold rotational symmetry of the finding solutions.

Majorana modes $b^{\prime}$ and $b^{\prime \prime}$ with zero energy are localized
on different cores of the pair of vortices in Fig.~\ref{m_modes_4vort}a. It is seen that the vortices in this pair are more distant from each other than vortices in the other pairs in the four-vortex system. Further, vortex bound states corresponding to the second excitation energy $\varepsilon_2 \sim 10^{-4}$ are formed on the another pair of vortices ($\sqrt{\text{PrD}}_2$ for the second excitation is presented in Fig.~\ref{m_modes_4vort}b). The nonzero energy of these bound states is caused by their overlapping. With increasing intervortex distance determined by $n_{vd}$ both energies oscillate and decrease to zero as it is seen from Fig.~\ref{m_modes_4vort}c with the results for the lattice with $N_s = 153$ at $\mu = -0.5902$. Therefore the bound states shown in Fig.~\ref{m_modes_4vort}b become Majorana vortex modes in the limit of well-separated vortices, as it is known \cite{read-00, stern-04, nayak-08}.
With further increasing of  $n_{vd}$ in Fig.~\ref{m_modes_4vort}c the energies grow due to overlapping of vortex and edge bound states.

The energy difference between such near-zero-energy vortex bound states and higher-energy vortex bound states is still $\approx 0.05$ as in the case of the single vortex or the pair of vortices, although a number of nearby high-energy excitations corresponding to vortex bound states also grows. This result confirms that Majorana vortex modes in the presence of several pairs of vortices on the limited lattice locally are still well separated on the energy scale to be experimentally detected by methods dealing with LDOS.

\begin{figure} [b]
        \includegraphics[width=0.4\textwidth]{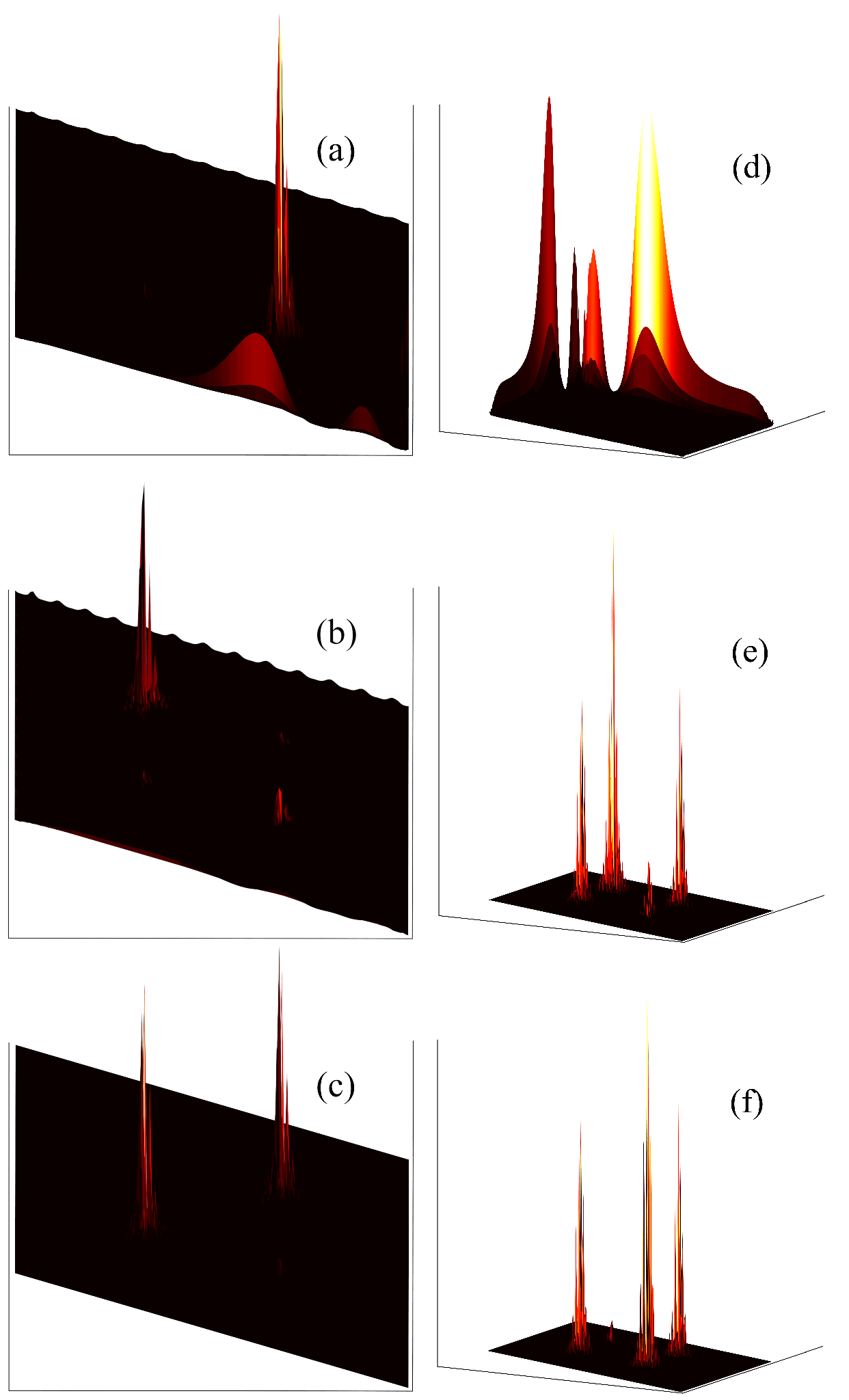}
        \caption{\label{diff_bs} (a)-(c) PrD$_{j=1, \, 2, \, 3}$ for the lowest three excitation energies ($\varepsilon_1 \to 0$ for (a); $\varepsilon_2 \approx 2.5 \, 10^{-6}$ for (b); $\varepsilon_3 \approx 1.1 \, 10^{-5}$ for (c)) in the lattice with $N_s = 153$ and four-vortex structure with $n_{vd} = (N_s+3)/6$ at $\mu = -0.669$. (e)-(f) Qualitatively different solutions for PrD$_{j=1, \, 2, \, 3}$ at $\mu = -0.7455$ ($\varepsilon_1 \to 0$ for (a); $\varepsilon_2 \approx 2.2 \, 10^{-5}$ for (b); $\varepsilon_3 \approx 4.5 \, 10^{-5}$ for (c)).}
\end{figure}

Note that Majorana vortex modes shown in Fig.~\ref{m_modes_4vort}a can be split by the edges. In this case one near-zero-energy solution is localized on the one vortex core from this pair and on the nearby edges, while the second solution is localized on the opposite vortex core and edges. The third excitation in this case is similar to the result shown in Fig.~\ref{m_modes_4vort}b. All these results are presented in Fig.~\ref{diff_bs}a-c.
Different zero modes are obtained by varying chemical potential $\mu$. There are also zero energy solutions corresponding to modes localized along the edges in the four-vortices case (see Fig.~\ref{diff_bs}d). In this case the second and third excitations involve vortex bound states (Fig.~\ref{diff_bs}e-f). These bound states are distributed predominantly over all four vortex cores or over three vortex cores in contrast to the localization on two cores presented in Fig.~\ref{m_modes_4vort}.

\subsection{\label{secE} Different lattice shapes}

To clarify the formation of Majorana vortex modes in limited lattices with different shapes
we consider the triangular lattice with the hexagon edges, preserving the rotational symmetry of triangular lattice, and the triangular lattice with disk-like shape.

\begin{figure} [b]
        \includegraphics[width=0.4\textwidth]{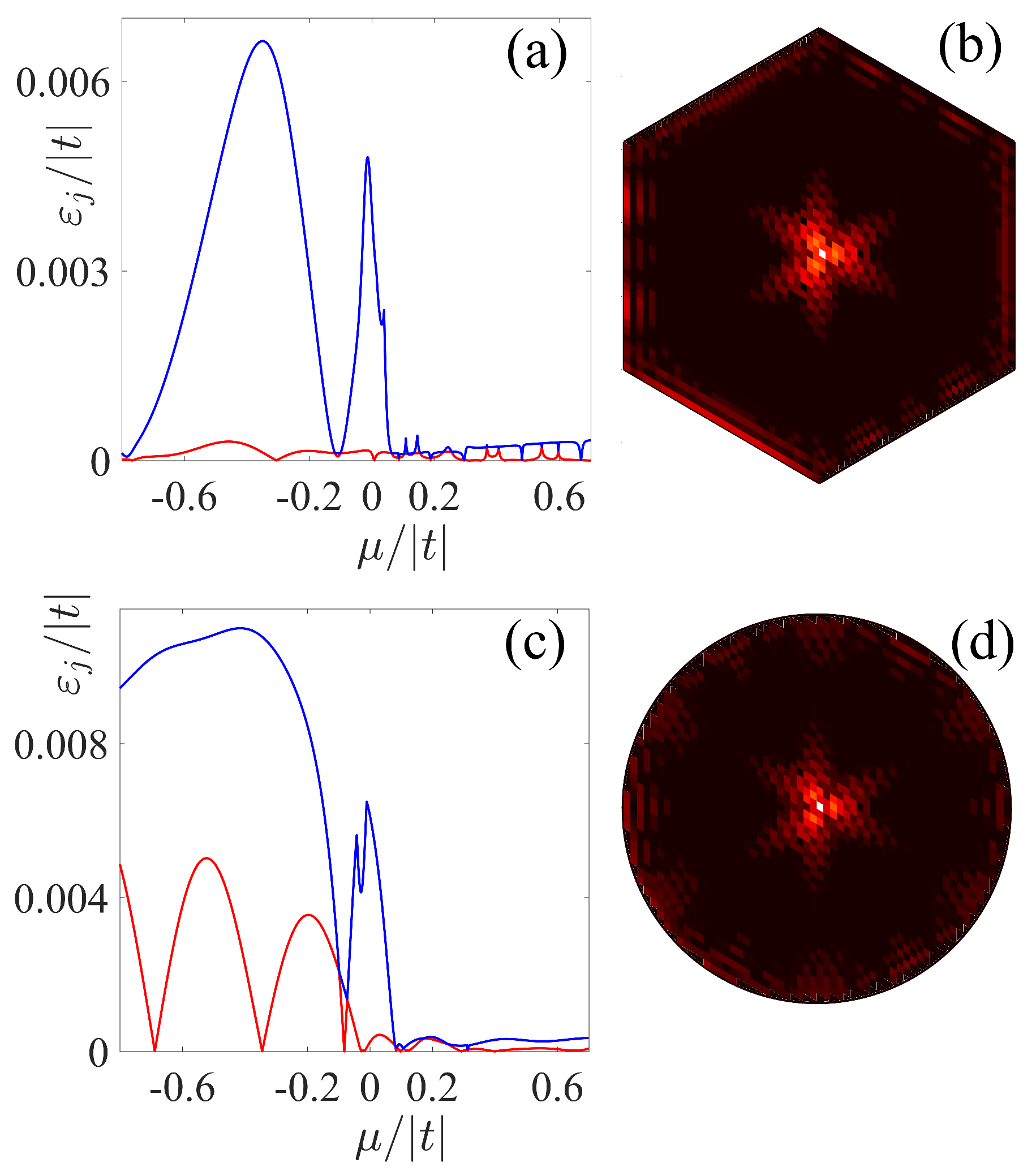}
        \caption{\label{hex_circle} (a) The chemical potential dependence of the lowest excitation
energies for the triangular lattice with the hexagon shape and the vortex at the lattice center. The parameters are as at Fig.~\ref{cond_zero}. (b) Distribution of Majorana modes ($\sqrt{\text{PrD}}_1$) on the lattice with the hexagon shape at $\mu = -0.3056$. (c) The dependencies $\varepsilon_{1,2}(\mu)$ for the triangular lattice with the disk shape containing one vortex. (d) $\sqrt{\text{PrD}}_1$ on the disk at $\mu = -0.3448$.}
\end{figure}

The hexagon shape contains $(3N_s^2+1)/4$ sites ($N_s = 63$ is taken). The same single vortex as in Sec.~\ref{secA} is considered at the center of such lattice. At Fig.~\ref{hex_circle}a the dependencies of the lowest excitation energies on chemical potential are demonstrated for the hexagon. It is seen that a more symmetric form of the lattice leads to the more regular dependencies of $\varepsilon_j(\mu)$ in the topological phase with $\tilde{N}_3 = 1$ in comparison with the parallelogram case described in Sec.~\ref{secA}. This regular dependence is similar to the case of periodic boundary conditions described in Sec.~\ref{sec22}.  However, the number of zero-energy solutions on the considered range in the $\tilde{N}_3 = 1$ phase decreases in the hexagon. As previously, the Majorana mode localized at the vortex core and the other mode near the edges appear for the zero energy solutions. The probability density $\sqrt{\text{PrD}}_1$ corresponding to zero energy at $\mu = -0.3056$ is shown at Fig.~\ref{hex_circle}b. The features of the excitation spectrum at $\mu = -0.3056$ such as the difference between nearby energies of edge states and the energies of vortex bound states are in quantitatively good agreement with the results presented in Fig.~\ref{exc_energy}.

To obtain the disk-like shape of the triangular lattice we take the circle with the radius $|{\bf R}_{c} - {\bf R}_{v}| = 25.24a$ and the vortex at the disk center. There are 36 sites of triangular lattice on the circle radius. Therefore, the obtained edge is not ideal circle. Such circle-like shape can be cut from the triangular lattice with the parallelogram shape and $N_s = 63$. The results on excitation energies depending on $\mu$ and Majorana modes formation in the disk-like shape are shown at Fig.~\ref{hex_circle}c,~d.

We argue that the obtained results do not qualitatively (and even quantitatively for the energy spectrum) depend on the shape of the limited triangular lattice. Nevertheless, the considered in this Section hexagon and disk shapes may be more difficult to obtain in experiments.

\section{\label{sec4} Discussion}

In the described results the focus is on the presence of few vortices in the lattice instead of the case of periodic vortex lattice. It is connected with the fact that there are novel experimental possibilities to create and manipulate such single vortices or vortex-antivortex pairs, such as far-field optical method with use of heating by a laser beam~\cite{veshchunov-16, rochet-20}, local stress with the tip in scanning SQUID microscopy~\cite{kremen-16}, and heating by the scanning tunneling microscope tip~\cite{ge-17}. Moreover such structures are more suitable to provide braiding of Majorana vortex modes.

It is known that for well-separated $2N_v$ ($N_v$ is integer) vortices there are $2N_v$ Majorana vortex modes \cite{read-00, stern-04, nayak-08}. For such vortices the phase term of the superconducting order parameter \eqref{SCOP} near the $i$th vortex can be replaced by
\begin{equation}
e^{i \arg(z({\bf R}_{ff_1}) - z({\bf R}_{v_i}))} e^{\Omega_i}
\end{equation}
with the relative phase
$\Omega_i = \sum_{j \ne i} i\arg(z({\bf R}_{v_j}) - z({\bf R}_{v_i}))$.

For periodic vortex lattices Majorana bands are flat and approach zero energy in the limit of a dilute vortex lattice~\cite{silaev-13, liu-15, pathak-21}. As an intervortex hybridization increases the Majorana bands are broadened. These results are obtained with taking into account the orbital effects of the applied magnetic field as a necessary component to describe stable vortex lattices~\cite{franz-00, takigawa-00}.

Increasing the excitation energy due to vortex modes hybridization can be traced in the presented results for the four-vortex case. It is shown that an overlapping of bound states of nearest vortices shifts the excitation energy from zero, as it shown in Fig. \ref{m_modes_4vort} (b).

It is expected that for topological superconductors with noncollinear spin ordering the results in the presence of well-separated vortices or periodic vortex lattice will remain qualitatively the same as in other topologically nontrivial systems. Nevertheless, such an analysis is beyond the scope of this work focusing on the small structures of vortices. We suppose that the obtained results, when Majorana modes may not exist on each vortex pair and be disturbed by edges and other neighboring vortices, should be taken into account in experimental studies of a structure of few vortices.

The results for LDOS show that the energy gap for Majorana vortex modes is determined by the excitation energy of vortex bound states. In s-wave superconductors these states are Caroli-de Gennes-Matricon states with the energy $\varepsilon_{\nu}^{s} \sim \nu \Delta/k_F \xi_s$~\cite{caroli-64}, where $\xi_s$ is the coherence length, $k_F$ is the radius of Fermi circle, and $\nu$ is a half integer. The coherence length for s-wave superconductor is found as $\xi_s = \hbar v_F/2 \Delta$. Therefore the energy $\varepsilon^{s}_{\nu}$ becomes $\sim \nu \Delta^2/\varepsilon_F$. It should be noted that the additional spin mixing mechanisms in spin-singlet topological superconductors, such as spin-orbit coupling or noncollinear magnetic order, significantly complicates an analytical estimation of the energy of bound states. However, the numerical solution of this problem is often used (for example, \cite{iskin-12}).

Using the results of Lee and Schnyder \cite{lee-16}, for the tight-binding model of chiral d-wave superconductor on the triangular lattice in the long-wave limit (near the bottom of the initial electron band) the excitation energy of vortex bound states can be expressed as

\begin{equation}
\label{CdGM_did}
\varepsilon^{d+id}_{\nu} = \frac{2 \nu |\tilde{\Delta}| k_Fa}{\xi/a} C.
\end{equation}
Here, $a$ is a lattice parameter, $\tilde{\Delta} = 3\Delta/4$ ($\Delta$ is an amplitude of the superconducting order parameter in \eqref{SCOP}), and $C$ is almost the same factor that appeared in the s-wave case (it has order of 1, but also can have slow linear dependence on $\Delta$). The additional details
are presented in Appendix B. The difference of this result from the case of s-wave superconductor can be explained by the fact that the bulk $d_1+id_2$ superconducting gap is $g= |\tilde{\Delta}| (k_Fa)^2$ in the considered approximation. Then the coherence length can be estimated as $\xi = 1/(2|\tilde{\Delta}|k_F/\tilde{t})$ with $\tilde{t} = 3t$, and $t$ is the hopping parameter between nearest neighbors in the lattice. Thus, finally we get $\varepsilon^{d+id}_{\nu} \sim l |\tilde{\Delta}|^2(k_F a)^2/\tilde{t}$. The Fermi energy now is $\varepsilon_F = \varepsilon_0 + \mu = \tilde{t}(k_F a)^2/2$. It is seen that the energy of vortex bound states in chiral $d_1+id_2$ superconducting state linearly increases (decreases) with growth of chemical potential near the bottom (top) of the band.

\begin{figure} [b]
        \includegraphics[width=0.4\textwidth]{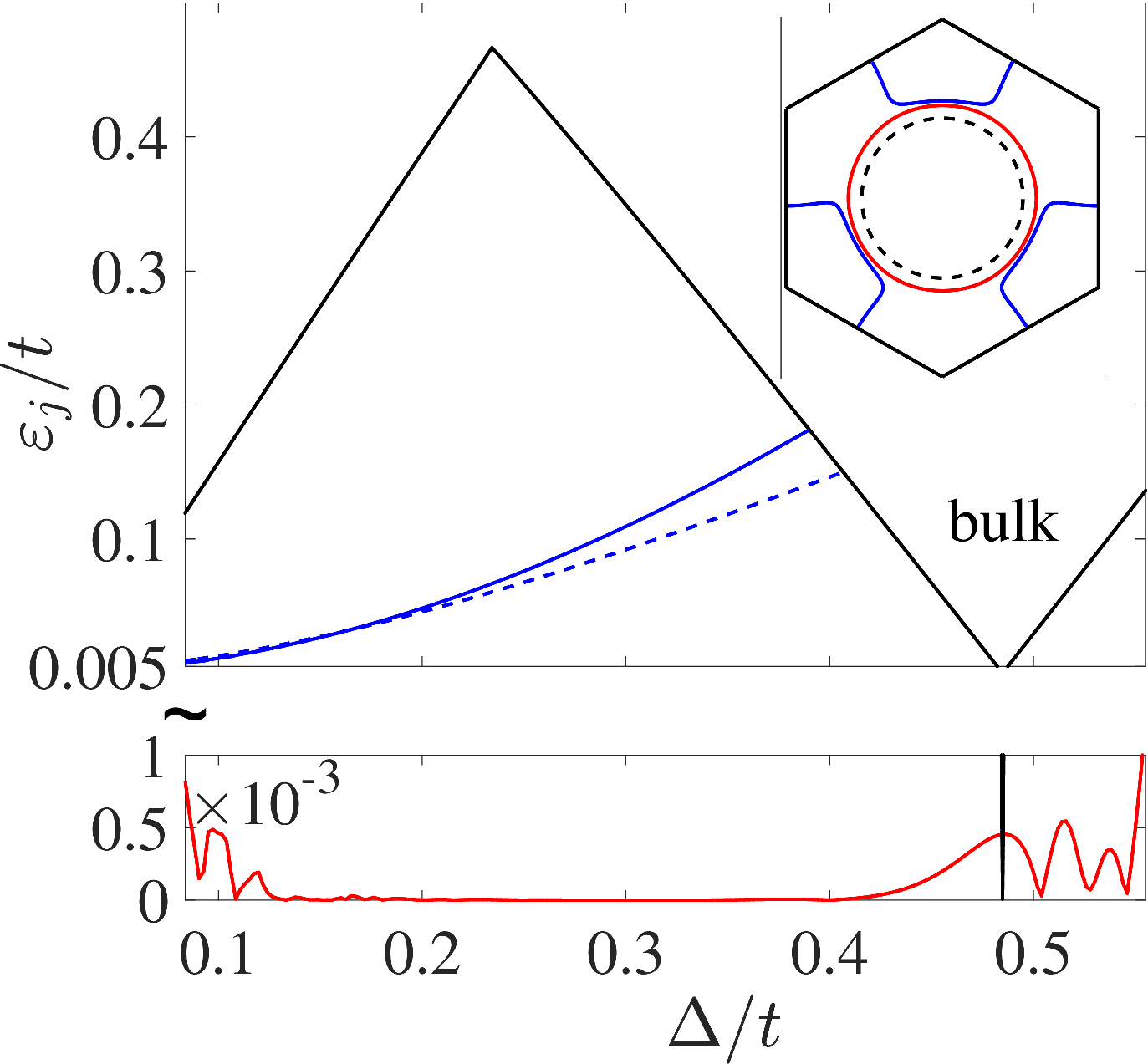}
        \caption{\label{Del_sq} The dependence of the excitation energy for Majorana vortex modes (red solid line) and the nearest non-zero energy of vortex bound state (blue solid line) on the amplitude $\Delta$ of the superconducting order parameter for the single vortex (see Sec. \ref{secA}, other excitation energies are omitted). The black solid line shows the dependence of the superconducting bulk gap above which bulk excitations appear. The dashed line is the analytical result (see \eqref{CdGM_did} and \cite{lee-16}) for the energy of vortex bound states in chiral d-wave superconductor with the triangular lattice. In the inset the hexagonal Brillouin is presented with the Fermi circle in the long-wave approximation (dashed line), the real Fermi contour in the nonmagnetic case (red solid line), and the Fermi contour in the presence of 120$^{\circ}$ spin ordering (blue solid line) for the used parameters $\mu = -0.736$, $h = 1$.}
\end{figure}

We argue that the obtained dependence of $\varepsilon^{d+id}_{\nu}$ can not be spread for free electrons, since in this case the change $\Delta \to \Delta/k_F^2$ must be made~\cite{kallin-16}. Nevertheless, the increasing of energy of bound states with $k_F$ increase is confirmed by the numerical solution of the Bogoliubov-de Gennes equations for the model \eqref{Ham1} in the absence of noncollinear spin ordering. In the studied chemical potential range $\mu \in [-0.8 \, - \, 0.7]$ the long-wave approximation does not hold and the energy is reached it's maximum value $\sim |\Delta|^2/|t|$. It can be seen in Fig. \ref{PBC} where the excitation energies of vortex bound states in the presence of a vortex-antivortex pair in the $d_1+id_2$ nonmagnetic superconductor are presented. The energies oscillate due to the presence of two vortices near almost the same value. The excitation energy in the presence of a single vortex in the limited triangular lattice has the same order.

Appearance of noncollinear spin ordering leads to formation of Majorana modes, but the above mentioned behavior of the energy of vortex bound states remains the same. There is a rise of excitation energies of bound states with increasing chemical potential (it can be seen in Fig.~\ref{PBC}). As shown in Fig.~\ref{Del_sq} for the case of the single vortex (Sec. \ref{secA}) in the center of the triangular lattice with $N_s = 99$, the energy of the first vortex bound state with nonzero energy (blue solid line) grows nearly quadratically with increasing the amplitude $\Delta$ of the superconducting order parameter while the other parameters are fixed ($\mu = -0.736$, $h = 1$). To obtain this result the coherence length $\xi$ in \eqref{SCOP} is considered inversely proportional to $\Delta$, such as $\xi/a = 0.8t/\Delta$. We note that for the previous results shown in Figs. \ref{PBC} and \ref{cond_zero} the coherence length has been approximately considered to be constant. The Fermi momentum $k_F$ unperturbed by the exchange filed increases on only 20 \% at the presented chemical potential range. Therefore the dependence $\xi$ on $k_F$ can be neglected on the qualitative level.

The excitation energy for Majorana vortex modes shown in Fig.~\ref{Del_sq} by the red solid line is zero with good accuracy on the wide range of $\Delta$. The black solid line is for the bulk superconducting gap. It seen that in the beginning the bulk energy increases linearly with $\Delta$. In this region the minimal bulk gap $g$ is realized on the Fermi contour in the Brillouin zone with regard to the long-range magnetic order and is equal to the $|\Delta_{k_{Fm}}|$, as in the nonmagnetic case. Here $k_{Fm}$ describes momentums of the Fermi contour in the presence of the spin ordering (see Appendix A). It is also seen that with further increasing $\Delta$ the bulk gap is linearly decreased. This behavior is caused by the noncollinear spin ordering, which drives the system to the topological phase transition similarly to that shown in the inset of Fig. \ref{cond_zero}(b).  On this curve the minimal gap is implemented at the symmetric points $(\pi/3, \pi/3)$, $(-2\pi/3, \pi/3)$, $(\pi/3, -2\pi/3)$ of the Brillouin zone \cite{zlotnikov-21} and $g = \left| \sqrt{(t-\mu)^2 + 4 \Delta^2} - h \right|$. It should be noted that the energy of vortex bound states does not follow to the bulk gap behavior in this case and it increases up to the bulk energy. We do not show excitation energies above the gap. For simplicity, here we also neglect contributions of the Zeeman term $h_z$, since $h_z \ll \Delta, h$.

The blue dashed line in Fig.~\ref{Del_sq} is the energy $\varepsilon_{1/2}^{d+id}$ from \eqref{CdGM_did} presented for comparison. The agreement of the numerical results with the analytical estimation is quite well. In the inset of Fig. \ref{Del_sq} the hexagonal Brillouin zone is presented, where the Fermi circle with the dashed line is obtained from the relation $\varepsilon_0 + \mu = \tilde{t}(k_F a)^2/2$, while the real almost-circle Fermi contour shown by the solid line is obtained from $t_{k_{Fc}} = \mu$ (the definition of $t_k$ is given in Appendix A). For the better agreement with the numerical results $k_{Fc}$ is used in Eq. \eqref{CdGM_did}. The other solid lines on the Brillouin zone are for the Fermi contour in the presence of the noncollinear magnetic ordering described by $k_{Fm}$. It is seen that for the used parameters some parts of the Fermi contour influenced  by the exchange field almost coincide with the Fermi circle $k_{Fc}$ unperturbed by spin ordering. Therefore, it may prove applicability of the estimation \eqref{CdGM_did} for the energy of vortex bound states even in the presence of noncollinear spin ordering.

The noncollinear spin ordering influences on the energy of vortex bound states in two different ways. First, growth of the exchange field in the $\tilde{N}_3 = 1$ phase increases the energy of vortex bound states. Qualitatively it can be explained by the fact that $k_{Fc}$ must be also increased in this case due to the modification of the initial bands by magnetic order. This effect is obtained both for fixed chemical potential or fixed average electron concentration while $h$ increases. On the other hand, the self-consistent calculation shows that the superconducting amplitude $\Delta$ slightly decreases in the presence of 120$^{\circ}$ spin ordering. We argue that noncollinear spin order does not destroy superconductivity and the previous results are consistent with the solutions of the self-consistent equation on $\Delta$. Decreasing $\Delta$ leads to reduction of the energy of vortex bound states. It is shown that the both effects almost fully compensate each other. To conclude noncollinear spin ordering in the considered system leads to formation of Majorana vortex modes, while the higher energy vortex bound states remains almost the same as in the chiral d-wave superconductor due to the competing effects. However, the problem of the analytical derivation of the bound states energy in the presence of noncollinear magnetism is an important issue for future research which may reveal effects hidden in the present study.

It also can be concluded that the ratio $\Delta/|t|$ in a superconducting structure with noncollinear spin ordering on the triangular lattice should exceed the experimental resolution of STM ($\approx 50 \, \mu \text{eV}$) to detect Majorana vortex modes studied here. It is believed that this limit can be reached in FeTe$_{1-x}$Se$_{x}$ and (Li$_{0.84}$Fe$_{0.16}$)OHFeSe with large enough superconducting gap ($0.7-10 \, \text{meV}$) on the background of small $\varepsilon_F$ ($2-50 \, \text{meV}$) \cite{chen-18, liu-18, wang-18, zhu-20, pathak-21}. Therefore an appearance of new materials and structures with the same features can be expected. However, the obtained increase of the vortex bound states energy with increasing chemical potential in the chiral d-wave superconducting state in contrast to the s-wave case may give an additional contribution when $k_Fa>1$ to achieve the necessary energy gap for Majorana vortex modes.

\section{\label{sec5} Conclusions}

In the present article structures with two and four vortices, as well as vortex-antivortex pair, in the 2D chiral $d_{x^2-y^2}+id_{xy}$ topological superconductor on the triangular lattice in the presence of 120$^{\circ}$ spin ordering are considered. Due to the presence of edges in real topological materials or existence of neighboring regions of a sample with various topology of the ground state (topologically trivial or nontrivial) the case of open boundary conditions is studied. It is shown that there appear many subgap states including zero energy modes in the Shubnikov phase under the simultaneous consideration of vortices and edges. The majority of finite-energy states are edge bound states, while the rest are vortex bound states (like Caroli-de Gennes-Matricon states).

It is shown that in the presence of a single vortex in the magnetic topological superconductor one zero mode is localized at the vortex core, while the other mode propagates along a closed path around the core, like in the other 2D topological superconductors. In the considered model the path coincides with the lattice edges. The zero modes for even number of vortices in the large enough lattice correspond either Majorana vortex modes or zero edge modes.

It is proved that namely the noncollinear magnetic order leads to formation of Majorana vortex modes with zero energy, since in the absence of the ordering vortex bound states can have only finite energy. This feature distinguishes Majorana vortex modes from zero edge modes, because the last can have zero (or near zero) energy both in the case of the presence of magnetic order and in the initial topologically nontrivial phase of the nonmagnetic chiral d-wave superconductor.

The exact numerical solutions of the Bogoliubov-de Gennes equations allow to study a hybridization of Majorana vortex modes between themselves and with the edges by changing the intervortex distance, lattice size, and position of the vortex relative to the edges. These effects are clearly manifested for the four-vortex structure. In the considered geometry the pair of Majorana vortex modes selects the pair of vortices which are most distant from each other. At the same time, for the remaining pair of vortices with the decreased intervortex distance the energy of vortex bound states is shifted from zero. We argue that simultaneous increasing of the lattice size and intervortex distance leads to formation of Majorana vortex modes on each vortex pair. However, such effects can be important on the limited scale.

Finite-energy vortex bound states can also have nearby excitation energies forming a cluster, and the number of such energies corresponds to the number of vortices. In turn, the energy difference between various clusters of vortex bound states in the excitation spectrum is an order of magnitude higher than the difference between the adjacent energies for edge states. This result shows that in principle the zero energy Majorana vortex modes can be distinguished from other higher energy vortex bound states in experiments, while the edge states can not. It is supposed that this can be done by tools measuring local density of states near the vortex core, such as scanning tunneling microscopy.

The studied system has distinct properties from the other topological superconductors. Firstly, the noncollinear spin ordering plays a role of the synthetic spin-orbit coupling and leads to formation of the topologically nontrivial phase supporting Majorana modes. Secondly, in the considered part  of the topological phase diagram there exist two different topologically nontrivial phases. The phase with topological invariant $\tilde{N}_3 = 1$ is induced by spin ordering and supports Majorana vortex modes, while the phase with $\tilde{N}_3 = 4$ is realized initially in the chiral d-wave superconductor in the absence of magnetic order. It is shown that vortex modes with zero energy appeared in the last phase in the presence of the noncollinear ordering are not Majorana modes, since they can not be separated on isolated modes. Studying the topological phase transition we demonstrate that in the limited lattice such phases differ by the excitation spectrum. Thirdly, it is shown that Majorana vortex modes and Majorana antivortex modes induced by noncollinear magnetism have different features in energy and spatially resolved density of states and can be distinguished from each other. This result is connected with the chiral symmetry of the superconducting order parameter.
Fourthly, it is found that although the energy of vortex bound states is renormalized by the exchange field, but it occurs in two competitive ways and finally the energy agrees well with the bound state energy in the nonmagnetic chiral d-wave superconductor. The considered energy determines the gap separating Majorana vortex modes from other bound states. We note that in order to experimentally detect the obtained Majorana vortex modes the ratio of the superconducting order parameter to hopping parameter (or Fermi energy) should exceed the experimental energy resolution as in other superconducting systems. Nevertheless, in the chiral d-wave state the energy gap for Majorana vortex modes can increases with growth of chemical potential in contrast to the case of s-wave superconducting state.

\begin{acknowledgments}
I am grateful for fruitful discussions with participants of the seminar of Laboratory of Theoretical Physics in Kirensky Institute of Physics, especially with V.V. Val'kov, S.V. Aksenov, A.D. Fedoseev, and M.S. Shustin. The reported study was funded by the Theoretical Physics and Mathematics Advancement Foundation ``BASIS''.
\end{acknowledgments}


\appendix
\section{\label{apxA}}

Hamiltonian \eqref{Ham1} under periodic boundary conditions and with the homogeneous superconducting order parameter in the absence of vortices can be written in k-space as
\begin{eqnarray}
\label{Ham1_PBC}
\mathscr{H} & = & \sum_{k \sigma} \xi_{k \sigma} c_{k \sigma}^{\dag}c_{k \sigma} + h \sum_{k} \left( c_{k \uparrow}^{\dag}c_{k-Q \downarrow} + c_{k-Q \downarrow}^{\dag}c_{k \uparrow}  \right) +
\nonumber \\
& + & \sum_{k} \left( \Delta_{k} c_{k \uparrow}^{\dag}c_{-k \downarrow}^{\dag}  + \Delta_{k}^*  c_{-k \downarrow} c_{k \uparrow} \right),
\end{eqnarray}
where $\xi_{k \sigma} = -\mu + t_k - \eta_{\sigma} h_z$,  the function $\eta_{\sigma}= +1$ for $\sigma = \uparrow$ and $-1$, otherwise for $\sigma = \downarrow$. It is seen, that in the momentum representation the space of the electron
states is limited by the subspace of the states $\left(k, \sigma = \uparrow \right)$ and $\left(k-Q, \sigma = \downarrow \right)$ invariant under the action of Hamiltonian for
all $k$. Therefore, the calculations in the presence of long-range magnetic ordering and with the periodic boundary conditions can be carried out on the whole first Brillouin zone in the nonmagnetic case instead of using many-sublattice representation. The mean-field Hamiltonian \eqref{Ham1_PBC} can be diagonalized by using Bogoliubov transformations.

In the nearest neighbors approximation
\begin{eqnarray}
t_{k} & = & 2t \left( \cos k_1 + \cos(k_1+k_2) + \cos k_2 \right),
\nonumber \\
\Delta_{k} & = & 2\Delta \left( \cos k_1 + e^{i2\pi/3} \cos(k_1+k_2) + e^{i4\pi/3} \cos k_2 \right),
\nonumber \\
\end{eqnarray}
and $k_i$ are determined the quasimomentum vector ${\bf k} = k_1{\bf b_1} + k_2{\bf b_2}$ in the units of the basic vectors ${\bf b_i}$ of the reciprocal lattice.

To calculate the topological invariant $\tilde{N}_3$ \eqref{Z_inv} the normal and the anomalous Matsubara Green's functions are introduced:
\begin{eqnarray}
\label{FGR_perp}
G_{0 \sigma, 0 s} \left(p, p'; \tau-\tau'\right) & = & - \left\langle T_{\tau} \tilde{c}_{p \sigma}\left(\tau\right) \tilde{c}_{p' s}^{\dag}\left(\tau'\right) \right\rangle,
\nonumber \\
F_{\sigma 0, 0 s} \left(p, p'; \tau-\tau'\right) & = & - \left\langle T_{\tau} \tilde{c}_{p \sigma}^{\dag}\left(\tau\right) \tilde{c}_{p' s}^{\dag}\left(\tau'\right) \right\rangle,
\nonumber \\
G_{\sigma 0, s 0} \left(p, p'; \tau-\tau'\right) & = & - \left\langle T_{\tau} \tilde{c}_{p \sigma}^{\dag}\left(\tau\right) \tilde{c}_{p' s}\left(\tau'\right) \right\rangle,
\nonumber \\
F_{0 \sigma, s 0} \left(p, p'; \tau-\tau'\right) & = & - \left\langle T_{\tau} \tilde{c}_{p \sigma}\left(\tau\right) \tilde{c}_{p' s}\left(\tau'\right) \right\rangle.
\end{eqnarray}
The Gorkov equations for coexistence of superconductivity and noncollinear spin ordering are obtained in the form
\begin{eqnarray}
\label{G_min_1}
&& \left[
  \begin{array}{cccc}
    i\omega_n-\xi_{p\sigma} & -\eta_{\sigma} \Delta_p & -h & 0 \\
    -\eta_{\sigma}\Delta_p^{*} & i\omega_n+\xi_{p \bar{\sigma}} & 0 & h \\
    -h & 0 & i\omega_n-\xi_{p-\eta_{\sigma}Q} & \eta_{\sigma}\Delta_{-p+\eta_{\sigma}Q} \\
    0 & h & \eta_{\sigma}\Delta^*_{-p+\eta_{\sigma}Q} & i\omega_n+\xi_{-p+\eta_{\sigma}Q} \\
  \end{array}
\right]
\nonumber \\
&&\left[
  \begin{array}{c}
   G_{0 \sigma, 0 \sigma} \left(p, p; i\omega_n \right)\\
   F_{\bar{\sigma} 0, 0 \sigma} \left(-p, p; i\omega_n \right)\\
   G_{0 \bar{\sigma}, 0 \sigma} \left(p-\eta_{\sigma}Q, p; i\omega_n \right)\\
   F_{\sigma 0, 0 \sigma} \left(-p+\eta_{\sigma}Q, p; i\omega_n \right)\\
  \end{array}
\right] = \left[
  \begin{array}{c}
   1\\
   0\\
   0\\
   0\\
  \end{array}
\right].
\end{eqnarray}
Here, $\bar{\sigma}$ denotes the opposite direction of
the spin moment $\sigma$.

It can be easily found that the obtained
matrix of the system of Gorkov equations is the inverse matrix of the matrix Green's function $\widehat{G}_{\sigma}$ included in \eqref{Z_inv} after the replacement $\omega_n \to \omega$. The definition of this matrix is
\begin{equation}
\label{Def_D}
\widehat{G}_{\sigma} = \left[
  \begin{array}{cc}
    \widehat{G}_{1\sigma}(p; i\omega_n) & \widehat{G}_{2 \sigma}(p, p-\eta_{\sigma}Q; i\omega_n) \\
    \widehat{G}_{2 \bar{\sigma}}(p-\eta_{\sigma}Q, p; i\omega_n) & \widehat{G}_{1 \bar{\sigma}}(p-\eta_{\sigma}Q; i\omega_n) \\
  \end{array}
\right],
\end{equation}
and $\widehat{G}_{1 \sigma} (p; i\omega_n) =$
\begin{eqnarray}
\left[
    \begin{array}{cc}
    G_{0 \sigma, 0 \sigma} (p, p; i\omega_n) & F_{0 \sigma, \bar{\sigma} 0} (p,-p; i\omega_n) \\
    F_{\bar{\sigma} 0, 0 \sigma} (-p,p; i\omega_n) & G_{\bar{\sigma} 0, \bar{\sigma} 0} (-p, -p, i\omega_n) \\
    \end{array}
\right], \nonumber
\end{eqnarray}
$\widehat{G}_{2 \sigma} (p, p'; i\omega_n) = $
\begin{eqnarray}
\left[
    \begin{array}{cc}
    G_{0 \sigma, 0 \bar{\sigma}} (p, p'; i\omega_n) & F_{0 \sigma, \sigma 0} (p, -p'; i\omega_n) \\
    F_{\bar{\sigma} 0, 0\bar{\sigma}} (-p,p'; i\omega_n) & G_{\bar{\sigma}0, \sigma 0} (-p,-p'; i\omega_n) \\
    \end{array}
\right]. \nonumber
\end{eqnarray}
From the equations $\text{det}\left( \widehat{G}_{\sigma}^{-1} \right) = 0$ the branches $\pm E_{ip}$ $(i=1, \,2, \,3, \,4)$ of the energy spectrum in the coexistence phase of superconductivity and noncollinear spin ordering with Zeeman splitting are obtained. The bulk gap is determined as a minimal gap (counting from the Fermi energy) of the energy spectrum in the Brillouin zone.

The energy spectrum in the coexistence phase of superconductivity and noncollinear spin ordering in the absence of Zeeman splitting $h_z$ is simplified to
\begin{eqnarray}
\label{spectr_sc_nco}
E_{1,2 \, k} & = & \sqrt{\frac{1}{2} \left( \xi_{k}^2 + \xi_{k-Q}^2 + |\Delta_k|^2 + |\Delta_{-k+Q}|^2  \right) + h^2  \mp \lambda_{k}},
\nonumber \\
\end{eqnarray}
where
\begin{eqnarray}
\lambda_{k} = \left\{ \frac{1}{4} \left( \xi_{k}^2 - \xi_{k-Q}^2 + |\Delta_k|^2 - |\Delta_{-k+Q}|^2 \right)^{2} + \right.
\nonumber \\
\left. + h^2 \left[ \left( \xi_{k} + \xi_{k-Q} \right)^2 + |\Delta_k+\Delta_{-k+Q}|^2 \right] \right\}^{1/2},
\nonumber
\end{eqnarray}
and $\xi_k = -\mu + t_k$.

The Fermi contour in the presence of noncollinear spin
ordering is found from the relation
\begin{eqnarray}
\label{cond_cont_Fermi}
\xi_{k_{Fm}}\xi_{k_{Fm}-Q} = h^2.
\end{eqnarray}
The bulk gap on the Fermi contour can be obtained from \eqref{spectr_sc_nco} by excluding the chemical potential according to \eqref{cond_cont_Fermi}. If the greatest terms are $h$ and $|t_{k_{Fm}} - t_{k_{Fm}-Q}|$, then the k-dependent bulk gap can be estimated as
\begin{eqnarray}
\label{gap_kFm}
g_{k_{Fm}} =  \left|\Delta_{k_{Fm}}\right| \, \, \, \text{for} \, \, \, t_{k_{Fm}}<t_{k_{Fm}-Q},
\\
g_{k_{Fm}} =  \left|\Delta_{k_{Fm}-Q}\right| \, \, \, \text{for} \, \, \, t_{k_{Fm}}>t_{k_{Fm}-Q}.
\end{eqnarray}
This value determines the bulk gap on the Fermi contour. Nevertheless, the minimal bulk gap can be realized in other points of the Brillouin zone due to noncollinear spin ordering.

\section{\label{apxB}}

By using the results of \cite{lee-16} the energy of vortex bound states of the chiral $d_1 + id_2$ superconductor on the triangular lattice can be written as
\begin{eqnarray}
\label{CdGM_did_apx}
&& \varepsilon^{d+id}_{\nu} = \nu \frac{|\tilde{\Delta}|^2 \left( \varepsilon_0 + \mu \right)}{\tilde{t}^2} 2\sqrt{2} L/a \times
\nonumber \\
&& \frac{\int_{a}^{\infty} dr/r \sinh(r/\xi) \cosh(r/\xi)^{-2\sqrt{2}\xi/L-1}}{\cosh(a/\xi)^{-2\sqrt{2}\xi/L} + \int_{a_{+}}^{\infty} dr/a \cosh(r/\xi)^{-2\sqrt{2}\xi/L}},
\nonumber \\
\end{eqnarray}
where the parameters $\tilde{\Delta}$, $\tilde{t}$, $\varepsilon_0$ can be obtained in the long-wave approximation expanding $t_k$ and $\Delta_k$ in the vicinity of $k_1 = k_2 =0$ (the bottom of $t_k$) or $k_1 = k_2 = -2\pi/3$ (the top of $t_k$). Near the bottom of the band they are
$|\tilde{\Delta}| = 3 |\Delta|/4$, $\tilde{t} = 3 |t|$, $\varepsilon_0 = 6 |t|$. $\xi$ is the coherence length and the effective length $L$ introduced in \cite{lee-16} is
\begin{equation}
L = \frac{a}{\tilde{|\Delta|}/\tilde{t} \sqrt{(\varepsilon_0+\mu)/\tilde{t}}}.
\end{equation}
The Fermi energy now is $\varepsilon_F = \varepsilon_0 + \mu = \tilde{t}(k_F a)^2/2$. Similarly to the s-wave superconducting state the coherence length in the chiral d-wave superconductor can be
estimated as
\begin{equation}
\xi = \frac{1}{2|\tilde{\Delta}|/\tilde{t}k_F}.
\end{equation}
Then expressing the chemical potential through $k_F$, it can be seen that $2\sqrt{2}\xi/L = 1$.
Therefore the expression \eqref{CdGM_did_apx} is simplified to
\begin{eqnarray}
\label{CdGM_did_2}
\varepsilon^{d+id}_{\nu} = \frac{2 \nu |\tilde{\Delta}| k_Fa}{\xi/a} C,
\end{eqnarray}
where
\begin{eqnarray}
C = \frac{\int_{a_{+}}^{\infty} dr/r \sinh(r/\xi) \cosh(r/\xi)^{-2}}{a/\xi\cosh(a/\xi)^{-1} + \pi - 2\arctan\left( e^{a/\xi} \right)}.
\end{eqnarray}
This formula is presented in the main text (see \eqref{CdGM_did}). It should be noted that the initial expression \eqref{CdGM_did_apx} for arbitrary coherence length leads to the qualitatively same results.

\bibliographystyle{aipnum4-1} 
\bibliography{bib_Zlotnikov_2022}

\providecommand{\noopsort}[1]{}\providecommand{\singleletter}[1]{#1}%
\begin{thebibliography}{71}%
\makeatletter
\providecommand \@ifxundefined [1]{%
 \@ifx{#1\undefined}
}%
\providecommand \@ifnum [1]{%
 \ifnum #1\expandafter \@firstoftwo
 \else \expandafter \@secondoftwo
 \fi
}%
\providecommand \@ifx [1]{%
 \ifx #1\expandafter \@firstoftwo
 \else \expandafter \@secondoftwo
 \fi
}%
\providecommand \natexlab [1]{#1}%
\providecommand \enquote  [1]{``#1''}%
\providecommand \bibnamefont  [1]{#1}%
\providecommand \bibfnamefont [1]{#1}%
\providecommand \citenamefont [1]{#1}%
\providecommand \href@noop [0]{\@secondoftwo}%
\providecommand \href [0]{\begingroup \@sanitize@url \@href}%
\providecommand \@href[1]{\@@startlink{#1}\@@href}%
\providecommand \@@href[1]{\endgroup#1\@@endlink}%
\providecommand \@sanitize@url [0]{\catcode `\\12\catcode `\$12\catcode
  `\&12\catcode `\#12\catcode `\^12\catcode `\_12\catcode `\%12\relax}%
\providecommand \@@startlink[1]{}%
\providecommand \@@endlink[0]{}%
\providecommand \url  [0]{\begingroup\@sanitize@url \@url }%
\providecommand \@url [1]{\endgroup\@href {#1}{\urlprefix }}%
\providecommand \urlprefix  [0]{URL }%
\providecommand \Eprint [0]{\href }%
\providecommand \doibase [0]{http://dx.doi.org/}%
\providecommand \selectlanguage [0]{\@gobble}%
\providecommand \bibinfo  [0]{\@secondoftwo}%
\providecommand \bibfield  [0]{\@secondoftwo}%
\providecommand \translation [1]{[#1]}%
\providecommand \BibitemOpen [0]{}%
\providecommand \bibitemStop [0]{}%
\providecommand \bibitemNoStop [0]{.\EOS\space}%
\providecommand \EOS [0]{\spacefactor3000\relax}%
\providecommand \BibitemShut  [1]{\csname bibitem#1\endcsname}%
\let\auto@bib@innerbib\@empty
\bibitem [{\citenamefont {Lutchyn}, \citenamefont {Sau},\ and\ \citenamefont
  {Das~Sarma}(2010)}]{lutchyn-10}%
  \BibitemOpen
  \bibfield  {author} {\bibinfo {author} {\bibfnamefont {R.~M.}\ \bibnamefont
  {Lutchyn}}, \bibinfo {author} {\bibfnamefont {J.~D.}\ \bibnamefont {Sau}}, \
  and\ \bibinfo {author} {\bibfnamefont {S.}~\bibnamefont {Das~Sarma}},\ }\href
  {\doibase 10.1103/PhysRevLett.105.077001} {\bibfield  {journal} {\bibinfo
  {journal} {Phys. Rev. Lett.}\ }\textbf {\bibinfo {volume} {105}},\ \bibinfo
  {pages} {077001} (\bibinfo {year} {2010})}\BibitemShut {NoStop}%
\bibitem [{\citenamefont {Oreg}, \citenamefont {Refael},\ and\ \citenamefont
  {von Oppen}(2010)}]{oreg-10}%
  \BibitemOpen
  \bibfield  {author} {\bibinfo {author} {\bibfnamefont {Y.}~\bibnamefont
  {Oreg}}, \bibinfo {author} {\bibfnamefont {G.}~\bibnamefont {Refael}}, \ and\
  \bibinfo {author} {\bibfnamefont {F.}~\bibnamefont {von Oppen}},\ }\href
  {\doibase 10.1103/PhysRevLett.105.177002} {\bibfield  {journal} {\bibinfo
  {journal} {Phys. Rev. Lett.}\ }\textbf {\bibinfo {volume} {105}},\ \bibinfo
  {pages} {177002} (\bibinfo {year} {2010})}\BibitemShut {NoStop}%
\bibitem [{\citenamefont {Shen}\ \emph {et~al.}(2021)\citenamefont {Shen},
  \citenamefont {Winkler}, \citenamefont {Borsoi}, \citenamefont {Heedt},
  \citenamefont {Levajac}, \citenamefont {Wang}, \citenamefont {van Driel},
  \citenamefont {Bouman}, \citenamefont {Gazibegovic}, \citenamefont
  {Op~Het~Veld}, \citenamefont {Car}, \citenamefont {Logan}, \citenamefont
  {Pendharkar}, \citenamefont {Palmstrøm}, \citenamefont {Bakkers},
  \citenamefont {Kouwenhoven},\ and\ \citenamefont {van Heck}}]{shen-21}%
  \BibitemOpen
  \bibfield  {author} {\bibinfo {author} {\bibfnamefont {J.}~\bibnamefont
  {Shen}}, \bibinfo {author} {\bibfnamefont {G.~W.}\ \bibnamefont {Winkler}},
  \bibinfo {author} {\bibfnamefont {F.}~\bibnamefont {Borsoi}}, \bibinfo
  {author} {\bibfnamefont {S.}~\bibnamefont {Heedt}}, \bibinfo {author}
  {\bibfnamefont {V.}~\bibnamefont {Levajac}}, \bibinfo {author} {\bibfnamefont
  {J.-Y.}\ \bibnamefont {Wang}}, \bibinfo {author} {\bibfnamefont
  {D.}~\bibnamefont {van Driel}}, \bibinfo {author} {\bibfnamefont
  {D.}~\bibnamefont {Bouman}}, \bibinfo {author} {\bibfnamefont
  {S.}~\bibnamefont {Gazibegovic}}, \bibinfo {author} {\bibfnamefont
  {R.~L.~M.}\ \bibnamefont {Op~Het~Veld}}, \bibinfo {author} {\bibfnamefont
  {D.}~\bibnamefont {Car}}, \bibinfo {author} {\bibfnamefont {J.~A.}\
  \bibnamefont {Logan}}, \bibinfo {author} {\bibfnamefont {M.}~\bibnamefont
  {Pendharkar}}, \bibinfo {author} {\bibfnamefont {C.~J.}\ \bibnamefont
  {Palmstrøm}}, \bibinfo {author} {\bibfnamefont {E.~P. A.~M.}\ \bibnamefont
  {Bakkers}}, \bibinfo {author} {\bibfnamefont {L.~P.}\ \bibnamefont
  {Kouwenhoven}}, \ and\ \bibinfo {author} {\bibfnamefont {B.}~\bibnamefont
  {van Heck}},\ }\href {\doibase 10.1103/PhysRevB.104.045422} {\bibfield
  {journal} {\bibinfo  {journal} {Phys. Rev. B}\ }\textbf {\bibinfo {volume}
  {104}},\ \bibinfo {pages} {045422} (\bibinfo {year} {2021})}\BibitemShut
  {NoStop}%
\bibitem [{\citenamefont {Val’kov}\ \emph {et~al.}(2022)\citenamefont
  {Val’kov}, \citenamefont {Shustin}, \citenamefont {Aksenov}, \citenamefont
  {Zlotnikov}, \citenamefont {Fedoseev}, \citenamefont {Mitskan},\ and\
  \citenamefont {Kagan}}]{valkov-22}%
  \BibitemOpen
  \bibfield  {author} {\bibinfo {author} {\bibfnamefont {V.~V.}\ \bibnamefont
  {Val’kov}}, \bibinfo {author} {\bibfnamefont {M.~S.}\ \bibnamefont
  {Shustin}}, \bibinfo {author} {\bibfnamefont {S.~V.}\ \bibnamefont
  {Aksenov}}, \bibinfo {author} {\bibfnamefont {A.~O.}\ \bibnamefont
  {Zlotnikov}}, \bibinfo {author} {\bibfnamefont {A.~D.}\ \bibnamefont
  {Fedoseev}}, \bibinfo {author} {\bibfnamefont {V.~A.}\ \bibnamefont
  {Mitskan}}, \ and\ \bibinfo {author} {\bibfnamefont {M.~Y.}\ \bibnamefont
  {Kagan}},\ }\href {\doibase 10.3367/ufne.2021.03.038950} {\bibfield
  {journal} {\bibinfo  {journal} {Phys.-Usp.}\ }\textbf {\bibinfo {volume}
  {65}},\ \bibinfo {pages} {2} (\bibinfo {year} {2022})}\BibitemShut {NoStop}%
\bibitem [{\citenamefont {Nayak}\ \emph {et~al.}(2008)\citenamefont {Nayak},
  \citenamefont {Simon}, \citenamefont {Stern}, \citenamefont {Freedman},\ and\
  \citenamefont {Das~Sarma}}]{nayak-08}%
  \BibitemOpen
  \bibfield  {author} {\bibinfo {author} {\bibfnamefont {C.}~\bibnamefont
  {Nayak}}, \bibinfo {author} {\bibfnamefont {S.~H.}\ \bibnamefont {Simon}},
  \bibinfo {author} {\bibfnamefont {A.}~\bibnamefont {Stern}}, \bibinfo
  {author} {\bibfnamefont {M.}~\bibnamefont {Freedman}}, \ and\ \bibinfo
  {author} {\bibfnamefont {S.}~\bibnamefont {Das~Sarma}},\ }\href {\doibase
  10.1103/RevModPhys.80.1083} {\bibfield  {journal} {\bibinfo  {journal} {Rev.
  Mod. Phys.}\ }\textbf {\bibinfo {volume} {80}},\ \bibinfo {pages} {1083}
  (\bibinfo {year} {2008})}\BibitemShut {NoStop}%
\bibitem [{\citenamefont {Read}\ and\ \citenamefont {Green}(2000)}]{read-00}%
  \BibitemOpen
  \bibfield  {author} {\bibinfo {author} {\bibfnamefont {N.}~\bibnamefont
  {Read}}\ and\ \bibinfo {author} {\bibfnamefont {D.}~\bibnamefont {Green}},\
  }\href {\doibase 10.1103/PhysRevB.61.10267} {\bibfield  {journal} {\bibinfo
  {journal} {Phys. Rev. B}\ }\textbf {\bibinfo {volume} {61}},\ \bibinfo
  {pages} {10267} (\bibinfo {year} {2000})}\BibitemShut {NoStop}%
\bibitem [{\citenamefont {Ivanov}(2001)}]{ivanov-01}%
  \BibitemOpen
  \bibfield  {author} {\bibinfo {author} {\bibfnamefont {D.~A.}\ \bibnamefont
  {Ivanov}},\ }\href {\doibase 10.1103/PhysRevLett.86.268} {\bibfield
  {journal} {\bibinfo  {journal} {Phys. Rev. Lett.}\ }\textbf {\bibinfo
  {volume} {86}},\ \bibinfo {pages} {268} (\bibinfo {year} {2001})}\BibitemShut
  {NoStop}%
\bibitem [{\citenamefont {Stern}, \citenamefont {von Oppen},\ and\
  \citenamefont {Mariani}(2004)}]{stern-04}%
  \BibitemOpen
  \bibfield  {author} {\bibinfo {author} {\bibfnamefont {A.}~\bibnamefont
  {Stern}}, \bibinfo {author} {\bibfnamefont {F.}~\bibnamefont {von Oppen}}, \
  and\ \bibinfo {author} {\bibfnamefont {E.}~\bibnamefont {Mariani}},\ }\href
  {\doibase 10.1103/PhysRevB.70.205338} {\bibfield  {journal} {\bibinfo
  {journal} {Phys. Rev. B}\ }\textbf {\bibinfo {volume} {70}},\ \bibinfo
  {pages} {205338} (\bibinfo {year} {2004})}\BibitemShut {NoStop}%
\bibitem [{\citenamefont {Gurarie}\ and\ \citenamefont
  {Radzihovsky}(2007)}]{gurarie-07}%
  \BibitemOpen
  \bibfield  {author} {\bibinfo {author} {\bibfnamefont {V.}~\bibnamefont
  {Gurarie}}\ and\ \bibinfo {author} {\bibfnamefont {L.}~\bibnamefont
  {Radzihovsky}},\ }\href {\doibase 10.1103/PhysRevB.75.212509} {\bibfield
  {journal} {\bibinfo  {journal} {Phys. Rev. B}\ }\textbf {\bibinfo {volume}
  {75}},\ \bibinfo {pages} {212509} (\bibinfo {year} {2007})}\BibitemShut
  {NoStop}%
\bibitem [{\citenamefont {Fu}\ and\ \citenamefont {Kane}(2008)}]{fu-08}%
  \BibitemOpen
  \bibfield  {author} {\bibinfo {author} {\bibfnamefont {L.}~\bibnamefont
  {Fu}}\ and\ \bibinfo {author} {\bibfnamefont {C.~L.}\ \bibnamefont {Kane}},\
  }\href {\doibase 10.1103/PhysRevLett.100.096407} {\bibfield  {journal}
  {\bibinfo  {journal} {Phys. Rev. Lett.}\ }\textbf {\bibinfo {volume} {100}},\
  \bibinfo {pages} {096407} (\bibinfo {year} {2008})}\BibitemShut {NoStop}%
\bibitem [{\citenamefont {Elliott}\ and\ \citenamefont
  {Franz}(2015)}]{elliott-15}%
  \BibitemOpen
  \bibfield  {author} {\bibinfo {author} {\bibfnamefont {S.~R.}\ \bibnamefont
  {Elliott}}\ and\ \bibinfo {author} {\bibfnamefont {M.}~\bibnamefont
  {Franz}},\ }\href {\doibase 10.1103/RevModPhys.87.137} {\bibfield  {journal}
  {\bibinfo  {journal} {Rev. Mod. Phys.}\ }\textbf {\bibinfo {volume} {87}},\
  \bibinfo {pages} {137} (\bibinfo {year} {2015})}\BibitemShut {NoStop}%
\bibitem [{\citenamefont {Yang}\ \emph {et~al.}(2016)\citenamefont {Yang},
  \citenamefont {Stano}, \citenamefont {Klinovaja},\ and\ \citenamefont
  {Loss}}]{yang-16}%
  \BibitemOpen
  \bibfield  {author} {\bibinfo {author} {\bibfnamefont {G.}~\bibnamefont
  {Yang}}, \bibinfo {author} {\bibfnamefont {P.}~\bibnamefont {Stano}},
  \bibinfo {author} {\bibfnamefont {J.}~\bibnamefont {Klinovaja}}, \ and\
  \bibinfo {author} {\bibfnamefont {D.}~\bibnamefont {Loss}},\ }\href {\doibase
  10.1103/PhysRevB.93.224505} {\bibfield  {journal} {\bibinfo  {journal} {Phys.
  Rev. B}\ }\textbf {\bibinfo {volume} {93}},\ \bibinfo {pages} {224505}
  (\bibinfo {year} {2016})}\BibitemShut {NoStop}%
\bibitem [{\citenamefont {Zlotnikov}, \citenamefont {Shustin},\ and\
  \citenamefont {Fedoseev}(2021)}]{zlotnikov-21}%
  \BibitemOpen
  \bibfield  {author} {\bibinfo {author} {\bibfnamefont {A.~O.}\ \bibnamefont
  {Zlotnikov}}, \bibinfo {author} {\bibfnamefont {M.~S.}\ \bibnamefont
  {Shustin}}, \ and\ \bibinfo {author} {\bibfnamefont {A.~D.}\ \bibnamefont
  {Fedoseev}},\ }\href {\doibase 10.1007/s10948-021-06029-z} {\bibfield
  {journal} {\bibinfo  {journal} {J. Supercond. Nov. Magn.}\ }\textbf {\bibinfo
  {volume} {34}},\ \bibinfo {pages} {3053} (\bibinfo {year}
  {2021})}\BibitemShut {NoStop}%
\bibitem [{\citenamefont {Volovik}(2010)}]{volovik-10}%
  \BibitemOpen
  \bibfield  {author} {\bibinfo {author} {\bibfnamefont {G.~E.}\ \bibnamefont
  {Volovik}},\ }\href {\doibase 10.1134/S0021364010040090} {\bibfield
  {journal} {\bibinfo  {journal} {JETP Lett.}\ }\textbf {\bibinfo {volume}
  {91}},\ \bibinfo {pages} {201} (\bibinfo {year} {2010})}\BibitemShut
  {NoStop}%
\bibitem [{\citenamefont {Zhu}(2018)}]{zhu-18}%
  \BibitemOpen
  \bibfield  {author} {\bibinfo {author} {\bibfnamefont {X.}~\bibnamefont
  {Zhu}},\ }\href {\doibase 10.1103/PhysRevB.97.205134} {\bibfield  {journal}
  {\bibinfo  {journal} {Phys. Rev. B}\ }\textbf {\bibinfo {volume} {97}},\
  \bibinfo {pages} {205134} (\bibinfo {year} {2018})}\BibitemShut {NoStop}%
\bibitem [{\citenamefont {Tewari}, \citenamefont {Das~Sarma},\ and\
  \citenamefont {Lee}(2007)}]{tewari-07}%
  \BibitemOpen
  \bibfield  {author} {\bibinfo {author} {\bibfnamefont {S.}~\bibnamefont
  {Tewari}}, \bibinfo {author} {\bibfnamefont {S.}~\bibnamefont {Das~Sarma}}, \
  and\ \bibinfo {author} {\bibfnamefont {D.-H.}\ \bibnamefont {Lee}},\ }\href
  {\doibase 10.1103/PhysRevLett.99.037001} {\bibfield  {journal} {\bibinfo
  {journal} {Phys. Rev. Lett.}\ }\textbf {\bibinfo {volume} {99}},\ \bibinfo
  {pages} {037001} (\bibinfo {year} {2007})}\BibitemShut {NoStop}%
\bibitem [{\citenamefont {Silaev}(2008)}]{silaev-08}%
  \BibitemOpen
  \bibfield  {author} {\bibinfo {author} {\bibfnamefont {M.~A.}\ \bibnamefont
  {Silaev}},\ }\href {\doibase 10.1134/S0021364008080110} {\bibfield  {journal}
  {\bibinfo  {journal} {Jetp Lett.}\ }\textbf {\bibinfo {volume} {87}},\
  \bibinfo {pages} {441} (\bibinfo {year} {2008})}\BibitemShut {NoStop}%
\bibitem [{\citenamefont {Kraus}\ \emph {et~al.}(2009)\citenamefont {Kraus},
  \citenamefont {Auerbach}, \citenamefont {Fertig},\ and\ \citenamefont
  {Simon}}]{kraus-09}%
  \BibitemOpen
  \bibfield  {author} {\bibinfo {author} {\bibfnamefont {Y.~E.}\ \bibnamefont
  {Kraus}}, \bibinfo {author} {\bibfnamefont {A.}~\bibnamefont {Auerbach}},
  \bibinfo {author} {\bibfnamefont {H.~A.}\ \bibnamefont {Fertig}}, \ and\
  \bibinfo {author} {\bibfnamefont {S.~H.}\ \bibnamefont {Simon}},\ }\href
  {\doibase 10.1103/PhysRevB.79.134515} {\bibfield  {journal} {\bibinfo
  {journal} {Phys. Rev. B}\ }\textbf {\bibinfo {volume} {79}},\ \bibinfo
  {pages} {134515} (\bibinfo {year} {2009})}\BibitemShut {NoStop}%
\bibitem [{\citenamefont {Silaev}(2013)}]{silaev-13}%
  \BibitemOpen
  \bibfield  {author} {\bibinfo {author} {\bibfnamefont {M.~A.}\ \bibnamefont
  {Silaev}},\ }\href {\doibase 10.1103/PhysRevB.88.064514} {\bibfield
  {journal} {\bibinfo  {journal} {Phys. Rev. B}\ }\textbf {\bibinfo {volume}
  {88}},\ \bibinfo {pages} {064514} (\bibinfo {year} {2013})}\BibitemShut
  {NoStop}%
\bibitem [{\citenamefont {Liu}\ and\ \citenamefont {Franz}(2015)}]{liu-15}%
  \BibitemOpen
  \bibfield  {author} {\bibinfo {author} {\bibfnamefont {T.}~\bibnamefont
  {Liu}}\ and\ \bibinfo {author} {\bibfnamefont {M.}~\bibnamefont {Franz}},\
  }\href {\doibase 10.1103/PhysRevB.92.134519} {\bibfield  {journal} {\bibinfo
  {journal} {Phys. Rev. B}\ }\textbf {\bibinfo {volume} {92}},\ \bibinfo
  {pages} {134519} (\bibinfo {year} {2015})}\BibitemShut {NoStop}%
\bibitem [{\citenamefont {Akzyanov}\ \emph {et~al.}(2016)\citenamefont
  {Akzyanov}, \citenamefont {Rakhmanov}, \citenamefont {Rozhkov},\ and\
  \citenamefont {Nori}}]{akzyanov-16}%
  \BibitemOpen
  \bibfield  {author} {\bibinfo {author} {\bibfnamefont {R.~S.}\ \bibnamefont
  {Akzyanov}}, \bibinfo {author} {\bibfnamefont {A.~L.}\ \bibnamefont
  {Rakhmanov}}, \bibinfo {author} {\bibfnamefont {A.~V.}\ \bibnamefont
  {Rozhkov}}, \ and\ \bibinfo {author} {\bibfnamefont {F.}~\bibnamefont
  {Nori}},\ }\href {\doibase 10.1103/PhysRevB.94.125428} {\bibfield  {journal}
  {\bibinfo  {journal} {Phys. Rev. B}\ }\textbf {\bibinfo {volume} {94}},\
  \bibinfo {pages} {125428} (\bibinfo {year} {2016})}\BibitemShut {NoStop}%
\bibitem [{\citenamefont {Iskin}(2012)}]{iskin-12}%
  \BibitemOpen
  \bibfield  {author} {\bibinfo {author} {\bibfnamefont {M.}~\bibnamefont
  {Iskin}},\ }\href {\doibase 10.1103/PhysRevA.85.013622} {\bibfield  {journal}
  {\bibinfo  {journal} {Phys. Rev. A}\ }\textbf {\bibinfo {volume} {85}},\
  \bibinfo {pages} {013622} (\bibinfo {year} {2012})}\BibitemShut {NoStop}%
\bibitem [{\citenamefont {Bj{\"o}rnson}\ and\ \citenamefont
  {Black-Schaffer}(2013)}]{bjornson-13}%
  \BibitemOpen
  \bibfield  {author} {\bibinfo {author} {\bibfnamefont {K.}~\bibnamefont
  {Bj{\"o}rnson}}\ and\ \bibinfo {author} {\bibfnamefont {A.~M.}\ \bibnamefont
  {Black-Schaffer}},\ }\href {\doibase 10.1103/PhysRevB.88.024501} {\bibfield
  {journal} {\bibinfo  {journal} {Phys. Rev. B}\ }\textbf {\bibinfo {volume}
  {88}},\ \bibinfo {pages} {024501} (\bibinfo {year} {2013})}\BibitemShut
  {NoStop}%
\bibitem [{\citenamefont {Chiu}, \citenamefont {Gilbert},\ and\ \citenamefont
  {Hughes}(2011)}]{chiu-11}%
  \BibitemOpen
  \bibfield  {author} {\bibinfo {author} {\bibfnamefont {C.-K.}\ \bibnamefont
  {Chiu}}, \bibinfo {author} {\bibfnamefont {M.~J.}\ \bibnamefont {Gilbert}}, \
  and\ \bibinfo {author} {\bibfnamefont {T.~L.}\ \bibnamefont {Hughes}},\
  }\href {\doibase 10.1103/PhysRevB.84.144507} {\bibfield  {journal} {\bibinfo
  {journal} {Phys. Rev. B}\ }\textbf {\bibinfo {volume} {84}},\ \bibinfo
  {pages} {144507} (\bibinfo {year} {2011})}\BibitemShut {NoStop}%
\bibitem [{\citenamefont {Akzyanov}\ \emph {et~al.}(2014)\citenamefont
  {Akzyanov}, \citenamefont {Rozhkov}, \citenamefont {Rakhmanov},\ and\
  \citenamefont {Nori}}]{akzyanov-14}%
  \BibitemOpen
  \bibfield  {author} {\bibinfo {author} {\bibfnamefont {R.~S.}\ \bibnamefont
  {Akzyanov}}, \bibinfo {author} {\bibfnamefont {A.~V.}\ \bibnamefont
  {Rozhkov}}, \bibinfo {author} {\bibfnamefont {A.~L.}\ \bibnamefont
  {Rakhmanov}}, \ and\ \bibinfo {author} {\bibfnamefont {F.}~\bibnamefont
  {Nori}},\ }\href {\doibase 10.1103/PhysRevB.89.085409} {\bibfield  {journal}
  {\bibinfo  {journal} {Phys. Rev. B}\ }\textbf {\bibinfo {volume} {89}},\
  \bibinfo {pages} {085409} (\bibinfo {year} {2014})}\BibitemShut {NoStop}%
\bibitem [{\citenamefont {Chen}\ \emph {et~al.}(2018)\citenamefont {Chen},
  \citenamefont {Chen}, \citenamefont {Yang}, \citenamefont {Du}, \citenamefont
  {Zhu}, \citenamefont {Wang},\ and\ \citenamefont {Wen}}]{chen-18}%
  \BibitemOpen
  \bibfield  {author} {\bibinfo {author} {\bibfnamefont {M.}~\bibnamefont
  {Chen}}, \bibinfo {author} {\bibfnamefont {X.}~\bibnamefont {Chen}}, \bibinfo
  {author} {\bibfnamefont {H.}~\bibnamefont {Yang}}, \bibinfo {author}
  {\bibfnamefont {Z.}~\bibnamefont {Du}}, \bibinfo {author} {\bibfnamefont
  {X.}~\bibnamefont {Zhu}}, \bibinfo {author} {\bibfnamefont {E.}~\bibnamefont
  {Wang}}, \ and\ \bibinfo {author} {\bibfnamefont {H.-H.}\ \bibnamefont
  {Wen}},\ }\href {\doibase 10.1038/s41467-018-03404-8} {\bibfield  {journal}
  {\bibinfo  {journal} {Nat Commun}\ }\textbf {\bibinfo {volume} {9}},\
  \bibinfo {pages} {970} (\bibinfo {year} {2018})}\BibitemShut {NoStop}%
\bibitem [{\citenamefont {Wang}\ \emph {et~al.}(2018)\citenamefont {Wang},
  \citenamefont {Kong}, \citenamefont {Fan}, \citenamefont {Chen},
  \citenamefont {Zhu}, \citenamefont {Liu}, \citenamefont {Cao}, \citenamefont
  {Sun}, \citenamefont {Du}, \citenamefont {Schneeloch}, \citenamefont {Zhong},
  \citenamefont {Gu}, \citenamefont {Fu}, \citenamefont {Ding},\ and\
  \citenamefont {Gao}}]{wang-18}%
  \BibitemOpen
  \bibfield  {author} {\bibinfo {author} {\bibfnamefont {D.}~\bibnamefont
  {Wang}}, \bibinfo {author} {\bibfnamefont {L.}~\bibnamefont {Kong}}, \bibinfo
  {author} {\bibfnamefont {P.}~\bibnamefont {Fan}}, \bibinfo {author}
  {\bibfnamefont {H.}~\bibnamefont {Chen}}, \bibinfo {author} {\bibfnamefont
  {S.}~\bibnamefont {Zhu}}, \bibinfo {author} {\bibfnamefont {W.}~\bibnamefont
  {Liu}}, \bibinfo {author} {\bibfnamefont {L.}~\bibnamefont {Cao}}, \bibinfo
  {author} {\bibfnamefont {Y.}~\bibnamefont {Sun}}, \bibinfo {author}
  {\bibfnamefont {S.}~\bibnamefont {Du}}, \bibinfo {author} {\bibfnamefont
  {J.}~\bibnamefont {Schneeloch}}, \bibinfo {author} {\bibfnamefont
  {R.}~\bibnamefont {Zhong}}, \bibinfo {author} {\bibfnamefont
  {G.}~\bibnamefont {Gu}}, \bibinfo {author} {\bibfnamefont {L.}~\bibnamefont
  {Fu}}, \bibinfo {author} {\bibfnamefont {H.}~\bibnamefont {Ding}}, \ and\
  \bibinfo {author} {\bibfnamefont {H.-J.}\ \bibnamefont {Gao}},\ }\href
  {\doibase 10.1126/science.aao1797} {\bibfield  {journal} {\bibinfo  {journal}
  {Science}\ }\textbf {\bibinfo {volume} {362}},\ \bibinfo {pages} {333}
  (\bibinfo {year} {2018})}\BibitemShut {NoStop}%
\bibitem [{\citenamefont {Zhu}\ \emph {et~al.}(2020)\citenamefont {Zhu},
  \citenamefont {Kong}, \citenamefont {Cao}, \citenamefont {Chen},
  \citenamefont {Papaj}, \citenamefont {Du}, \citenamefont {Xing},
  \citenamefont {Liu}, \citenamefont {Wang}, \citenamefont {Shen},
  \citenamefont {Yang}, \citenamefont {Schneeloch}, \citenamefont {Zhong},
  \citenamefont {Gu}, \citenamefont {Fu}, \citenamefont {Zhang}, \citenamefont
  {Ding},\ and\ \citenamefont {Gao}}]{zhu-20}%
  \BibitemOpen
  \bibfield  {author} {\bibinfo {author} {\bibfnamefont {S.}~\bibnamefont
  {Zhu}}, \bibinfo {author} {\bibfnamefont {L.}~\bibnamefont {Kong}}, \bibinfo
  {author} {\bibfnamefont {L.}~\bibnamefont {Cao}}, \bibinfo {author}
  {\bibfnamefont {H.}~\bibnamefont {Chen}}, \bibinfo {author} {\bibfnamefont
  {M.}~\bibnamefont {Papaj}}, \bibinfo {author} {\bibfnamefont
  {S.}~\bibnamefont {Du}}, \bibinfo {author} {\bibfnamefont {Y.}~\bibnamefont
  {Xing}}, \bibinfo {author} {\bibfnamefont {W.}~\bibnamefont {Liu}}, \bibinfo
  {author} {\bibfnamefont {D.}~\bibnamefont {Wang}}, \bibinfo {author}
  {\bibfnamefont {C.}~\bibnamefont {Shen}}, \bibinfo {author} {\bibfnamefont
  {F.}~\bibnamefont {Yang}}, \bibinfo {author} {\bibfnamefont {J.}~\bibnamefont
  {Schneeloch}}, \bibinfo {author} {\bibfnamefont {R.}~\bibnamefont {Zhong}},
  \bibinfo {author} {\bibfnamefont {G.}~\bibnamefont {Gu}}, \bibinfo {author}
  {\bibfnamefont {L.}~\bibnamefont {Fu}}, \bibinfo {author} {\bibfnamefont
  {Y.-Y.}\ \bibnamefont {Zhang}}, \bibinfo {author} {\bibfnamefont
  {H.}~\bibnamefont {Ding}}, \ and\ \bibinfo {author} {\bibfnamefont {H.-J.}\
  \bibnamefont {Gao}},\ }\href {\doibase 10.1126/science.aax0274} {\bibfield
  {journal} {\bibinfo  {journal} {Science}\ }\textbf {\bibinfo {volume}
  {367}},\ \bibinfo {pages} {189} (\bibinfo {year} {2020})}\BibitemShut
  {NoStop}%
\bibitem [{\citenamefont {Pathak}, \citenamefont {Plugge},\ and\ \citenamefont
  {Franz}(2021)}]{pathak-21}%
  \BibitemOpen
  \bibfield  {author} {\bibinfo {author} {\bibfnamefont {V.}~\bibnamefont
  {Pathak}}, \bibinfo {author} {\bibfnamefont {S.}~\bibnamefont {Plugge}}, \
  and\ \bibinfo {author} {\bibfnamefont {M.}~\bibnamefont {Franz}},\ }\href
  {\doibase 10.1016/j.aop.2021.168431} {\bibfield  {journal} {\bibinfo
  {journal} {Ann. Phys. (N. Y.)}\ }\textbf {\bibinfo {volume} {435}},\ \bibinfo
  {pages} {168431} (\bibinfo {year} {2021})}\BibitemShut {NoStop}%
\bibitem [{\citenamefont {Yan}, \citenamefont {Wu},\ and\ \citenamefont
  {Huang}(2020)}]{yan-20}%
  \BibitemOpen
  \bibfield  {author} {\bibinfo {author} {\bibfnamefont {Z.}~\bibnamefont
  {Yan}}, \bibinfo {author} {\bibfnamefont {Z.}~\bibnamefont {Wu}}, \ and\
  \bibinfo {author} {\bibfnamefont {W.}~\bibnamefont {Huang}},\ }\href
  {\doibase 10.1103/PhysRevLett.124.257001} {\bibfield  {journal} {\bibinfo
  {journal} {Phys. Rev. Lett.}\ }\textbf {\bibinfo {volume} {124}},\ \bibinfo
  {pages} {257001} (\bibinfo {year} {2020})}\BibitemShut {NoStop}%
\bibitem [{\citenamefont {Steffensen}, \citenamefont {Andersen},\ and\
  \citenamefont {Kotetes}(2021)}]{steffensen-21}%
  \BibitemOpen
  \bibfield  {author} {\bibinfo {author} {\bibfnamefont {D.}~\bibnamefont
  {Steffensen}}, \bibinfo {author} {\bibfnamefont {B.~M.}\ \bibnamefont
  {Andersen}}, \ and\ \bibinfo {author} {\bibfnamefont {P.}~\bibnamefont
  {Kotetes}},\ }\href {\doibase 10.1103/PhysRevB.104.174502} {\bibfield
  {journal} {\bibinfo  {journal} {Phys. Rev. B}\ }\textbf {\bibinfo {volume}
  {104}},\ \bibinfo {pages} {174502} (\bibinfo {year} {2021})}\BibitemShut
  {NoStop}%
\bibitem [{\citenamefont {Sato}\ and\ \citenamefont
  {Fujimoto}(2010)}]{sato-10}%
  \BibitemOpen
  \bibfield  {author} {\bibinfo {author} {\bibfnamefont {M.}~\bibnamefont
  {Sato}}\ and\ \bibinfo {author} {\bibfnamefont {S.}~\bibnamefont
  {Fujimoto}},\ }\href {\doibase 10.1103/PhysRevLett.105.217001} {\bibfield
  {journal} {\bibinfo  {journal} {Phys. Rev. Lett.}\ }\textbf {\bibinfo
  {volume} {105}},\ \bibinfo {pages} {217001} (\bibinfo {year}
  {2010})}\BibitemShut {NoStop}%
\bibitem [{\citenamefont {Zhou}\ and\ \citenamefont {Wang}(2008)}]{zhou-08}%
  \BibitemOpen
  \bibfield  {author} {\bibinfo {author} {\bibfnamefont {S.}~\bibnamefont
  {Zhou}}\ and\ \bibinfo {author} {\bibfnamefont {Z.}~\bibnamefont {Wang}},\
  }\href {\doibase 10.1103/PhysRevLett.100.217002} {\bibfield  {journal}
  {\bibinfo  {journal} {Phys. Rev. Lett.}\ }\textbf {\bibinfo {volume} {100}},\
  \bibinfo {pages} {217002} (\bibinfo {year} {2008})}\BibitemShut {NoStop}%
\bibitem [{\citenamefont {Val'kov}, \citenamefont {Val'kova},\ and\
  \citenamefont {Mitskan}(2017)}]{valkov-17}%
  \BibitemOpen
  \bibfield  {author} {\bibinfo {author} {\bibfnamefont {V.~V.}\ \bibnamefont
  {Val'kov}}, \bibinfo {author} {\bibfnamefont {T.~A.}\ \bibnamefont
  {Val'kova}}, \ and\ \bibinfo {author} {\bibfnamefont {V.~A.}\ \bibnamefont
  {Mitskan}},\ }\href {\doibase 10.1016/j.jmmm.2016.12.085} {\bibfield
  {journal} {\bibinfo  {journal} {J. Magn. Magn. Mater.}\ }\textbf {\bibinfo
  {volume} {440}},\ \bibinfo {pages} {129} (\bibinfo {year}
  {2017})}\BibitemShut {NoStop}%
\bibitem [{\citenamefont {Baskaran}(2003)}]{baskaran-03}%
  \BibitemOpen
  \bibfield  {author} {\bibinfo {author} {\bibfnamefont {G.}~\bibnamefont
  {Baskaran}},\ }\href {\doibase 10.1103/PhysRevLett.91.097003} {\bibfield
  {journal} {\bibinfo  {journal} {Phys. Rev. Lett.}\ }\textbf {\bibinfo
  {volume} {91}},\ \bibinfo {pages} {097003} (\bibinfo {year}
  {2003})}\BibitemShut {NoStop}%
\bibitem [{\citenamefont {Black-Schaffer}, \citenamefont {Wu},\ and\
  \citenamefont {Le~Hur}(2014)}]{black-schaffer-14}%
  \BibitemOpen
  \bibfield  {author} {\bibinfo {author} {\bibfnamefont {A.~M.}\ \bibnamefont
  {Black-Schaffer}}, \bibinfo {author} {\bibfnamefont {W.}~\bibnamefont {Wu}},
  \ and\ \bibinfo {author} {\bibfnamefont {K.}~\bibnamefont {Le~Hur}},\ }\href
  {\doibase 10.1103/PhysRevB.90.054521} {\bibfield  {journal} {\bibinfo
  {journal} {Phys. Rev. B}\ }\textbf {\bibinfo {volume} {90}},\ \bibinfo
  {pages} {054521} (\bibinfo {year} {2014})}\BibitemShut {NoStop}%
\bibitem [{\citenamefont {Kagan}, \citenamefont {Mitskan},\ and\ \citenamefont
  {Korovushkin}(2015)}]{kagan-15}%
  \BibitemOpen
  \bibfield  {author} {\bibinfo {author} {\bibfnamefont {M.~Y.}\ \bibnamefont
  {Kagan}}, \bibinfo {author} {\bibfnamefont {V.~A.}\ \bibnamefont {Mitskan}},
  \ and\ \bibinfo {author} {\bibfnamefont {M.~M.}\ \bibnamefont
  {Korovushkin}},\ }\href {\doibase 10.3367/UFNe.0185.201508a.0785} {\bibfield
  {journal} {\bibinfo  {journal} {Phys.-Usp.}\ }\textbf {\bibinfo {volume}
  {58}},\ \bibinfo {pages} {733} (\bibinfo {year} {2015})}\BibitemShut
  {NoStop}%
\bibitem [{\citenamefont {Volovik}(1997)}]{volovik-97}%
  \BibitemOpen
  \bibfield  {author} {\bibinfo {author} {\bibfnamefont {G.~E.}\ \bibnamefont
  {Volovik}},\ }\href {\doibase 10.1134/1.567563} {\bibfield  {journal}
  {\bibinfo  {journal} {JETP Lett.}\ }\textbf {\bibinfo {volume} {66}},\
  \bibinfo {pages} {522} (\bibinfo {year} {1997})}\BibitemShut {NoStop}%
\bibitem [{\citenamefont {Volovik}(1999)}]{volovik-99}%
  \BibitemOpen
  \bibfield  {author} {\bibinfo {author} {\bibfnamefont {G.~E.}\ \bibnamefont
  {Volovik}},\ }\href {\doibase 10.1134/1.568223} {\bibfield  {journal}
  {\bibinfo  {journal} {Jetp Lett.}\ }\textbf {\bibinfo {volume} {70}},\
  \bibinfo {pages} {609} (\bibinfo {year} {1999})}\BibitemShut {NoStop}%
\bibitem [{\citenamefont {Lee}\ and\ \citenamefont {Schnyder}(2016)}]{lee-16}%
  \BibitemOpen
  \bibfield  {author} {\bibinfo {author} {\bibfnamefont {D.}~\bibnamefont
  {Lee}}\ and\ \bibinfo {author} {\bibfnamefont {A.~P.}\ \bibnamefont
  {Schnyder}},\ }\href {\doibase 10.1103/PhysRevB.93.064522} {\bibfield
  {journal} {\bibinfo  {journal} {Phys. Rev. B}\ }\textbf {\bibinfo {volume}
  {93}},\ \bibinfo {pages} {064522} (\bibinfo {year} {2016})}\BibitemShut
  {NoStop}%
\bibitem [{\citenamefont {Volovik}(2016)}]{volovik-16}%
  \BibitemOpen
  \bibfield  {author} {\bibinfo {author} {\bibfnamefont {G.~E.}\ \bibnamefont
  {Volovik}},\ }\href {\doibase 10.1134/S0021364016150029} {\bibfield
  {journal} {\bibinfo  {journal} {Jetp Lett.}\ }\textbf {\bibinfo {volume}
  {104}},\ \bibinfo {pages} {201} (\bibinfo {year} {2016})}\BibitemShut
  {NoStop}%
\bibitem [{\citenamefont {Choy}\ \emph {et~al.}(2011)\citenamefont {Choy},
  \citenamefont {Edge}, \citenamefont {Akhmerov},\ and\ \citenamefont
  {Beenakker}}]{choy-11}%
  \BibitemOpen
  \bibfield  {author} {\bibinfo {author} {\bibfnamefont {T.-P.}\ \bibnamefont
  {Choy}}, \bibinfo {author} {\bibfnamefont {J.~M.}\ \bibnamefont {Edge}},
  \bibinfo {author} {\bibfnamefont {A.~R.}\ \bibnamefont {Akhmerov}}, \ and\
  \bibinfo {author} {\bibfnamefont {C.~W.~J.}\ \bibnamefont {Beenakker}},\
  }\href {\doibase 10.1103/PhysRevB.84.195442} {\bibfield  {journal} {\bibinfo
  {journal} {Phys. Rev. B}\ }\textbf {\bibinfo {volume} {84}},\ \bibinfo
  {pages} {195442} (\bibinfo {year} {2011})}\BibitemShut {NoStop}%
\bibitem [{\citenamefont {Nadj-Perge}\ \emph {et~al.}(2013)\citenamefont
  {Nadj-Perge}, \citenamefont {Drozdov}, \citenamefont {Bernevig},\ and\
  \citenamefont {Yazdani}}]{nadj-perge-13}%
  \BibitemOpen
  \bibfield  {author} {\bibinfo {author} {\bibfnamefont {S.}~\bibnamefont
  {Nadj-Perge}}, \bibinfo {author} {\bibfnamefont {I.~K.}\ \bibnamefont
  {Drozdov}}, \bibinfo {author} {\bibfnamefont {B.~A.}\ \bibnamefont
  {Bernevig}}, \ and\ \bibinfo {author} {\bibfnamefont {A.}~\bibnamefont
  {Yazdani}},\ }\href {\doibase 10.1103/PhysRevB.88.020407} {\bibfield
  {journal} {\bibinfo  {journal} {Phys. Rev. B}\ }\textbf {\bibinfo {volume}
  {88}},\ \bibinfo {pages} {020407(R)} (\bibinfo {year} {2013})}\BibitemShut
  {NoStop}%
\bibitem [{\citenamefont {Klinovaja}\ \emph {et~al.}(2013)\citenamefont
  {Klinovaja}, \citenamefont {Stano}, \citenamefont {Yazdani},\ and\
  \citenamefont {Loss}}]{klinovaja-13}%
  \BibitemOpen
  \bibfield  {author} {\bibinfo {author} {\bibfnamefont {J.}~\bibnamefont
  {Klinovaja}}, \bibinfo {author} {\bibfnamefont {P.}~\bibnamefont {Stano}},
  \bibinfo {author} {\bibfnamefont {A.}~\bibnamefont {Yazdani}}, \ and\
  \bibinfo {author} {\bibfnamefont {D.}~\bibnamefont {Loss}},\ }\href {\doibase
  10.1103/PhysRevLett.111.186805} {\bibfield  {journal} {\bibinfo  {journal}
  {Phys. Rev. Lett.}\ }\textbf {\bibinfo {volume} {111}},\ \bibinfo {pages}
  {186805} (\bibinfo {year} {2013})}\BibitemShut {NoStop}%
\bibitem [{\citenamefont {Martin}\ and\ \citenamefont
  {Morpurgo}(2012)}]{martin-12}%
  \BibitemOpen
  \bibfield  {author} {\bibinfo {author} {\bibfnamefont {I.}~\bibnamefont
  {Martin}}\ and\ \bibinfo {author} {\bibfnamefont {A.~F.}\ \bibnamefont
  {Morpurgo}},\ }\href {\doibase 10.1103/PhysRevB.85.144505} {\bibfield
  {journal} {\bibinfo  {journal} {Phys. Rev. B}\ }\textbf {\bibinfo {volume}
  {85}},\ \bibinfo {pages} {144505} (\bibinfo {year} {2012})}\BibitemShut
  {NoStop}%
\bibitem [{\citenamefont {Lu}\ and\ \citenamefont {Wang}(2013)}]{lu-13}%
  \BibitemOpen
  \bibfield  {author} {\bibinfo {author} {\bibfnamefont {Y.-M.}\ \bibnamefont
  {Lu}}\ and\ \bibinfo {author} {\bibfnamefont {Z.}~\bibnamefont {Wang}},\
  }\href {\doibase 10.1103/PhysRevLett.110.096403} {\bibfield  {journal}
  {\bibinfo  {journal} {Phys. Rev. Lett.}\ }\textbf {\bibinfo {volume} {110}},\
  \bibinfo {pages} {096403} (\bibinfo {year} {2013})}\BibitemShut {NoStop}%
\bibitem [{\citenamefont {Val{\textquoteright}kov}, \citenamefont {Zlotnikov},\
  and\ \citenamefont {Shustin}(2018)}]{valkov-18}%
  \BibitemOpen
  \bibfield  {author} {\bibinfo {author} {\bibfnamefont {V.~V.}\ \bibnamefont
  {Val{\textquoteright}kov}}, \bibinfo {author} {\bibfnamefont {A.~O.}\
  \bibnamefont {Zlotnikov}}, \ and\ \bibinfo {author} {\bibfnamefont {M.~S.}\
  \bibnamefont {Shustin}},\ }\href {\doibase 10.1016/j.jmmm.2017.11.115}
  {\bibfield  {journal} {\bibinfo  {journal} {J. Magn. Magn. Mater.}\ }\textbf
  {\bibinfo {volume} {459}},\ \bibinfo {pages} {112} (\bibinfo {year}
  {2018})}\BibitemShut {NoStop}%
\bibitem [{\citenamefont {Bedow}\ \emph {et~al.}(2020)\citenamefont {Bedow},
  \citenamefont {Mascot}, \citenamefont {Posske}, \citenamefont {Uhrig},
  \citenamefont {Wiesendanger}, \citenamefont {Rachel},\ and\ \citenamefont
  {Morr}}]{bedow-20}%
  \BibitemOpen
  \bibfield  {author} {\bibinfo {author} {\bibfnamefont {J.}~\bibnamefont
  {Bedow}}, \bibinfo {author} {\bibfnamefont {E.}~\bibnamefont {Mascot}},
  \bibinfo {author} {\bibfnamefont {T.}~\bibnamefont {Posske}}, \bibinfo
  {author} {\bibfnamefont {G.~S.}\ \bibnamefont {Uhrig}}, \bibinfo {author}
  {\bibfnamefont {R.}~\bibnamefont {Wiesendanger}}, \bibinfo {author}
  {\bibfnamefont {S.}~\bibnamefont {Rachel}}, \ and\ \bibinfo {author}
  {\bibfnamefont {D.~K.}\ \bibnamefont {Morr}},\ }\href {\doibase
  10.1103/PhysRevB.102.180504} {\bibfield  {journal} {\bibinfo  {journal}
  {Phys. Rev. B}\ }\textbf {\bibinfo {volume} {102}},\ \bibinfo {pages}
  {180504(R)} (\bibinfo {year} {2020})}\BibitemShut {NoStop}%
\bibitem [{\citenamefont {Rex}, \citenamefont {Gornyi},\ and\ \citenamefont
  {Mirlin}(2020)}]{rex-20}%
  \BibitemOpen
  \bibfield  {author} {\bibinfo {author} {\bibfnamefont {S.}~\bibnamefont
  {Rex}}, \bibinfo {author} {\bibfnamefont {I.~V.}\ \bibnamefont {Gornyi}}, \
  and\ \bibinfo {author} {\bibfnamefont {A.~D.}\ \bibnamefont {Mirlin}},\
  }\href {\doibase 10.1103/PhysRevB.102.224501} {\bibfield  {journal} {\bibinfo
   {journal} {Phys. Rev. B}\ }\textbf {\bibinfo {volume} {102}},\ \bibinfo
  {pages} {224501} (\bibinfo {year} {2020})}\BibitemShut {NoStop}%
\bibitem [{\citenamefont {Steffensen}\ \emph {et~al.}(2022)\citenamefont
  {Steffensen}, \citenamefont {Christensen}, \citenamefont {Andersen},\ and\
  \citenamefont {Kotetes}}]{steffensen-22}%
  \BibitemOpen
  \bibfield  {author} {\bibinfo {author} {\bibfnamefont {D.}~\bibnamefont
  {Steffensen}}, \bibinfo {author} {\bibfnamefont {M.~H.}\ \bibnamefont
  {Christensen}}, \bibinfo {author} {\bibfnamefont {B.~M.}\ \bibnamefont
  {Andersen}}, \ and\ \bibinfo {author} {\bibfnamefont {P.}~\bibnamefont
  {Kotetes}},\ }\href {\doibase 10.1103/PhysRevResearch.4.013225} {\bibfield
  {journal} {\bibinfo  {journal} {Phys. Rev. Research}\ }\textbf {\bibinfo
  {volume} {4}},\ \bibinfo {pages} {013225} (\bibinfo {year}
  {2022})}\BibitemShut {NoStop}%
\bibitem [{\citenamefont {Menzel}\ \emph {et~al.}(2012)\citenamefont {Menzel},
  \citenamefont {Mokrousov}, \citenamefont {Wieser}, \citenamefont {Bickel},
  \citenamefont {Vedmedenko}, \citenamefont {Bl{\"u}gel}, \citenamefont
  {Heinze}, \citenamefont {von Bergmann}, \citenamefont {Kubetzka},\ and\
  \citenamefont {Wiesendanger}}]{menzel-12}%
  \BibitemOpen
  \bibfield  {author} {\bibinfo {author} {\bibfnamefont {M.}~\bibnamefont
  {Menzel}}, \bibinfo {author} {\bibfnamefont {Y.}~\bibnamefont {Mokrousov}},
  \bibinfo {author} {\bibfnamefont {R.}~\bibnamefont {Wieser}}, \bibinfo
  {author} {\bibfnamefont {J.~E.}\ \bibnamefont {Bickel}}, \bibinfo {author}
  {\bibfnamefont {E.}~\bibnamefont {Vedmedenko}}, \bibinfo {author}
  {\bibfnamefont {S.}~\bibnamefont {Bl{\"u}gel}}, \bibinfo {author}
  {\bibfnamefont {S.}~\bibnamefont {Heinze}}, \bibinfo {author} {\bibfnamefont
  {K.}~\bibnamefont {von Bergmann}}, \bibinfo {author} {\bibfnamefont
  {A.}~\bibnamefont {Kubetzka}}, \ and\ \bibinfo {author} {\bibfnamefont
  {R.}~\bibnamefont {Wiesendanger}},\ }\href {\doibase
  10.1103/PhysRevLett.108.197204} {\bibfield  {journal} {\bibinfo  {journal}
  {Phys. Rev. Lett.}\ }\textbf {\bibinfo {volume} {108}},\ \bibinfo {pages}
  {197204} (\bibinfo {year} {2012})}\BibitemShut {NoStop}%
\bibitem [{\citenamefont {Kim}\ \emph {et~al.}(2018)\citenamefont {Kim},
  \citenamefont {Palacio-Morales}, \citenamefont {Posske}, \citenamefont
  {R{\'o}zsa}, \citenamefont {Palot{\'a}s}, \citenamefont {Szunyogh},
  \citenamefont {Thorwart},\ and\ \citenamefont {Wiesendanger}}]{kim-18}%
  \BibitemOpen
  \bibfield  {author} {\bibinfo {author} {\bibfnamefont {H.}~\bibnamefont
  {Kim}}, \bibinfo {author} {\bibfnamefont {A.}~\bibnamefont
  {Palacio-Morales}}, \bibinfo {author} {\bibfnamefont {T.}~\bibnamefont
  {Posske}}, \bibinfo {author} {\bibfnamefont {L.}~\bibnamefont {R{\'o}zsa}},
  \bibinfo {author} {\bibfnamefont {K.}~\bibnamefont {Palot{\'a}s}}, \bibinfo
  {author} {\bibfnamefont {L.}~\bibnamefont {Szunyogh}}, \bibinfo {author}
  {\bibfnamefont {M.}~\bibnamefont {Thorwart}}, \ and\ \bibinfo {author}
  {\bibfnamefont {R.}~\bibnamefont {Wiesendanger}},\ }\href {\doibase
  10.1126/sciadv.aar5251} {\bibfield  {journal} {\bibinfo  {journal} {Sci.
  Adv.}\ }\textbf {\bibinfo {volume} {4}},\ \bibinfo {pages} {eaar5251}
  (\bibinfo {year} {2018})}\BibitemShut {NoStop}%
\bibitem [{\citenamefont {Palacio-Morales}\ \emph {et~al.}(2019)\citenamefont
  {Palacio-Morales}, \citenamefont {Mascot}, \citenamefont {Cocklin},
  \citenamefont {Kim}, \citenamefont {Rachel}, \citenamefont {Morr},\ and\
  \citenamefont {Wiesendanger}}]{palacio-morales-19}%
  \BibitemOpen
  \bibfield  {author} {\bibinfo {author} {\bibfnamefont {A.}~\bibnamefont
  {Palacio-Morales}}, \bibinfo {author} {\bibfnamefont {E.}~\bibnamefont
  {Mascot}}, \bibinfo {author} {\bibfnamefont {S.}~\bibnamefont {Cocklin}},
  \bibinfo {author} {\bibfnamefont {H.}~\bibnamefont {Kim}}, \bibinfo {author}
  {\bibfnamefont {S.}~\bibnamefont {Rachel}}, \bibinfo {author} {\bibfnamefont
  {D.~K.}\ \bibnamefont {Morr}}, \ and\ \bibinfo {author} {\bibfnamefont
  {R.}~\bibnamefont {Wiesendanger}},\ }\href {\doibase 10.1126/sciadv.aav6600}
  {\bibfield  {journal} {\bibinfo  {journal} {Sci. Adv.}\ }\textbf {\bibinfo
  {volume} {5}},\ \bibinfo {pages} {eaav6600} (\bibinfo {year}
  {2019})}\BibitemShut {NoStop}%
\bibitem [{\citenamefont {Spethmann}\ \emph {et~al.}(2020)\citenamefont
  {Spethmann}, \citenamefont {Meyer}, \citenamefont {von Bergmann},
  \citenamefont {Wiesendanger}, \citenamefont {Heinze},\ and\ \citenamefont
  {Kubetzka}}]{spethmann-20}%
  \BibitemOpen
  \bibfield  {author} {\bibinfo {author} {\bibfnamefont {J.}~\bibnamefont
  {Spethmann}}, \bibinfo {author} {\bibfnamefont {S.}~\bibnamefont {Meyer}},
  \bibinfo {author} {\bibfnamefont {K.}~\bibnamefont {von Bergmann}}, \bibinfo
  {author} {\bibfnamefont {R.}~\bibnamefont {Wiesendanger}}, \bibinfo {author}
  {\bibfnamefont {S.}~\bibnamefont {Heinze}}, \ and\ \bibinfo {author}
  {\bibfnamefont {A.}~\bibnamefont {Kubetzka}},\ }\href {\doibase
  10.1103/PhysRevLett.124.227203} {\bibfield  {journal} {\bibinfo  {journal}
  {Phys. Rev. Lett.}\ }\textbf {\bibinfo {volume} {124}},\ \bibinfo {pages}
  {227203} (\bibinfo {year} {2020})}\BibitemShut {NoStop}%
\bibitem [{\citenamefont {Smirnov}\ \emph {et~al.}(2017)\citenamefont
  {Smirnov}, \citenamefont {Soldatov}, \citenamefont {Petrenko}, \citenamefont
  {Takata}, \citenamefont {Kida}, \citenamefont {Hagiwara}, \citenamefont
  {Shapiro},\ and\ \citenamefont {Zhitomirsky}}]{smirnov-17}%
  \BibitemOpen
  \bibfield  {author} {\bibinfo {author} {\bibfnamefont {A.~I.}\ \bibnamefont
  {Smirnov}}, \bibinfo {author} {\bibfnamefont {T.~A.}\ \bibnamefont
  {Soldatov}}, \bibinfo {author} {\bibfnamefont {O.~A.}\ \bibnamefont
  {Petrenko}}, \bibinfo {author} {\bibfnamefont {A.}~\bibnamefont {Takata}},
  \bibinfo {author} {\bibfnamefont {T.}~\bibnamefont {Kida}}, \bibinfo {author}
  {\bibfnamefont {M.}~\bibnamefont {Hagiwara}}, \bibinfo {author}
  {\bibfnamefont {A.~Y.}\ \bibnamefont {Shapiro}}, \ and\ \bibinfo {author}
  {\bibfnamefont {M.~E.}\ \bibnamefont {Zhitomirsky}},\ }\href {\doibase
  10.1103/PhysRevLett.119.047204} {\bibfield  {journal} {\bibinfo  {journal}
  {Phys. Rev. Lett.}\ }\textbf {\bibinfo {volume} {119}},\ \bibinfo {pages}
  {047204} (\bibinfo {year} {2017})}\BibitemShut {NoStop}%
\bibitem [{\citenamefont {Soldatov}\ \emph {et~al.}(2020)\citenamefont
  {Soldatov}, \citenamefont {Sakhratov}, \citenamefont {Svistov},\ and\
  \citenamefont {Smirnov}}]{soldatov-20}%
  \BibitemOpen
  \bibfield  {author} {\bibinfo {author} {\bibfnamefont {T.~A.}\ \bibnamefont
  {Soldatov}}, \bibinfo {author} {\bibfnamefont {Y.~A.}\ \bibnamefont
  {Sakhratov}}, \bibinfo {author} {\bibfnamefont {L.~E.}\ \bibnamefont
  {Svistov}}, \ and\ \bibinfo {author} {\bibfnamefont {A.~I.}\ \bibnamefont
  {Smirnov}},\ }\href {\doibase 10.1134/S1063776120070122} {\bibfield
  {journal} {\bibinfo  {journal} {J. Exp. Theor. Phys.}\ }\textbf {\bibinfo
  {volume} {131}},\ \bibinfo {pages} {62} (\bibinfo {year} {2020})}\BibitemShut
  {NoStop}%
\bibitem [{\citenamefont {Caroli}, \citenamefont {De~Gennes},\ and\
  \citenamefont {Matricon}(1964)}]{caroli-64}%
  \BibitemOpen
  \bibfield  {author} {\bibinfo {author} {\bibfnamefont {C.}~\bibnamefont
  {Caroli}}, \bibinfo {author} {\bibfnamefont {P.~G.}\ \bibnamefont
  {De~Gennes}}, \ and\ \bibinfo {author} {\bibfnamefont {J.}~\bibnamefont
  {Matricon}},\ }\href {\doibase 10.1016/0031-9163(64)90375-0} {\bibfield
  {journal} {\bibinfo  {journal} {Physics Letters}\ }\textbf {\bibinfo {volume}
  {9}},\ \bibinfo {pages} {307} (\bibinfo {year} {1964})}\BibitemShut {NoStop}%
\bibitem [{\citenamefont {Ishikawa}\ and\ \citenamefont
  {Matsuyama}(1987)}]{ishikawa-87}%
  \BibitemOpen
  \bibfield  {author} {\bibinfo {author} {\bibfnamefont {K.}~\bibnamefont
  {Ishikawa}}\ and\ \bibinfo {author} {\bibfnamefont {T.}~\bibnamefont
  {Matsuyama}},\ }\href {\doibase 10.1016/0550-3213(87)90160-X} {\bibfield
  {journal} {\bibinfo  {journal} {Nucl. Phys. B}\ }\textbf {\bibinfo {volume}
  {280}},\ \bibinfo {pages} {523} (\bibinfo {year} {1987})}\BibitemShut
  {NoStop}%
\bibitem [{\citenamefont {Volovik}(2009)}]{volovik-09}%
  \BibitemOpen
  \bibfield  {author} {\bibinfo {author} {\bibfnamefont {G.~E.}\ \bibnamefont
  {Volovik}},\ }\href
  {http://www.oxfordscholarship.com/view/10.1093/acprof:oso/9780199564842.001.0001/acprof-9780199564842}
  {\emph {\bibinfo {title} {The {Universe} in a {Helium} {Droplet}}}}\
  (\bibinfo  {publisher} {Oxford University Press},\ \bibinfo {year}
  {2009})\BibitemShut {NoStop}%
\bibitem [{\citenamefont {Val'kov}\ and\ \citenamefont
  {Zlotnikov}(2019)}]{valkov-19}%
  \BibitemOpen
  \bibfield  {author} {\bibinfo {author} {\bibfnamefont {V.~V.}\ \bibnamefont
  {Val'kov}}\ and\ \bibinfo {author} {\bibfnamefont {A.~O.}\ \bibnamefont
  {Zlotnikov}},\ }\href {\doibase 10.1134/S0021364019110158} {\bibfield
  {journal} {\bibinfo  {journal} {JETP Lett.}\ }\textbf {\bibinfo {volume}
  {109}},\ \bibinfo {pages} {736} (\bibinfo {year} {2019})}\BibitemShut
  {NoStop}%
\bibitem [{\citenamefont {Kitaev}(2001)}]{kitaev-01}%
  \BibitemOpen
  \bibfield  {author} {\bibinfo {author} {\bibfnamefont {A.~Y.}\ \bibnamefont
  {Kitaev}},\ }\href {\doibase 10.1070/1063-7869/44/10S/S29} {\bibfield
  {journal} {\bibinfo  {journal} {Phys.-Usp.}\ }\textbf {\bibinfo {volume}
  {44}},\ \bibinfo {pages} {131} (\bibinfo {year} {2001})}\BibitemShut
  {NoStop}%
\bibitem [{\citenamefont {Takigawa}, \citenamefont {Ichioka},\ and\
  \citenamefont {Machida}(2000)}]{takigawa-00}%
  \BibitemOpen
  \bibfield  {author} {\bibinfo {author} {\bibfnamefont {M.}~\bibnamefont
  {Takigawa}}, \bibinfo {author} {\bibfnamefont {M.}~\bibnamefont {Ichioka}}, \
  and\ \bibinfo {author} {\bibfnamefont {K.}~\bibnamefont {Machida}},\ }\href
  {\doibase 10.1143/JPSJ.69.3943} {\bibfield  {journal} {\bibinfo  {journal}
  {J. Phys. Soc. Jpn.}\ }\textbf {\bibinfo {volume} {69}},\ \bibinfo {pages}
  {3943} (\bibinfo {year} {2000})}\BibitemShut {NoStop}%
\bibitem [{\citenamefont {Franz}\ and\ \citenamefont {Te{\v
  s}anovi{\'c}}(2000)}]{franz-00}%
  \BibitemOpen
  \bibfield  {author} {\bibinfo {author} {\bibfnamefont {M.}~\bibnamefont
  {Franz}}\ and\ \bibinfo {author} {\bibfnamefont {Z.}~\bibnamefont {Te{\v
  s}anovi{\'c}}},\ }\href {\doibase 10.1103/PhysRevLett.84.554} {\bibfield
  {journal} {\bibinfo  {journal} {Phys. Rev. Lett.}\ }\textbf {\bibinfo
  {volume} {84}},\ \bibinfo {pages} {554} (\bibinfo {year} {2000})}\BibitemShut
  {NoStop}%
\bibitem [{\citenamefont {Hegde}\ and\ \citenamefont
  {Vishveshwara}(2016)}]{hegde-16}%
  \BibitemOpen
  \bibfield  {author} {\bibinfo {author} {\bibfnamefont {S.~S.}\ \bibnamefont
  {Hegde}}\ and\ \bibinfo {author} {\bibfnamefont {S.}~\bibnamefont
  {Vishveshwara}},\ }\href {\doibase 10.1103/PhysRevB.94.115166} {\bibfield
  {journal} {\bibinfo  {journal} {Phys. Rev. B}\ }\textbf {\bibinfo {volume}
  {94}},\ \bibinfo {pages} {115166} (\bibinfo {year} {2016})}\BibitemShut
  {NoStop}%
\bibitem [{\citenamefont {Nijholt}\ and\ \citenamefont
  {Akhmerov}(2016)}]{nijholt-16}%
  \BibitemOpen
  \bibfield  {author} {\bibinfo {author} {\bibfnamefont {B.}~\bibnamefont
  {Nijholt}}\ and\ \bibinfo {author} {\bibfnamefont {A.~R.}\ \bibnamefont
  {Akhmerov}},\ }\href {\doibase 10.1103/PhysRevB.93.235434} {\bibfield
  {journal} {\bibinfo  {journal} {Phys. Rev. B}\ }\textbf {\bibinfo {volume}
  {93}},\ \bibinfo {pages} {235434} (\bibinfo {year} {2016})}\BibitemShut
  {NoStop}%
\bibitem [{\citenamefont {Veshchunov}\ \emph {et~al.}(2016)\citenamefont
  {Veshchunov}, \citenamefont {Magrini}, \citenamefont {Mironov}, \citenamefont
  {Godin}, \citenamefont {Trebbia}, \citenamefont {Buzdin}, \citenamefont
  {Tamarat},\ and\ \citenamefont {Lounis}}]{veshchunov-16}%
  \BibitemOpen
  \bibfield  {author} {\bibinfo {author} {\bibfnamefont {I.~S.}\ \bibnamefont
  {Veshchunov}}, \bibinfo {author} {\bibfnamefont {W.}~\bibnamefont {Magrini}},
  \bibinfo {author} {\bibfnamefont {S.~V.}\ \bibnamefont {Mironov}}, \bibinfo
  {author} {\bibfnamefont {A.~G.}\ \bibnamefont {Godin}}, \bibinfo {author}
  {\bibfnamefont {J.-B.}\ \bibnamefont {Trebbia}}, \bibinfo {author}
  {\bibfnamefont {A.~I.}\ \bibnamefont {Buzdin}}, \bibinfo {author}
  {\bibfnamefont {P.}~\bibnamefont {Tamarat}}, \ and\ \bibinfo {author}
  {\bibfnamefont {B.}~\bibnamefont {Lounis}},\ }\href {\doibase
  10.1038/ncomms12801} {\bibfield  {journal} {\bibinfo  {journal} {Nat Commun}\
  }\textbf {\bibinfo {volume} {7}},\ \bibinfo {pages} {12801} (\bibinfo {year}
  {2016})}\BibitemShut {NoStop}%
\bibitem [{\citenamefont {Rochet}\ \emph {et~al.}(2020)\citenamefont {Rochet},
  \citenamefont {Vadimov}, \citenamefont {Magrini}, \citenamefont {Thakur},
  \citenamefont {Trebbia}, \citenamefont {Melnikov}, \citenamefont {Buzdin},
  \citenamefont {Tamarat},\ and\ \citenamefont {Lounis}}]{rochet-20}%
  \BibitemOpen
  \bibfield  {author} {\bibinfo {author} {\bibfnamefont {A.}~\bibnamefont
  {Rochet}}, \bibinfo {author} {\bibfnamefont {V.}~\bibnamefont {Vadimov}},
  \bibinfo {author} {\bibfnamefont {W.}~\bibnamefont {Magrini}}, \bibinfo
  {author} {\bibfnamefont {S.}~\bibnamefont {Thakur}}, \bibinfo {author}
  {\bibfnamefont {J.-B.}\ \bibnamefont {Trebbia}}, \bibinfo {author}
  {\bibfnamefont {A.}~\bibnamefont {Melnikov}}, \bibinfo {author}
  {\bibfnamefont {A.}~\bibnamefont {Buzdin}}, \bibinfo {author} {\bibfnamefont
  {P.}~\bibnamefont {Tamarat}}, \ and\ \bibinfo {author} {\bibfnamefont
  {B.}~\bibnamefont {Lounis}},\ }\href {\doibase 10.1021/acs.nanolett.0c02166}
  {\bibfield  {journal} {\bibinfo  {journal} {Nano Lett.}\ }\textbf {\bibinfo
  {volume} {20}},\ \bibinfo {pages} {6488} (\bibinfo {year}
  {2020})}\BibitemShut {NoStop}%
\bibitem [{\citenamefont {Kremen}\ \emph {et~al.}(2016)\citenamefont {Kremen},
  \citenamefont {Wissberg}, \citenamefont {Haham}, \citenamefont {Persky},
  \citenamefont {Frenkel},\ and\ \citenamefont {Kalisky}}]{kremen-16}%
  \BibitemOpen
  \bibfield  {author} {\bibinfo {author} {\bibfnamefont {A.}~\bibnamefont
  {Kremen}}, \bibinfo {author} {\bibfnamefont {S.}~\bibnamefont {Wissberg}},
  \bibinfo {author} {\bibfnamefont {N.}~\bibnamefont {Haham}}, \bibinfo
  {author} {\bibfnamefont {E.}~\bibnamefont {Persky}}, \bibinfo {author}
  {\bibfnamefont {Y.}~\bibnamefont {Frenkel}}, \ and\ \bibinfo {author}
  {\bibfnamefont {B.}~\bibnamefont {Kalisky}},\ }\href {\doibase
  10.1021/acs.nanolett.5b04444} {\bibfield  {journal} {\bibinfo  {journal}
  {Nano Lett.}\ }\textbf {\bibinfo {volume} {16}},\ \bibinfo {pages} {1626}
  (\bibinfo {year} {2016})}\BibitemShut {NoStop}%
\bibitem [{\citenamefont {Ge}\ \emph {et~al.}(2017)\citenamefont {Ge},
  \citenamefont {Gladilin}, \citenamefont {Tempere}, \citenamefont {Devreese},\
  and\ \citenamefont {Moshchalkov}}]{ge-17}%
  \BibitemOpen
  \bibfield  {author} {\bibinfo {author} {\bibfnamefont {J.-Y.}\ \bibnamefont
  {Ge}}, \bibinfo {author} {\bibfnamefont {V.~N.}\ \bibnamefont {Gladilin}},
  \bibinfo {author} {\bibfnamefont {J.}~\bibnamefont {Tempere}}, \bibinfo
  {author} {\bibfnamefont {J.}~\bibnamefont {Devreese}}, \ and\ \bibinfo
  {author} {\bibfnamefont {V.~V.}\ \bibnamefont {Moshchalkov}},\ }\href
  {\doibase 10.1021/acs.nanolett.7b02180} {\bibfield  {journal} {\bibinfo
  {journal} {Nano Lett.}\ }\textbf {\bibinfo {volume} {17}},\ \bibinfo {pages}
  {5003} (\bibinfo {year} {2017})}\BibitemShut {NoStop}%
\bibitem [{\citenamefont {Kallin}\ and\ \citenamefont
  {Berlinsky}(2016)}]{kallin-16}%
  \BibitemOpen
  \bibfield  {author} {\bibinfo {author} {\bibfnamefont {C.}~\bibnamefont
  {Kallin}}\ and\ \bibinfo {author} {\bibfnamefont {J.}~\bibnamefont
  {Berlinsky}},\ }\href {\doibase 10.1088/0034-4885/79/5/054502} {\bibfield
  {journal} {\bibinfo  {journal} {Rep. Prog. Phys.}\ }\textbf {\bibinfo
  {volume} {79}},\ \bibinfo {pages} {054502} (\bibinfo {year}
  {2016})}\BibitemShut {NoStop}%
\bibitem [{\citenamefont {Liu}\ \emph {et~al.}(2018)\citenamefont {Liu},
  \citenamefont {Chen}, \citenamefont {Zhang}, \citenamefont {Peng},
  \citenamefont {Yan}, \citenamefont {Wen}, \citenamefont {Lou}, \citenamefont
  {Huang}, \citenamefont {Tian}, \citenamefont {Dong}, \citenamefont {Wang},
  \citenamefont {Bao}, \citenamefont {Wang}, \citenamefont {Yin}, \citenamefont
  {Zhao},\ and\ \citenamefont {Feng}}]{liu-18}%
  \BibitemOpen
  \bibfield  {author} {\bibinfo {author} {\bibfnamefont {Q.}~\bibnamefont
  {Liu}}, \bibinfo {author} {\bibfnamefont {C.}~\bibnamefont {Chen}}, \bibinfo
  {author} {\bibfnamefont {T.}~\bibnamefont {Zhang}}, \bibinfo {author}
  {\bibfnamefont {R.}~\bibnamefont {Peng}}, \bibinfo {author} {\bibfnamefont
  {Y.-J.}\ \bibnamefont {Yan}}, \bibinfo {author} {\bibfnamefont {C.-H.-P.}\
  \bibnamefont {Wen}}, \bibinfo {author} {\bibfnamefont {X.}~\bibnamefont
  {Lou}}, \bibinfo {author} {\bibfnamefont {Y.-L.}\ \bibnamefont {Huang}},
  \bibinfo {author} {\bibfnamefont {J.-P.}\ \bibnamefont {Tian}}, \bibinfo
  {author} {\bibfnamefont {X.-L.}\ \bibnamefont {Dong}}, \bibinfo {author}
  {\bibfnamefont {G.-W.}\ \bibnamefont {Wang}}, \bibinfo {author}
  {\bibfnamefont {W.-C.}\ \bibnamefont {Bao}}, \bibinfo {author} {\bibfnamefont
  {Q.-H.}\ \bibnamefont {Wang}}, \bibinfo {author} {\bibfnamefont {Z.-P.}\
  \bibnamefont {Yin}}, \bibinfo {author} {\bibfnamefont {Z.-X.}\ \bibnamefont
  {Zhao}}, \ and\ \bibinfo {author} {\bibfnamefont {D.-L.}\ \bibnamefont
  {Feng}},\ }\href {\doibase 10.1103/PhysRevX.8.041056} {\bibfield  {journal}
  {\bibinfo  {journal} {Phys. Rev. X}\ }\textbf {\bibinfo {volume} {8}},\
  \bibinfo {pages} {041056} (\bibinfo {year} {2018})}\BibitemShut {NoStop}%
\end{thebibliography}%

\end{document}